\documentclass[apjl]{emulateapj}

\usepackage{graphicx,natbib}
\usepackage{amsmath}
\usepackage{apjfonts}

\newcommand{\bootes}{Bo\"otes}
\newcommand{\bootess}{Bo\"otes }

\newcommand{\s}{$\sim$}

\begin{document}

\author{
Richard Beare\altaffilmark{1, 4},
Michael J. I. Brown\altaffilmark{1},
Kevin Pimbblet\altaffilmark{2, 1},
Edward N. Taylor\altaffilmark{3}.
}

\altaffiltext{1}{Monash Centre for Astrophysics, School of Physics and Astronomy, Monash University, Clayton, Victoria 3800, Australia. }
\altaffiltext{2}{E.A.Milne Centre for Astrophysics, University of Hull, Cottingham Road, Kingston-upon-Hull, HU6 7RX, UK.}
\altaffiltext{3}{Centre for Astrophysics and Supercomputing, Swinburne University of Technology, Hawthorn, Victoria 3122, Australia}
\altaffiltext{4}{Email: richard@beares.net}

\slugcomment{Accepted for publication in the Astrophysical Journal}

\date{\today}

\title{EVOLUTION OF THE STELLAR MASS FUNCTION AND INFRARED LUMINOSITY FUNCTION OF GALAXIES SINCE $z = 1.2$}

\shorttitle{SMF EVOLUTION}

\shortauthors{Beare \it {et al.}}

\begin{abstract}

We measured evolution of the $K$-band luminosity function and stellar mass function for red and blue galaxies  at $z<1.2$ using a sample of 353\,594 $I<24$ galaxies in 8.26 square degrees of \bootes.  We addressed several sources of systematic and random error in measurements of total galaxy light, photometric redshift and absolute magnitude. We have found that the $K$-band luminosity density for both red and blue galaxies increased by a factor of 1.2 from $z\sim1.1$ to $z\sim0.3$, while the most luminous red (blue) galaxies decreased in luminosity by 0.19\;(0.33) mag or $\times0.83\;(0.74)$. These results are consistent with $z<0.2$ studies while our large sample size and area result in smaller Poisson and cosmic variance uncertainties than most $z >0.4$ luminosity and mass function measurements. Using an evolving relation for $K$-band mass to light ratios as a function of  $(B-V)$ color, we found  a slowly decreasing rate of growth in red galaxy stellar mass density of $\times2.3$ from $z\sim1.1$ to $z\sim0.3$, indicating a slowly decreasing rate of migration from the blue cloud to the red sequence. Unlike some studies of the stellar mass function, we find that massive red galaxies grow by a factor of $\times1.7$ from $z\sim1.1$ to $z\sim0.3$, with the rate of growth due to mergers decreasing with time. These results are comparable with measurements of merger rates and clustering, and they are also consistent with the red galaxy stellar mass growth implied by comparing $K$-band luminosity evolution with the fading of passive stellar population models.
\end{abstract}


\section{INTRODUCTION}
\label{sec:intro}

Measurements of the optical and infrared luminosity functions (LFs) and the stellar mass function (SMF) at different redshifts provide important observational tests of large scale simulations of galaxy formation and evolution. Such simulations can be either cosmological hydrodynamical models, which attempt to model the detailed physical and chemical processes involved in galaxy formation, such as ILLUSTRIS \citep[][]{vogel14} and EAGLE \citep[][]{schay15}, or semi-analytic models \citep[SAMs, e.g.][]{ croto06, guo08, lacey16}. SAMs include simple empirical representations of physical processes and these are `added onto' the dark matter merger trees resulting from hierarchical N-body simulations \citep[e.g. the Millennium Simulation,][]{sprin05}. Both types of simulation make predictions of LF and SMF evolution that can be tested against observational measurements. Discrepancies between observations and simulations then provide the motivation for refining the models incorporated in the simulations.  In this way, measurements of optical and infrared LF evolution have in the past motivated significant improvements in our understanding of the physical processes occurring in galaxy formation and evolution \citep[e.g.][]{croto06, lacey16}.  

As benchmarks for testing and calibrating simulations, LFs have the advantage over SMFs that they have only limited model dependencies. SMFs require the determination of stellar masses from photometry and this involves use of a number of physical models, notably  stellar population synthesis (SPS) models \citep[e.g.][]{fioc97, bruzu03, maras05}, stellar initial mass functions \citep[IMFs; e.g.][]{salpe55, kenni83, chabr03}, and dust attenuation laws \citep[e.g.][]{calze00}. SMFs are therefore subject to significant model uncertainties in addition to the observational uncertainties inherent in LFs.

Table \ref{tab:LF_literature} and Table \ref{tab:SMF_literature} (respectively) summarise several recent measurements of the near-infrared LF and the SMF and their evolution.  Note that different studies have differentiated quiescent and star forming galaxies in different ways, e.g. by color, morphology or emission line strengths. Also included in these tables are low redshift ($z \leq 0.2$) studies that provide an accurate low redshift ``anchor'' for evolutionary studies.

A number of the studies in Table \ref{tab:SMF_literature} derived SMFs by fitting theoretical stellar population synthesis (SPS) models to available photometry, e.g. using the \textit{kcorrect} software of \citet{blant07} which fits observed photometry with combinations of five template SEDs that are derived from several hundred SPS models. Others used stellar mass to light $M/L$ ratios given as  a function of observed color \citep[e.g.][]{bell03, taylo11} or redshift for red/quiescent and blue/active galaxies \citep[e.g.][]{arnou07}.  It should be noted that empirical $M/L$ ratios are also derived with the aid of SPS models.

\begin{deluxetable*}{ccccccccc}								
\tablewidth{0pt}								
\tablecolumns{9}								
\tabletypesize{\scriptsize}								
\tablecaption {Some previous measurements of the near infrared luminosity function and its evolution}								
\tablehead{								
\colhead{reference} & 	\colhead{surveys} & 	\colhead{approx.} & 	\colhead{redshift} & 	\colhead{LF} & 	\colhead{approx} & 	\colhead{sample} & 	\colhead{approx.} & 	\colhead{subsamples} \tabularnewline 
\colhead{} & 	\colhead{used} & 	\colhead{redshift} & 	\colhead{type\tablenotemark{a}} & 	\colhead{wavebands} & 	\colhead{faint} & 	\colhead{size} & 	\colhead{sample area} & 	\colhead{} \tabularnewline 
\colhead{} & 	\colhead{} & 	\colhead{range} & 	\colhead{(s or p)} & 	\colhead{} & 	\colhead{limit (AB)\tablenotemark{b}} & 	\colhead{} & 	\colhead{(deg$^2$)} & 	\colhead{} 
}								
 \startdata															
\multicolumn{9}{l}{LOW REDSHIFT STUDIES}\\[4pt]														
\citet{loved00} & 	Cerro Tololo 1.5 m & 	$<0.04$ & 	s &	$K$ &	$K=13.8$ &	$345$ & 	$4\,270$ & 	ELG/non ELG\tablenotemark{c} \\[4pt]
\citet{kocha01} & 	2MASS & 	$<0.04$ & 	s &	$K$ &	$K=13.1$ &	$3\,878$ & 	$6\,960$ & 	morphology \\[4pt]
\citet{cole01} & 	2MASS, 2dFGRS & 	$<0.04$ & 	s &	$J, K$ &	$K=15$ &	$5\,683$ & 	$619$ & 	- \\[4pt]
\citet{bell03} & 	2MASS, SDSS & 	$<z>=0.078$ & 	s &	$ugrizK$ &	$K=15.3$ &	$6\,282$ & 	$414$ & 	morphology, \\
 & 	 & 	 & 	&	&	&	 & 	 & 	color \\[4pt]
\citet{huang03} & 	Hawaii+AAO & 	$<z>=0.136$ & 	s &	$K$ &	$K=16.8$ &	$1\,056$ & 	$8.22$ & 	morphology \\
 & 	K-band GRS & 	 & 	&	&	&	 & 	 & 	 \\[4pt]
\citet{eke05} & 	2MASS, 2dFGRS & 	$<0.12$ & 	s &	$J, K$ &	$K=15.2$ &	$15\,664$ & 	- & 	cluster or \\
 & 	 & 	 & 	&	&	&	 & 	 & 	group size \\[4pt]
\citet{jones06} & 	6dFGS & 	$0.054$ & 	s &	$b_JrJHK$ &	$K=14.6$ &	$60\,869$ & 	$9\,075$ & 	- \\[4pt]
\cite{dever09} & 	2MASS & 	$<0.01$ & 	s &	$K$ &	$K=11.8$ &	$1\,613$ & 	$\sim15\,000$ & 	morphology \\[4pt]
\citet{smith09} & 	UKIDSS, SDSS & 	$0.01<z<0.3$ & 	s &	$r, K$ &	$K=17.8$ &	$40\,111$ & 	$619$ & 	color \\[4pt]
\citet{hill10} & 	 MGC, UKIDSS, SDSS & 	$<0.1$ & 	s &	$ugrizYJHK$ &	$K=17.5$ & 	$1\,785$ & 	$28$ & 	- \\[4pt]
\citet{drive12} & 	GAMA, GALEX, & 	$<0.1$ & 	s &	FUV, NUV, &	$K=19.9$ & 	$7\,638$ & 	$125$ & 	morphology \\
 & 	SDSS, UKIDSS & 	 & 	&	$ugrizYJHK$ &	&	 & 	 & 	 \\[4pt]
\citet{kelvi14} & 	GAMA & 	$0.025<z<0.06$ & 	s &	$ugrizYJHK$ &	$r=19.4$ & 	$3\,727$ & 	$144$ & 	morphology \\[4pt]
\citet{bonne15} & 	2MASS & 	$<0.05$ & 	p &	$K$ &	$K=12.6$ &	$13\,325$ & 	all but & 	morphology, \\
 & 	 & 	 & 	&	&	&	 & 	GP\tablenotemark{d} & 	color \\[2pt]							
\multicolumn{9}{l}{STUDIES OF LF EVOLUTION}\\[4pt]													
\citet{drory03} & 	MUNICS & 	$0.4<z<1.2$ & 	p &	$K$ &	$K=21.3$ &	$\sim5\,000$ & 	$0.28$ & 	- \\[4pt]
\citet{pozze03} & 	K20 & 	$z=0.5, 1.0, 1.5$ & 	s &	$J, K$ &	$K=21.8$ &	$489$ & 	$0.014$ & 	ELG/non ELG\tablenotemark{c} \\
 & 	 & 	 & 	&	&	&	 & 	 & 	color \\[4pt]
\citet{ciras07} & 	UKIDSS UDS & 	$0.25<z<2.25$ & 	p &	$K$ &	$K=22.5$ &	$22\,000$ & 	$0.6$ & 	color \\[4pt]
\citet{arnou07} & 	SWIRE, VVDS, & 	$0.2<z<2.0$ & 	p &	$K$ &	$K=22.0$ &	$21\,200$ & 	$0.76$ & 	SED fits \\
 & 	CFHTLS, UKIDSS UDS & 	 & 	&	&	&	 & 	 & 	 \\[4pt]
\citet{ciras10} & 	UKIDSS UDS, SXDS & 	$0.2<z<4.0$ & 	p &	$K$ &	$K=23$ &	$\sim50\,000$ & 	$0.7$ & 	- \\[4pt]
\citet{mortl17} & 	UltraVISTA, & 	$0.25<z<3.75$ & 	p &	$K$ &	$K=22.8$ &	$88\,484$ & 	$1.0$ & 	- \\
 & 	CANDELS, HUDF & 	 & 	&	&	&	 & 	 & 	 \\[4pt]
This work & 	NDWFS, SDWFS,& 	$0.2 < z < 1.2$ & 	p &		$K$  &       $I=24,$ &    $359 \, 802$ & 8.26 & color\\
& NEWFIRM	 & 	&   &	   &   $[3.6 \mu\textrm{m}]$  &	 &      & 		 \\
& 	 & 	&   &	   &   $ = 23.3$  &	&	 & 	
\enddata								
\label{tab:LF_literature}
\tablenotetext{a}{\footnotesize{Spectroscopic (s) or photometric (p) redshifts.}}	
\tablenotetext{b}{\footnotesize{These faint limits are only intended to provide approximate depth comparisons between different surveys, as different authors quote survey depths to different completeness (typically $5\sigma$).}}
\tablenotetext{c}{\footnotesize{ELG = emission line galaxy.}}
\tablenotetext{d}{\footnotesize{GP = Galactic plane.}}					
\end{deluxetable*}

\begin{deluxetable*}{ccccccccc}								
\tablewidth{0pt}								
\tablecolumns{9}								
\tabletypesize{\scriptsize}								
\tablecaption {Some previous measurements of the stellar mass function and its evolution}								
\tablehead{								
\colhead{reference} & 	\colhead{surveys} & 	\colhead{approx.} & 	\colhead{redshift} & 	\colhead{approx} & 	\colhead{sample} & 	\colhead{approx.} & 	\colhead{subsamples} & 	\colhead{stellar} \tabularnewline 
\colhead{} & 	\colhead{used} & 	\colhead{redshift} & 	\colhead{type\tablenotemark{a}} & 	\colhead{faint} & 	\colhead{size} & 	\colhead{sample area} & 	\colhead{} & 	\colhead{mass} \tabularnewline 
\colhead{} & 	\colhead{} & 	\colhead{range} & 	\colhead{(s or p)} & 	\colhead{limit (AB)\tablenotemark{b}} & 	\colhead{} & 	\colhead{(deg$^2$)} & 	\colhead{} & 	\colhead{based on} 
}								
 \startdata																
\multicolumn{9}{l}{LOW REDSHIFT STUDIES}\\[4pt]																
\citet{cole01} & 	2MASS, 2dFGRS & 	$<0.04$ & 	s &	$K=13.1$ &	$5\,683$ & 	$6\,19$ & 	- & 	SED fit \\[4pt]
\citet{bell03} & 	2MASS, SDSS & 	$<z>=0.078$ & 	s &	$r=17.5$, &	$18\,714$ & 	$414$ & 	concn. index, & 	$M/L_g$, \\
 & 	 & 	 & 	&	$K=13.5$ &	 & 	 & 	color & 	$M/L_K$ \\[4pt]
\citet{eke05} & 	2MASS, 2dFGRS & 	$<0.12$ & 	s &	$K=15.2$ &	$15\,664$ & 	- & 	cluster or & 	SED fit \\
 & 	 & 	 & 	&	&	 & 	 & 	group size & 	 \\[4pt]
\citet{li09} & 	SDSS & 	$0.001<z<0.5$ & 	s &	$r=17.6$ &	$486\,840$ & 	$6\,437$ & 	- & 	kcorrect \\[4pt]
\citet{smith09} & 	UKIDSS, SDSS & 	$0.01<z<0.3$ & 	s &	$K=20.1$ &	$40\,111$ & 	$6\,19$ & 	color & 	kcorrect \\[4pt]
\citet{baldr12} & 	GAMA & 	$0.002<z<0.06$ & 	s &	$r=19.8$ &	$5\,210$ & 	$143$ & 	- & 	$M/L_i$ \\
 & 	 & 	 & 	&	&	 & 	 & 	 & 	 \\[4pt]
\multicolumn{9}{l}{STUDIES OF SMF EVOLUTION}\\[4pt]
\citet{drory05} & 	GOODS-S, FORS & 	$<0.5$ & p &	$I=26.8$, &	$5\,557$, & 	$0.025$ & 	- & 	SED fit \\
 & 	 & 	 & 	&	$K=25.4$ &	$3\,367$ & 	 & 	 & 	 \\[4pt]
\citet{arnou07} & 	SWIRE, VVDS, & 	$0.2<z<2.0$ & 	p &	$[3.6 \mu\textrm{m}]$ &	$\sim21\,200$ & 	$0.76$ & 	SED fits & 	$M/L_K$ \\
 & 	CFHTLS, & 	 & 	&	$=21.5$ &	 & 	 & 	 & 	 \\
  & 	UKIDSS UDS & 	 & 	&	&	 & 	 & 	 & 	 \\[4pt]
\citet{bundy06} & 	DEEP2 & 	$0.4<z<1.4$ & 	s &	$K=21.5$ &	$\sim8\,000$ & 	$1.5$ & 	morphology, & 	$M/L_K$ \\
 & 	Palomar NIR & 	 & 	&	&	 & 	 & 	color, ELW & 	 \\[4pt]
\citet{borch06} & 	COMBO17 & 	$0.2<z<1.0$ & 	p &	$R=24$ &	$\sim25\,000$ & 	$0.78$ & 	color & 	$M/L$ \\
 & 	 & 	 & 	&	&	 & 	 & 	 & 	by SED fit \\[4pt]
\citet{perez08} & 	IRAC, MIPS, & 	$0.0<z<4.0$ & 	p &	$[3.6 \mu\textrm{m}]$ &	$28\,000$ & 	$0.18$ & 	SFR & 	SED fit \\
 & 	Subaru, other & 	 & 	&	$=23.4$ &	 & 	 & 	 & 	$\textrm{for}\,[3.6 \mu\textrm{m}]$ \\
 & 	optical + IR & 	 & 	&	&	 & 	 & 	 & 	 \\[4pt]
 \citet{drory09} & 	COSMOS & 	$0.2<z<1.0$ & p &	$i^+=25$, &	$138\,001$ & 	$1.73$ & 	SED fits, & 	SED fit \\
 & 	 & 	 & 	&	$K=24$ &  & 	 & color	 & 	 \\[4pt]
\citet{ilber10} & 	COSMOS & 	$0.2<z<2.0$ & 	p &	$[3.6 \mu\textrm{m}]$ &	$196\,000$ & 	$2.3$ & 	morphology, & 	$M/L$ \\
 & 	 & 	 & 	&	$=23.9$ &	 & 	 & 	color & 	by SED fit \\[4pt]
\citet{bramm11} & 	NEWFIRM MBS & 	$0.4<z<2.2$ & 	p &	$K=22.8$ &	$\sim27\,000$ & 	$0.39$ & 	color & 	SED fit \\[4pt]
\citet{gonza11} & 	IRAC, HST ACS, & 	$4<z<7$ & 	p &	$J\sim28$ &	$437$ & 	0.011 & 	- & 	SED fit \\
 & 	HST WFC3/IR & 	 & 	&	&	 & 	 & 	 & 	 \\[4pt]
\citet{mortl11} & 	HST NICMOS & 	$1.0<z<3.5$ & 	p &	$H=26.8$ &	$8\,298$ & 	$0.73$ & 	color, SFR & 	SED fit \\[4pt]
\citet{ilber13} & 	UltraVISTA & 	$0.2<z<4.0$ & 	p &	$K=24$ &	$220\,000$ & 	$1.52$ & 	 color, SFR  & 	SED fit \\[4pt]
\citet{moust13} & 	PRIMUS (+SDSS) & 	$0<z<1.0$ & 	s &	$i=23$ &	$40\,430$ & 	$5.5$ & 	SFR & 	SED fit \\[4pt]
\citet{muzzi13b} & 	UltraVISTA & 	$0<z<4.0$ & 	p &	$K_s=23.4$ &	$26\,000$ & 	$1.62$ & 	color & 	SED fit \\[4pt]
\citet{maras13} & 	BOSS & 	$0.2<z<0.7$ & 	s &	$r=17.6$ &	$\sim400\,000$ & 	$3275$ & 	- & 	SED fit \\
 & 	 & 	 & 	&	&	(massive only) & 	 & 	 & 	 \\[4pt]
\citet{david13} & 	VIPERS & 	$0.5<z<1.3$ & 	s &	$i=22.5$ &	$53\,608$ & 	$10.31$ & 	color & 	SED fit \\
 & 	(+CFHT, GALEX) & 	 & 	&	&	 & 	 & 	 & 	 \\[4pt]
\citet{tomcz14} & 	ZFOURGE/ & 	$0.2<z<3.0$ & 	p &	$K\sim25$ &	$76\,505$ & 	$0.09$ & 	color & 	SED fit \\
 & 	CANDELS & 	 & 	&	&	 & 	 & 	 & 	 \\[4pt]
This work & 	NDWFS, SDWFS, & 	$0.2 < z < 1.2$ & 	p &       $I=24,$ &    $359 \, 802$ & 8.26 & color& 	$M/L_K$\\
& NEWFIRM	 & 	&   &	$[3.6 \mu\textrm{m}]$  &	&	 &      & 		 \\
& 	 & 	&   &	$ = 23.3$  &	&	 &      & 
\enddata								
\label{tab:SMF_literature}	
\tablenotetext{a}{\footnotesize{Spectroscopic (s) or photometric (p) redshifts.}}
\tablenotetext{b}{\footnotesize{These faint limits are only intended to provide approximate depth comparisons between different surveys, as different authors quote survey depths to different completeness (typically $5\sigma$).}}							
\end{deluxetable*}

Near infrared (especially $K$-band) $M/L$ ratios have often been preferred to optical ones for determining stellar masses because they are a much weaker function of stellar population color than optical ones. For example, \citet{bell01} found that $K$-band $M/L$ ratios amongst spiral galaxies varied by a factor of $\sim2$ while those in the $B$-band varied by a factor of $\sim7$. In this work we base our  measurements of SMF evolution on stellar masses calculated using $M/L_K$ given as a function of $(B-V)$ color.																																																	
Both the LF and the SMF can be approximated by \citet{schec76} functions, although the faint end power-law index is poorly constrained when survey magnitude limits are close to the characteristic magnitude or log stellar mass ($M_K^*$ or $\log_{10} M^*$ respectively). Given the steep faint-end slope for star forming galaxies, uncertainty in the power-law index has a significant impact on estimates of the luminosity and mass densities of these galaxies.

There is general agreement in the literature that the evolution of quiescent stellar mass density (SMD) has been more rapid than that of star-forming SMD, \citep[e.g.][]{bramm11, muzzi13b,  tomcz14, moust13}. There is  agreement that the rate of increase of the space density of quiescent galaxies has depended strongly on galaxy mass, with smaller galaxies increasing more rapidly in numbers than larger ones \citep[e.g.][]{muzzi13b}. The  most massive quiescent galaxies have increased relatively slowly in stellar mass, implying that most of their stellar mass was already in place at higher redshift with very little stellar mass being added since \citep[e.g.][]{brown07, moust13, muzzi13b}.

In this study we have provided precise  measurements of the evolution of the $K$-band LF and the galaxy SMF between $z = 0.2$ and $z = 1.2$. Our very large sample of  353\,594 galaxies spanning an area of 8.26 deg$^2$ in \bootess enabled us to significantly reduce Poisson errors and the effects of cosmic variance. The significant depth $I = 24.0$ of our survey allowed us to measure precisely the evolution of red and blue galaxies, both combined and separately, from $z = 1.2$.  For this work we used the low redshift \citet{kocha01} LF and the \citet{cole01} SMF as $z\sim0$ reference points.

We have employed the same data and similar methods to those that \citet{beare15} used to measure optical $B$-band LF evolution over the redshift range $0.2 < z < 1.2$.  We used the same optical and near-infrared photometry, our apparent magnitudes and colors were based on the same measurements of total flux, and we used the same method of calculating absolute magnitudes \citep{beare14}. For this paper we have used improved  photometric redshifts, employing the Bayesian EAZY code to match our photometry to the 129 empirical galaxy templates of \citet {brown14}. 

The structure of this paper is as follows. Section \ref{sec:data}  describes our imaging and catalogs, Section \ref{sec:photoz} our photometric redshifts, Sections \ref{sec:absmags} and \ref{sec:masses} how we measured galaxy luminosities and stellar masses, and Section \ref{sec:sample} our sample selection.  Sections \ref{sec:KLF} and \ref{sec:M} describe how we measured evolution of the $K$-band LF and the SMF. Sections \ref{sec:K_results} and \ref{sec:M_results} present our results for evolution of $K$-band luminosities and stellar masses and discuss their significance. Finally we summarise our work and conclusions in Section \ref{sec:summary}.

Our results were determined assuming a cosmology\footnote{Conversions to other cosmologies can be made as described in \citet{croto13}.} with $\Omega_0 = 0.3$, $\Omega_{\textrm{k}}=0$, $H_0=70 \,\, \textrm{km s}^{-1} \,  \textrm{Mpc}^{-1}$ which is similar to that measured by WMAP  \citep{benne13}, and presented using AB-based magnitudes and units in which $h_{70} = H_0 /70$.

\newpage

\section{Imaging and catalogs}
\label{sec:data}

The images, catalogs and photometry were identical to those used in \citet{beare15}. They provided a very large sample of 353\,594 galaxies covering a substantial area of 8.26 deg$^2$ in \bootes, surveyed to a depth of $I = 24.0$. Our sample was an excellent one for measuring LF and SMF evolution because it was deep enough to provide precise photometry out to $z = 1.2$. and because its large size minimised random (Poisson) error and the effects of cosmic variance. Furthermore, we were able to utilise photometry in 13 optical and near infrared wavebands, and this enabled us to obtain precise photometric redshifts, to calculate precise K-corrections \citep{beare14} for determining restframe magnitudes and colors, and to apply cuts to exclude stars and AGN. Below we  provide only a brief summary of our data. We refer the reader to  \citet{beare15} for a more thorough description.

We used an update of the galaxy catalog produced by \citet{brown07} which includes additional images, photometry and minor refinements to the photometry code. \citet{brown07}  detected sources in the \bootess field using SExtractor 2.3.2 \citep{berti96} run on $I$-band images from the NWDFS third data release.   They removed regions surrounding very extended galaxies and saturated stars in order to minimize contamination and the final sample covered an area of 8.26 deg\({}^2\) over a \( 2.9^\circ \times 3.6^\circ \) field of view. 

Optical photometry was based on $B_WRI$-band imaging from the NOAO Deep Wide Field Survey \citep[NDWFS,][]{jannu99}. Near-infrared photometry was derived from $J$, $H$ and $K_s$-band imaging from the NEWFIRM  \bootess Imaging Survey (Gonzalez et al. 2011, in prep), $u$ and $y$-band images were from the $2 \times 8.4$ m Large Binocular Telescope \citep[LBT;][]{bian13}, $z$-band data were from the 8.2 m Subaru Telescope \citep{miyaz12}, and 3.6, 4.5, 5.8 and 8.0 $\mu$m imaging was from the IRAC camera of the Spitzer Deep Wide Field Survey \citep[SDWFS;][]{ashby09, eisen08}.  

We used the method described in \citet{beare15} to measure the total flux from each galaxy. This employs magnitude dependent aperture sizes and applies corrections based on growth curves of apparent magnitude with aperture size to precisely account for flux falling outside the photometric aperture.

\begin{figure}
 	\centering
		\includegraphics[width=0.49\textwidth]{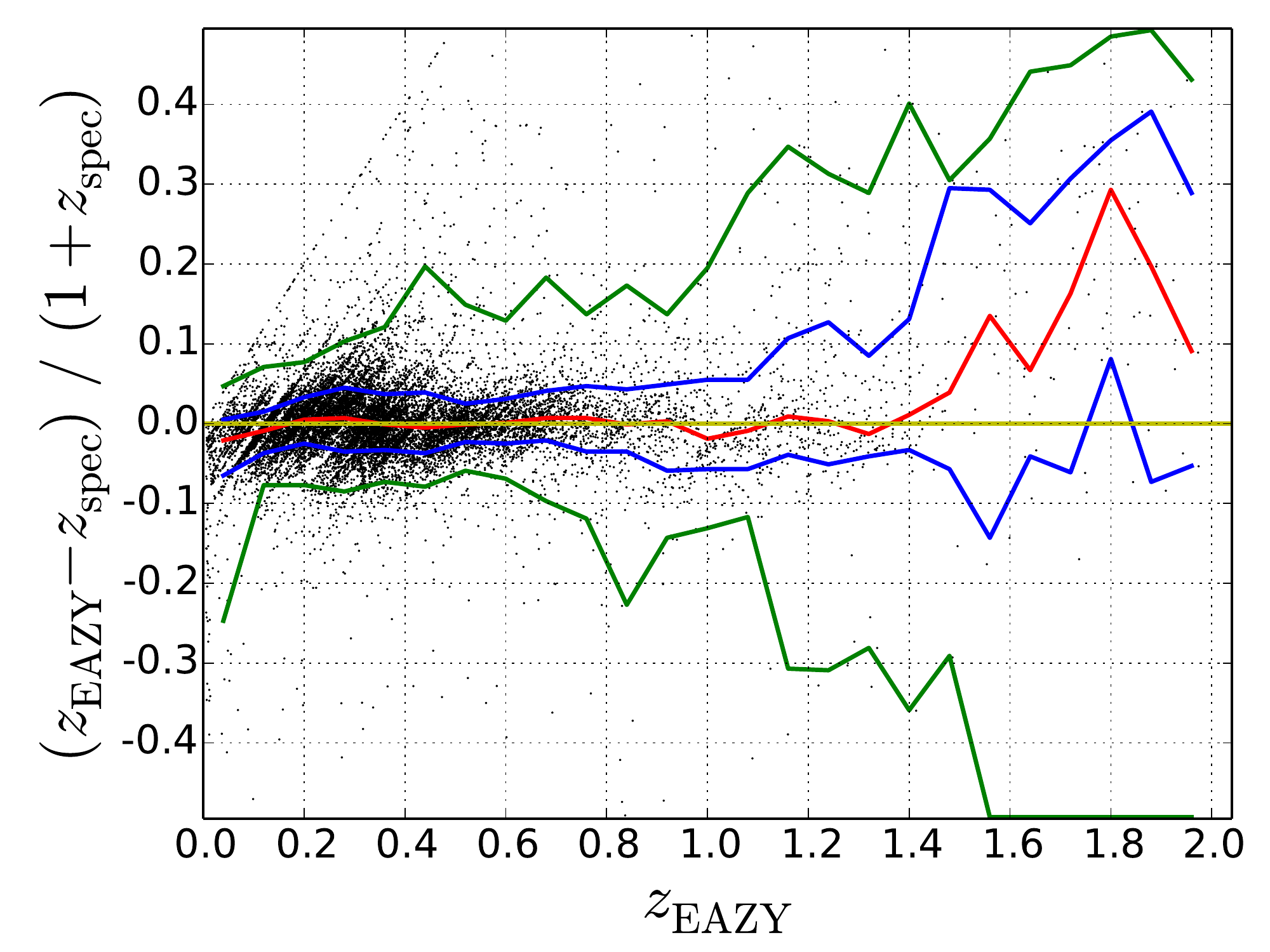}
		\caption{Fractional errors $(z_{\rm{phot}}-z_{\rm{spec}}) \, / \, (1 + z_{\rm{phot}})$ in our Bayesian photometric redshifts as calculated using EAZY. The systematic (median) error is shown by the red line. 68\% of errors lie between the blue 1-$\sigma$ lines and 95\% of errors between the green 2-$\sigma$ lines. Across the whole redshift range random errors are less than \s0.05, while systematic errors are less than 0.01 at $0.2 < z_{\rm{phot}} < 1.0$ and less than 0.02 at $1.0 < z_{\rm{phot}} < 1.2$.}
		\label{fig:photoz}
\end{figure}

\section{Determining photometric redshifts}
\label{sec:photoz}

Measurements of galaxy distances, and hence  restframe magnitudes, rely on accurate redshifts. For the \bootess field we had to rely on photometric redshifts for the vast majority of our galaxies, as spectroscopic redshifts were only available for 3.4\% of galaxies in the redshift range $0.2 \leq z  \leq 1.2$ (12\,191 in all). Our photometric redshifts were determined using the Bayesian EAZY code \citep{bramm08, taylo09} to model photometry in 13 optical to near-infrared wavebands using the 129 empirical template SEDs from \citet{brown14}. Figure \ref{fig:photoz} shows that our photometric redshifts have $(z_{\textrm{phot}}-z_{\textrm{spec}}) \, / \, (1 + z_{\textrm{specz}})$ systematic errors of less than 0.01 at $z_{\rm{phot}} < 1.0$ while at $1.0 < z_{\rm{phot}} < 1.2$ the systematic errors are less than 0.02. Our 1-$\sigma$ random errors were less than $\sim0.05$ over the whole of our redshift range. The percentage of catastrophic errors,  defined using the $\vert{z_{\textrm{phot}}-z_{\textrm{spec}}\vert \, / \, (1 + z_{\textrm{spec}}}) > 0.15$ criterion of  \citet[][]{ilber13} was 1.50 (0.75, 2.17)\% for all (red, blue) galaxies.

We chose to use Bayesian photometric redshifts because these do not exhibit the significant aliasing or bunching at specific redshifts seen in photometric redshifts based on frequentist least-squares fitting \citep[as in][]{beare15}. Both sets of photometric redshifts exhibit similar systematic and random errors but the even distribution of the Bayesian redshifts ensures that bunching around preferred values does not affect the numbers of galaxies allocated to different redshift bins.

When available we used spectroscopic redshifts in preference to photometric redshifts. These were mainly from the AGN and Galaxy Evolution Survey (AGES, Kochanek et al. 2012), with several hundred additional redshifts from SDSS and a variety of programmes with the Gemini, Keck and Kitt Peak National Observatory telescopes.

\section{Measuring absolute magnitudes}
\label{sec:absmags}

\begin{figure}
 	\centering
		\includegraphics[width=0.49\textwidth]{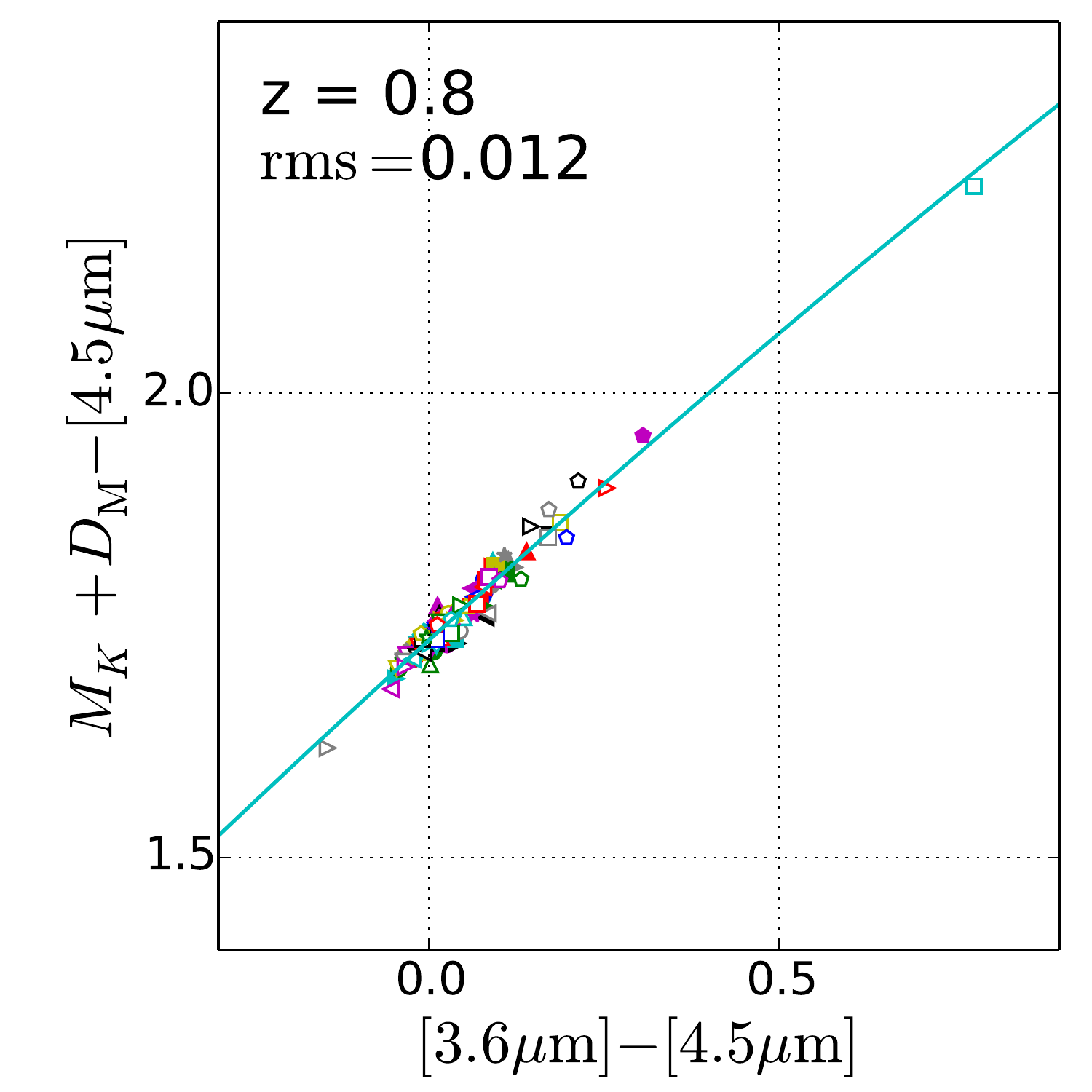}
		\caption{Example plot showing how we determined absolute magnitudes from observed colors using the method of \citet{beare14}. The colored markers plot computed values of ${(M_K + D_{\rm{M}}) -  [4.5 \mu \textrm{m}]}$  against ${[3.6 \mu \textrm{m}] - [4.5 \mu \textrm{m}])}$ for the 129 template SEDs from \citet{brown14} at $z=0.8$. $ D_{\rm{M}}$ is the distance modulus.  The curve is the best fit second order polynomial to the template data points and enables absolute magnitudes $M_K$ to be determined from apparent $[3.6 \mu \textrm{m}]$ and $[4.5 \mu \textrm{m}]$ magnitudes. The RMS offset from the template points is shown in the top left corner. Outliers offset by more than 0.2 mag from the polynomials were excluded from the polynomial fitting.}
		\label{fig:calibration_example}
\end{figure}

In order to measure evolution of the $K$-band LF we first used the method of \citet{beare14} to determine the absolute magnitudes of our galaxies. This method enables the absolute magnitude $M_W$ in a waveband $W$ to be precisely measured using a single, carefully chosen, observed color  $(m_Y - m_Z)$ and apparent magnitude $m_Z$.  As Figure \ref{fig:calibration_example} shows, $M_W$ is determined from the second degree polynomial best fit to a plot of $(M_W + D_M) - m_Z$ against $(m_Y - m_Z)$ for the 129 template galaxies of \citet{brown14}, $D_M$ being the distance modulus: 
\begin{equation}
\label{eq:absmag_polynomial}
	(M_W + D_M) - m_Z = a (m_Y - m_Z)^2 + b (m_Y - m_Z) + c.
\end{equation}

Table \ref{tab:observed_colors} lists the observed colors we used at different redshifts to determine absolute magnitudes in different restframe wavebands. The method  allows precise determination of the uncertainties due to photometric error, redshift error and intrinsic galaxy variability.  (The $Y$ and $Z$-band filter transmission functions we used took account of atmospheric absorption but the $W$-band restframe filter did not.) 

We used the same method to determine absolute magnitudes in the $B$ and $V$-bands as these were needed for calculating galaxy masses based on stellar mass to light ratios given as a function of restframe  $(B-V)$ color.

The template galaxy substantially redder in $([3.6 \mu m]-[4.5 \mu m])$ color than the others in Figure \ref{fig:calibration_example} is the ultra-luminous infrared galaxy UGC 5101. This exhibits substantial emission from hot (\s a few 100 K) dust in the near-infrared, indicating the presence of a powerful AGN. This outlier breaks the assumption that infrared light from galaxies is always dominated by stellar emission. However, it does in fact play a useful role in ``anchoring'' the polynomial used to determine absolute $K$-band magnitudes, preventing it being poorly constrained.

\section{Measuring galaxy stellar masses}
\label{sec:masses}

\begin{figure}
 	\centering
	\includegraphics[width=0.49\textwidth]{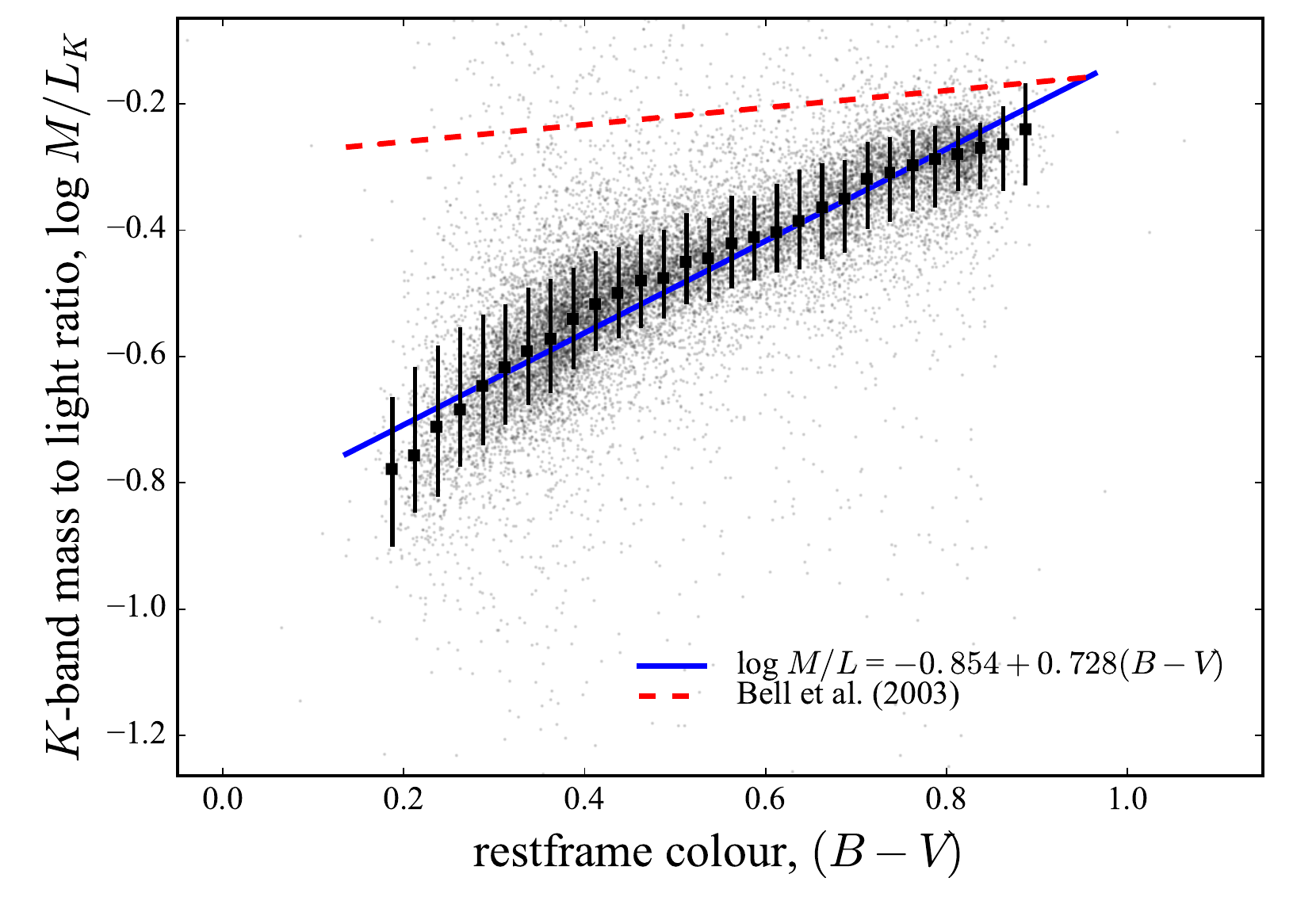}
	\caption{Stellar mass to light ratios for $z < 0.4$ GAMA galaxies. The grey points represent individual galaxies in the GAMA sample. Biweight means for $\log_{10}M/L_K$  in narrow bins of  $(M_B-M_V)$ restframe color are shown by the large data points with error bars.  The blue line is a best fit to these means as a function of color. The dashed red line shows the relationship from \citet{bell03} which overestimates galaxy masses relative to the more recent relationship based on GAMA galaxies.}
	\label{fig:MLK_GAMA}
\end{figure}

\citet{taylo11} showed that optical and near infrared stellar mass to light ratios ($M/L$) can be determined to within 0.2 dex using a linear function of a single restframe color. For example, Figure \ref{fig:MLK_GAMA} for GAMA galaxies shows how $M/L_K$ can be determined from restframe $(B-V)$ color using the following relation (in Solar units):
\begin{equation}
\label{eq:ML_GAMA}
	\log_{10}(M/L_K) = -0.854 + 0.728(M_B - M_V).
\end{equation}
Determining stellar masses from color dependent mass to light ratios is simpler than determining them from detailed SPS modelling of individual galaxies. However, it must be remembered that relationships such as Equation \eqref{eq:ML_GAMA} are derived by averaging the results of detailed SPS modelling and consequently suffer from similar problems while at the same time increasing uncertainty as a result of the loss of detailed color information.

The stellar masses in Figure \ref{fig:MLK_GAMA} were derived using \citet{bruzu03} SPS models fitted to $ugrizYJHK$ optical to near infrared photometry. The advantage of a relation such as Equation \eqref{eq:ML_GAMA} over full SED fitting is that photometry is required in many fewer wavebands - in this case just those required to compute the absolute magnitudes $M_B$ and $M_V$ using K-corrections (Equation \ref{eq:absmag_polynomial}).

Galaxies in the $z < 0.4$ GAMA survey have redshifts with a median value of 0.2 \citep{taylo11} and evolution of the relation Equation \eqref{eq:ML_GAMA} is not apparent in the data. However our data extend to $z = 1.2$ and it cannot be assumed that evolution of $\log_{10}(M/L_K)$ as a function of restframe $(B-V)$ is not significant over this larger redshift range. Indeed the SFHs of the stellar populations that lead to Equation \eqref{eq:ML_GAMA}  at $z \sim 0.2$ may not be the same as those of a comparable sample at $z \sim 1.2$ so some evolution in Equation \eqref{eq:ML_GAMA} is to be expected.

We therefore endeavoured to measure evolution of the dependence of $\log_{10}(M/L_K)$ on $(B-V)$ using the G10 catalog, which contains consistent total flux measurements across 38 far-UV to far-IR for sources in a ~1 deg$^2$ subset of the COSMOS region \citep{davie15a, andre17}. The G10 catalog employed the LAMBDAR code \citep{wrigh16} to largely eliminate systematic error arising from the different flux measurements and reduction methods used by constituent COSMOS surveys. It aimed to produce a catalog extending to $z = 1$ which was consistent with the low redshift ($z < 0.4$) spectroscopic GAMA survey \citep{drive11, liske15}.

\begin{figure}
 	\centering
		\includegraphics[width=0.49\textwidth]{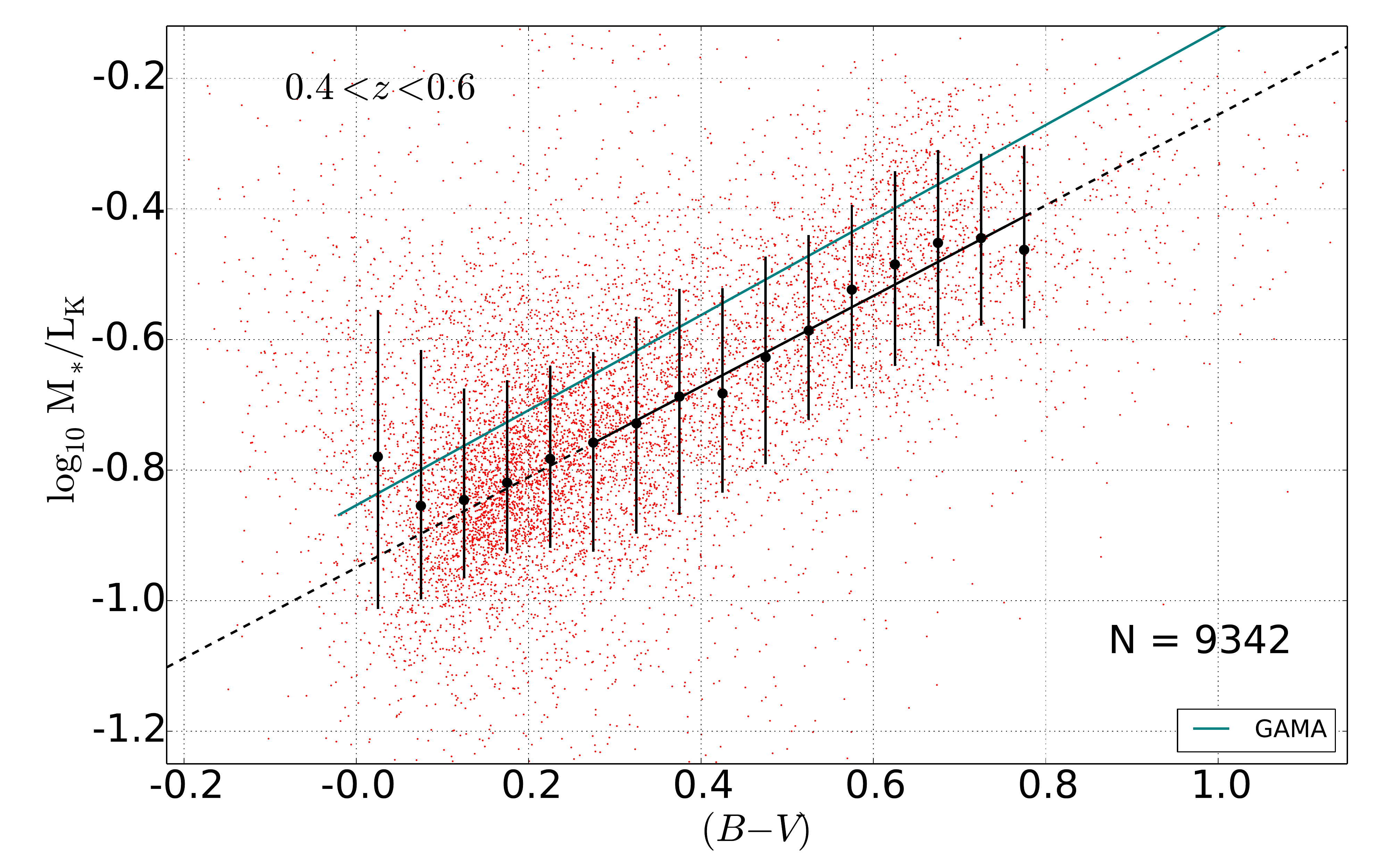}
		\caption{Example plot for G10 galaxies of $\log_{10}(M/L_K)$ against restframe $(B-V)$ color for one redshift bin. Galaxies are shown by red points. Median values of $\log_{10}(M/L_K)$ in color bins of width 0.5 mag are shown by filled circles and the 1-$\sigma$ errors by vertical bars. The linear best fit to the median $\log_{10}(M/L_K)$ values for colors between 0.3 and 0.7 is shown by the thick slanting line (extended beyond this range by  dashed line). For comparison, the thin sloping green line shows the non-evolving $z < 0.4$ GAMA relationship Equation \eqref{eq:ML_GAMA}.}
		\label{fig:logmass-colour_example}
\end{figure}

Stellar masses for galaxies in the G10 catalog were calculated using MAGPHYS \citep{dacun08}. As shown by the example in Figure \ref{fig:logmass-colour_example}, for each redshift bin, we plotted G10 values of $\log_{10}(M/L_K)$  and 1-$\sigma$ deviations against restframe $(B-V)$ color and determined the median y-axis values of $\log_{10}(M/L_K)$ for different color bins on the x-axis (solid points and vertical bars).

\begin{figure}
 	\centering
		\includegraphics[width=0.49\textwidth]{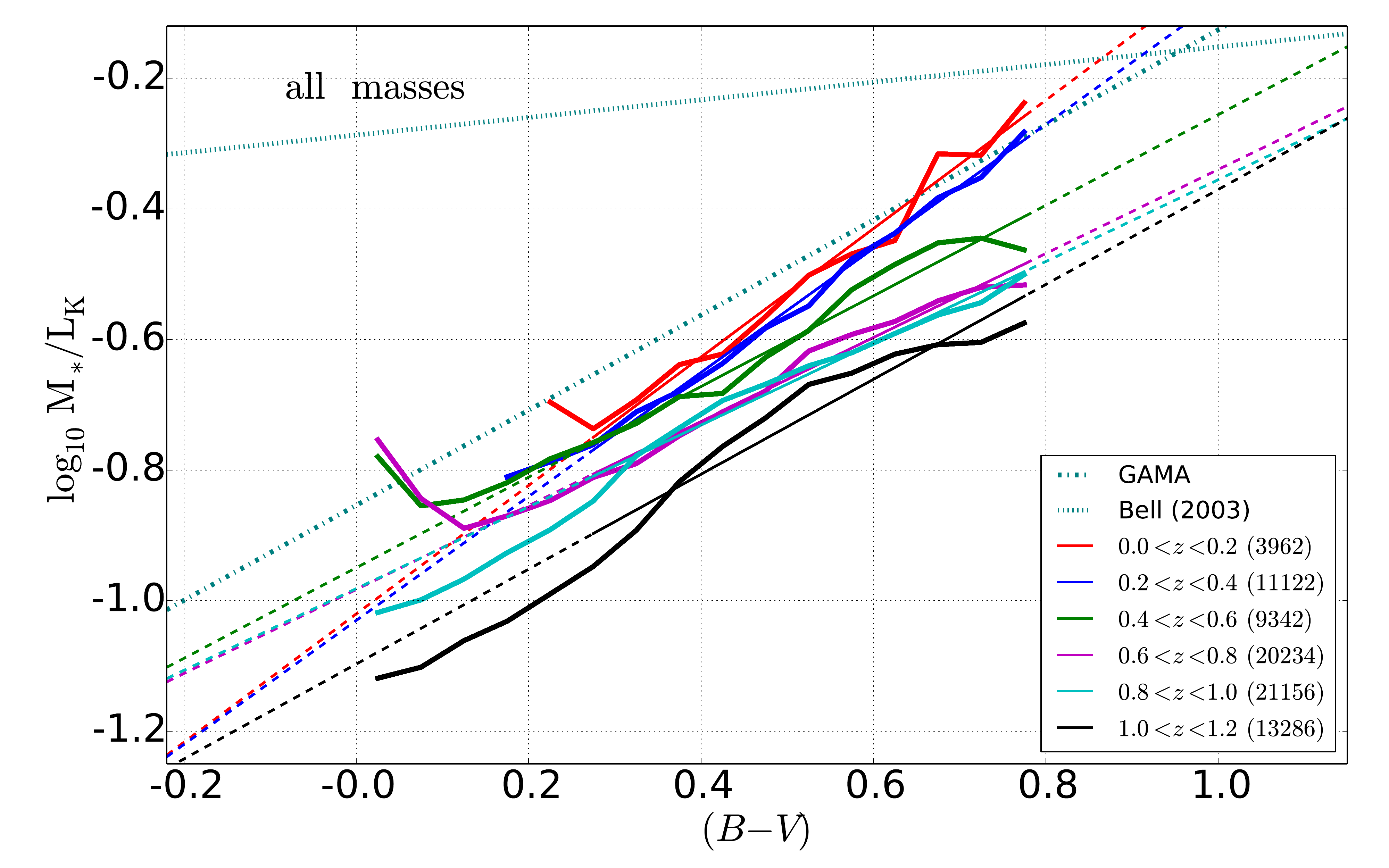}
		\caption{Median values of $\log_{10}(M/L_K)$ for G10 galaxies plotted against restframe $(B-V)$ color for redshifts in bins of width 0.2 mag between $z = 0$ and $z = 1.2$.  Linear best fits to the median $\log_{10}(M/L_K)$ values for colors between 0.3 and 0.7 are shown by thick slanting lines (extended beyond this range by  dashed lines). For comparison, the thin sloping green dot-dash and dotted lines shows the non-evolving $z < 0.4$ GAMA and $z < 0.2$ \citet{bell03} relationships respectively.}
		\label{fig:logmass-colour_all}
\end{figure}

We performed a linear best fit to the median $\log_{10}(M/L_K)$ values for colors in the range 0.3 to 0.7. Figure \ref{fig:logmass-colour_all} shows how these median $\log_{10}(M/L_K)$ values and best fit relationships have evolved with redshift.

\begin{figure}
 	\centering
		\includegraphics[width=0.49\textwidth]{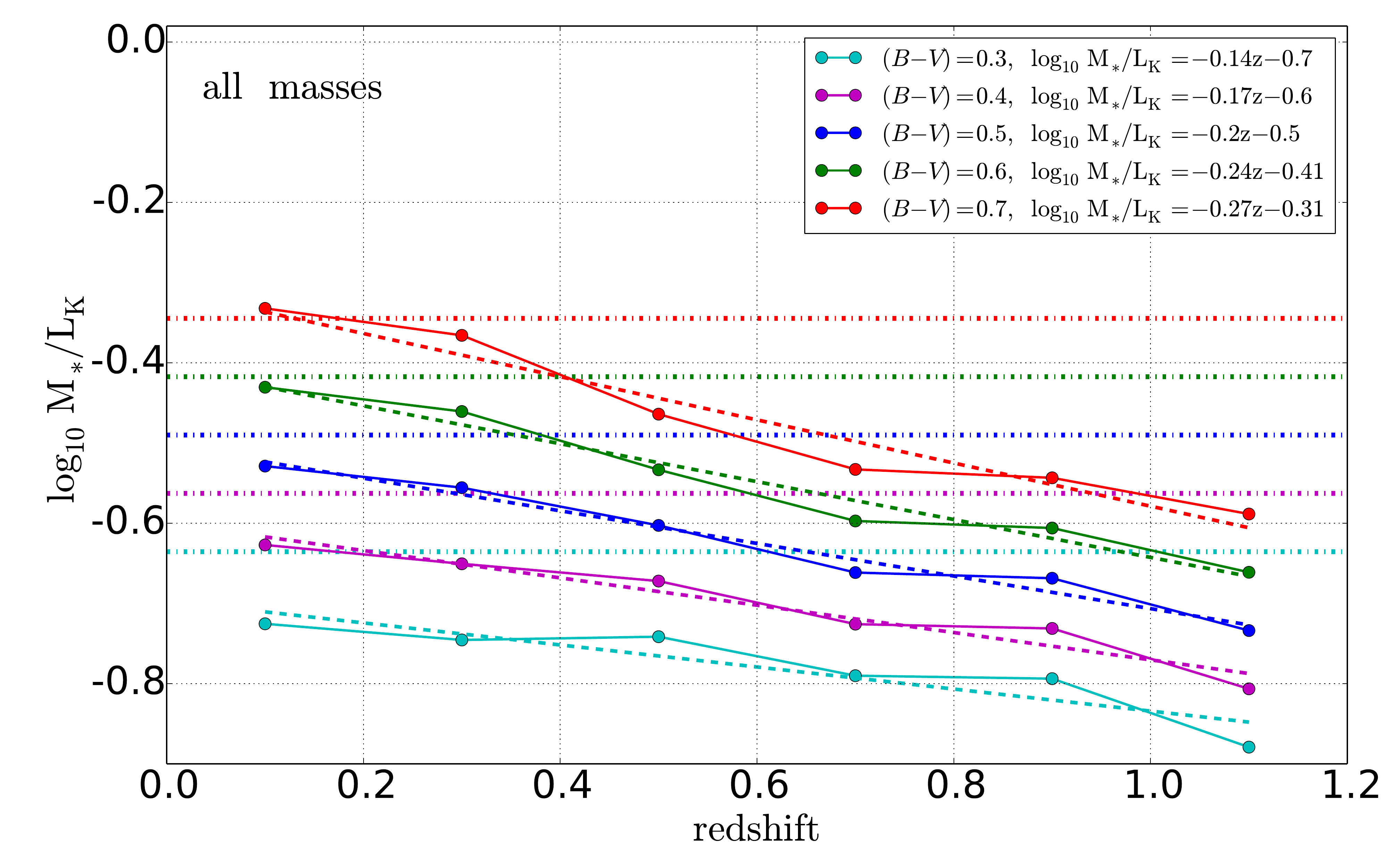}
		\caption{Evolution of the best fit values of $\log_{10}(M/L_K)$  in Figure \ref{fig:logmass-colour_all} for G10 galaxies with $(B-V)$ restframe colors of 0.3, 0.4, 0.5, 0.6 and 0.7.  Linear best fits are shown by dashed lines and their equations are given at upper right. These equations can all be summarised to within 0.01 dex by the single Equation \eqref{eq:ML_G10}: $\log_{10}(M/L_K) = - [0.05 + 0.3(M_B - M_V)] z + (M_B - M_V)  -  1.0$. This enables evolving mass to light ratios to be computed simply from redshifts and restframe colors. For comparison, the horizontal dash-dot lines show the non-evolving $z < 0.4$ GAMA relationship Equation \eqref{eq:ML_GAMA}. }
		\label{fig:logmass-colour_evolution}
\end{figure}

To measure evolution of $\log_{10}(M/L_K)$ as a function of restframe color, the best fit median $\log_{10}(M/L_K)$ values corresponding to $(B-V)$ colors of 0.3, 0.4, 0.5, 0.6 and 0.7 were  plotted against redshift, as in Figure \ref{fig:logmass-colour_evolution}. (Each curve in this plot corresponds to the intersections of a vertical line of constant color with the  best fit lines in the preceding plot.)

Equations for the linear best fits to the five plots in Figure \ref{fig:logmass-colour_evolution} are shown in the top right hand corner. It turns out that they can all be described to within 0.01 dex by just one function of $(B-V)$ and redshift, namely:
\begin{equation}
\label{eq:ML_G10}
	\log_{10}(M/L_K) = - [0.05 + 0.30(M_B - M_V)] z + (M_B - M_V)  -  1.00.
\end{equation}

We  derived Equation \eqref{eq:ML_G10} using $(B-V)$ colors between 0.3 and 0.7 because a primary focus of this paper is to study evolution of the stellar mass in red galaxies and this restframe color range is sufficient to encompass all red sequence galaxies at $z < 1.2$. Also, a close linear fit was apparent for colors between 0.3 and 0.7, random error being $\sim 0.2$ dex and systematic error being up to $\sim 0.03$ dex. For galaxies bluer than $(B-V) = 0.3$, the random error is similar but the systematic error greater - up to $\sim 0.06$ dex.

As the example plot in Figure \ref{fig:logmass-colour_example} illustrates, the 1-$\sigma$ scatter in $\log_{10}(M/L_K)$ values for given restframe color is generally \s0.2 dex.  However, some of this variation is due to random error in the G10 photometry rather than just intrinsic variability amongst galaxies. \citet{taylo11} found a smaller variation of only \s0.1 dex for GAMA galaxies at $z < 0.4$, as can be seen in Figure \ref{fig:MLK_GAMA}. We take their value $\sigma = 0.1$ dex as intrinsic scatter of $\log_{10}M_K/L$ at fixed $(B-V)$.

As Figure \ref{fig:logmass-colour_all} shows, our evolving relationship, Equation \eqref{eq:ML_G10}, exhibits a broadly similar dependence on restframe color to the GAMA one, but it evolves at a rate that depends on color, this rate being slightly faster for red galaxies than for blue. This is more clearly seen in Figure \ref{fig:logmass-colour_evolution}.

As is clear from Figure \ref{fig:logmass-colour_all}, the 2MASS based $\log_{10}(M/L_K)$ values from \citet{bell03} vary little with $(M_B - M_V)$ color, whereas our values and those based on GAMA are a strong function of color. \citet{bell03} used the observed SDSS+2MASS colors of galaxies and masses determined by SED fitting, and we suspect that the difference is due to 2MASS underestimating the fluxes of star forming galaxies causing their $M/L_K$ values to be greater than ours by \s0.4 dex for the bluest galaxies but less than \s0.1 dex for the reddest galaxies. When converted for a Chabrier IMF, the \citet{bell03} relation is:
\begin{equation}
\label{eq:ML_bell03}
	\log_{10}(M/L_K) = -0.287 + 0.135 (M_B - M_V).
\end{equation}

Equation \eqref{eq:ML_G10} is based on the mean properties of an ever-changing population of red G10 galaxies, i.e. a population which is continually being augmented by blue galaxies that have ceased star formation, and one in which major and minor mergers and possibly bursts of star formation are changing the demographics of the population.

By contrast, in the case of massive red galaxies, we are predominantly measuring how individual massive galaxies fade and grow via mergers. As the evolutionary history of these massive red galaxies is different to that of the majority of red galaxies we expect their dependence of $\log_{10}(M/L_K)$ on restframe $(B-V)$ color to evolve differently with time. We therefore investigated this separately using a subsample with $\log_{10}M > 10.75$. This mass cut-off was chosen to ensure that a range of masses was included on both sides of the mass corresponding to the fixed space density of $\widetilde{\phi} = 2.5 \times 10^{-4.0} {h_{70}}^3 {\rm{Mpc}}^{-3} \textrm{dex}^{-1}$ that we used to measure evolution of massive galaxies (see Section \ref{sec:M} below).

At $z > 0.8$, only the reddest ($B-V>0.6$) massive galaxies exist in significant numbers in the G10 data to permit a reliable analysis, so we confined ourselves to studying massive red galaxies with a single restframe color of $(B-V) = 0.7$. For massive red galaxies we found:
\begin{equation}
\label{eq:ML_G10_massive}
	\log_{10}(M/L_K) = - [0.11 + 0.3(M_B - M_V)] z + (M_B - M_V)  -  0.9.
\end{equation}

We used Equation \eqref{eq:ML_G10} for our measurement of red galaxy SMD evolution but Equation \eqref{eq:ML_G10_massive} for our measurement of the evolution of the massive red galaxies  ($\log_{10}M > 10.75$). As explained later in Section \ref{sec:M}, these two measurements are based on entirely separate calculations involving Schechter function fitting with different mass ranges and different $\alpha$ constraints (fixed and variable respectively). We  used only Equation \eqref{eq:ML_G10} when measuring evolution of red galaxy SMD, and did not treat massive red galaxies separately. To do so would have introduced a discontinuity into our stellar mass measurements, and would not in any case have affected our red galaxy SMD measurements, because the presence of a small number of massive galaxies contributes a negligible fraction to overall SMD.

We note that massive red galaxies of restframe color $B-V=0.7$ evolve more rapidly in $(M/L_K)$ (0.32 dex per unit redshift) than red galaxies of the same restframe color as a whole (0.26 dex per unit redshift). This is to be expected because, to some extent, new arrivals from the blue cloud start off their red sequence life with similar properties, whether they arrive at z = 1.1 or z = 0.1. We therefore expect the properties of red galaxies as a whole to evolve less rapidly than those of individual massive red galaxies that are not being ``diluted'' by new arrivals and affected by major mergers, and this is what we observe.

It is instructive to compare the rate of $(M/L_K)$ evolution for individual massive red galaxies to those of \citet{bruzu03} SPS models with different metallicities and formation redshifts. For solar metallicity simple stellar populations (SSPs) the rates of evolution in $M/L_K$ from $z = 1.2$ to $z = 0$ are -0.20, -0.24 and -0.51 dex per unit redshift for formation redshifts $z_f = 10.0$, 4.1 and 1.4 respectively.

From Equation \eqref{eq:ML_G10_massive}, calculated rates of evolution of   $\log_{10}(M/L_K)$ for individual galaxies depend on redshift and the evolving restframe $(B-V)$ color. If we assume reddening of 0.1 mag per unit redshift for the most luminous red galaxies, as observed in our \bootess data and as predicted by SSPs, the net rate of change in  $\log_{10}(M/L_K)$ is $\sim-0.4$ dex per unit redshift. This lies between the rates of change for solar metallicity SSPs with $z_f = 4.1$ and $z_f = 1.4$. We conclude therefore that the empirical Equation \eqref{eq:ML_G10_massive} is broadly consistent with the predictions of \citet{bruzu03} SSP models.

Our models of an evolving $M/L_K$ relation that is a function of rest-frame $(B-V)$ extend the work of \citet{taylo11}, showing $M/L$ relations can be determined with 0.2 dex accuracy using a single restframe color out to $z \sim 1$.  It should be mentioned that \citet{lopez18} reached a conflicting conclusion, i.e. that the dependence of $M/L_i$ on restframe $(g-i)$ color has not evolved since $z = 1.5$.

We find in this study that it is essential to include evolution of $M/L_K$ as a function of $(B-V)$ if our measurements of SMF evolution since $z = 1.2$ are to agree with previous studies that derive stellar masses directly from SPS modelling rather than via mass to light ratios. An evolving $M/L_K$ versus $(B-V)$ relationship is also essential if our measurements of rates of red galaxy stellar mass evolution are to agree with those inferred from comparisons with passively evolving stellar populations.

\begin{deluxetable}{cccc}	
\tablewidth{0pt}
\tablecolumns{4}
\tabletypesize{\scriptsize}
\tablecaption {The observed colors used to determine absolute magnitudes.}
\tablehead{
\colhead{restframe} & \colhead{effective} & \colhead{redshift} & \colhead{color}
\\
\colhead{waveband} & \colhead{wavelength} & \colhead{range} & \colhead{$(m_Y - m_Z)$}
 \\
\colhead{} & \colhead{$\mu m$} & \colhead{ } & \colhead{}
}
 \startdata
 $U\,$\tablenotemark{a} & 0.361 & 0.0 to 0.8 & $B_w - R$ \\
$U\,$\tablenotemark{a} & 0.361 & 0.8 to 1.2 & $R - I$ \\
\\
$B\,$\tablenotemark & 0.441 & 0.0 to 0.4 & $B_w - I$ \\
$B\,$\tablenotemark & 0.441 & 0.4 to 1.2 & $R - J$ \\
\\
$V$ & 0.551 & 0.2 to 0.4 & $R - K_s$ \\
$V$ & 0.551& 0.4 to 1.2 & $I - J$ \\
\\
$K$ & 2.195 & 0.2 to 0.6 & $K_s - [3.6 \mu \textrm{m}]$ \\
$K$ & 2.195 & 0.6 to 1.0 & $[3.6 \mu \textrm{m}] - [4.5 \mu  \textrm{m}]$ \\
$K$ & 2.195 & 1.0 to 1.2 & $J - [4.5 \mu  \textrm{m}]$ \\
\enddata
\label{tab:observed_colors}
\tablecomments{Absolute magnitudes in a waveband $W$ were calculated using the method of \citet{beare14}. Given two suitably chosen observed magnitudes $m_Y$ and $m_Z$, $(M_W - m_Z)$ is given by a second degree polynomial in the color $(m_Y - m_Z)$.}
\tablenotetext{a}{As calculated in \citet{beare15}}
\end{deluxetable}

\newpage

\section{Sample selection}
\label{sec:sample}

The same cuts as in \citet{beare15} were applied to limit the sample to objects brighter than $I = 24.0$ and $[3.6 \mu \textrm{m}] = 23.3$, and to exclude stars. We used the modified \citet{stern05} mid-infrared selection criteria that we  used previously in \citet{beare15} to exclude AGN, with additional AGN identifications being made with SDSS and AGES spectroscopy.  The AGN cuts removed less that 1\% of our galaxy sample. Quiescent and star-forming galaxies were separated using the same evolving red-blue color cut in restframe color-magnitude space:
\begin{equation}\label{eq:redbluecut}
	(M_{\rm{U}} - M_{\rm{B}}) > 1.074 - 0.18z  - 0.03(M_{\rm{B}} + 19.4). \\
\end{equation}
The final sample size for galaxies at redshifts $0.2 \leq z < 1.2$ with absolute magnitudes $-24 \leq M_B < 14$ was 353\,594. The sample was 85\% complete at our faint limit of $I = 24.0$ and $\simeq100\%$ complete at  $I = 21.5$. We corrected for the small degree of incompleteness over the range $21.5 < I < 24.0$.

\section{Determining the $K$-band luminosity function}
\label{sec:KLF}

$K$-band LFs were determined for red and blue galaxy sub-samples separately, as well as for the total sample. In each case, galaxies in the redshift range $0.2 \leq z < 1.2$ were allocated to five redshift bins of equal width $\Delta z = 0.2$. For each redshift bin, empirical binned LFs, $\Phi(M_K)$,  were obtained by dividing the (completeness corrected) numbers of galaxies $N$ in $K$-band absolute magnitude bins of width $\Delta M_K$ by the comoving volume $\Delta V$ corresponding to the given redshift range $\Delta z$, i.e. $\Phi(M_K) = N / \Delta V$.

For each redshift bin, we used the maximum likelihood method \citep{marsh83} to fit  \citet{schec76} functions to the (unbinned) absolute magnitude data over a magnitude range no wider than that over which the sample was complete to $I = 24.0$. This range was determined from a plot of $(M_K + D_{\rm{M}} - I)$ against redshift. In terms of luminosities, the Schechter function is:
\begin{equation} \label{eq:schechterL}
\phi_K \left(L_K\right) dL_K = \left( \frac{\phi_K^*}{L_K^*}\right) \left(\frac{L_K}{L_K^*}\right)^\alpha  \exp\left( \frac{-L_K}{L_K^*} \right) dL_K.
\end{equation}
Here $\phi_K$ is the comoving space density per unit increment in luminosity $L_K$, $\phi_K^*$ is a normalising factor (the characteristic space density), $L_K^*$  is the characteristic luminosity corresponding roughly to the transition from a power law LF to an exponential one, and $\alpha$ determines the slope of the power law variation at the faint end.  The characteristic space density $\phi_K^*$ provides an approximate measure of the space density close to the characteristic magnitude (more specifically $\phi_K = 1.086\phi_K^*$ at $M_K = M_K^ *$). The maximum likelihood method provides an estimate of Schechter fit parameter uncertainties.

It is often more useful to write the Schechter function in terms of absolute magnitudes $M_K$:
\begin{multline} \phi_K \left( M_K \right) dM_K \label{eq:schechter_M}
	\\
	= -0.4 \ln{10} \phi_K^* 10^{-0.4 (\alpha + 1) (M_K - M_K^*) } \exp ({ -10^{-0.4(M_K - M_K^*) } } )dM_K.
\end{multline}
Integrating over all luminosities gives the total luminosity density $j_K$ in waveband $K$:
\begin{equation} \label{eq:lumdens}
	j_K = \phi_K^* L_K^* \Gamma(\alpha + 2).
\end{equation}

Because $\alpha$ becomes increasingly poorly constrained as redshift increases, we kept $\alpha$ fixed at values corresponding to those for the lowest two redshift bins ($0.2 \leq z < 0.4$ and $0.4 \leq z < 0.6$) where $\alpha$ was a free parameter, namely -0.5, ~-1.2 and -1.0 for red, blue and all galaxies respectively.

To measure luminosity evolution of the most luminous galaxies we determined how the restframe magnitude $\widetilde{M}_K$ corresponding to a fixed space density of $\widetilde{\phi} = 10^{-4.0} {h_{70}}^3 {\rm{Mpc}}^{-3} {\rm{mag}}^{-1}$ had evolved. In order to do this as precisely as possible, we fitted a Schechter function with variable $\alpha$ parameter just to the bright end of the LF (i.e. using only data points brighter than $K$-band magnitude -23.0). This procedure effectively uses the Schechter function as a convenient measuring tool, the precise values of the Schechter parameters not having any other relevance).

\begin{figure}
 	\centering
	\includegraphics[width=0.49\textwidth]{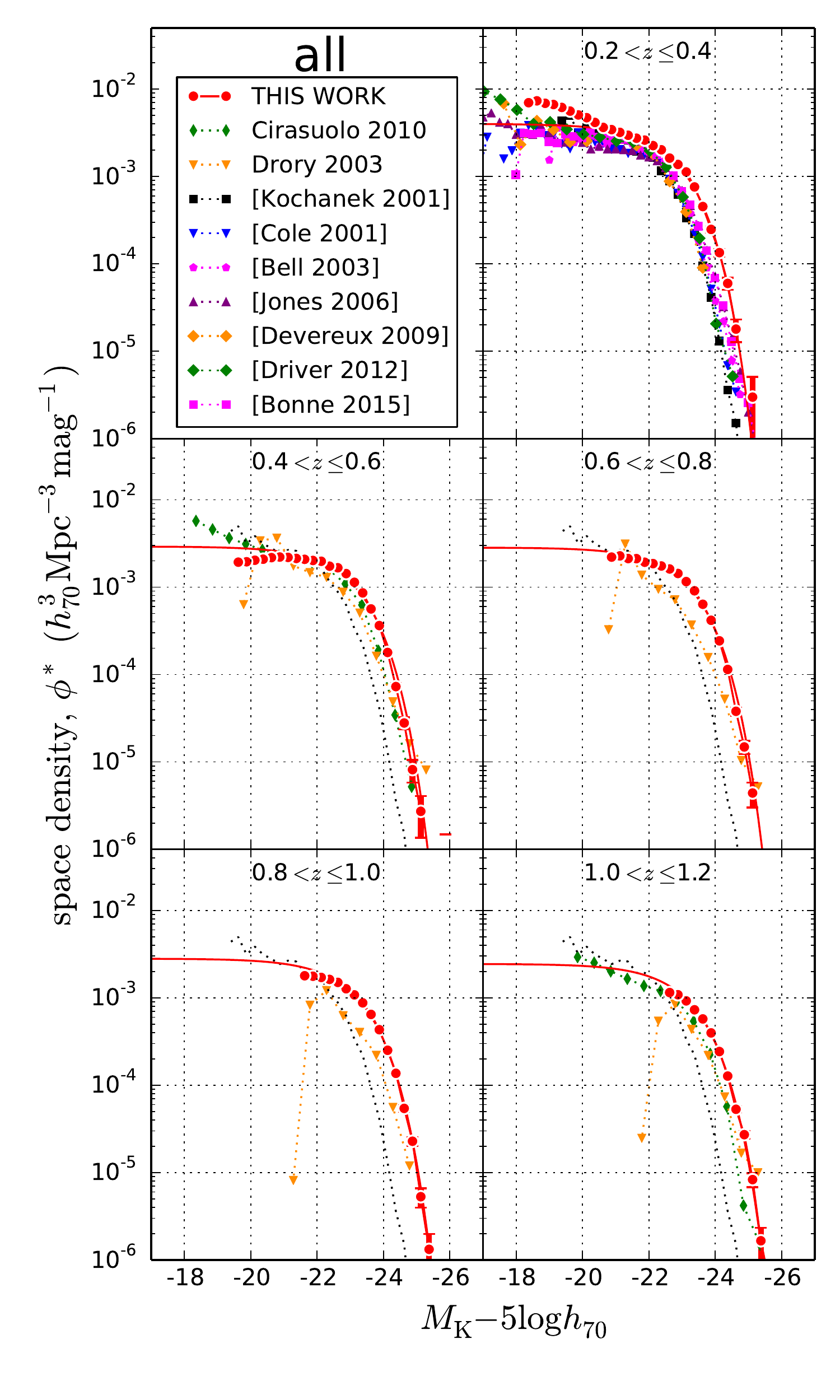}
	\caption{Binned $K$-band LFs for all galaxies in bins of width 0.2 mag  with $1-\sigma$ Poisson uncertainties shown for the \bootess data.  Overplotted in red are maximum likelihood fits to the (unbinned) data. LFs for the low redshift Universe are labelled using square brackets.  To provide a fixed reference, $z\sim0$ results from \citet{kocha01} are shown as black dotted lines in the lower four panels. Our LFs (red points) are brighter than some of the literature with offsets of up to $\sim0.5$ mag at $z\sim0.3$.}
	\label{fig:K_LF_binned_redandblue}
\end{figure}

\begin{figure}
 	\centering
	\includegraphics[width=0.49\textwidth]{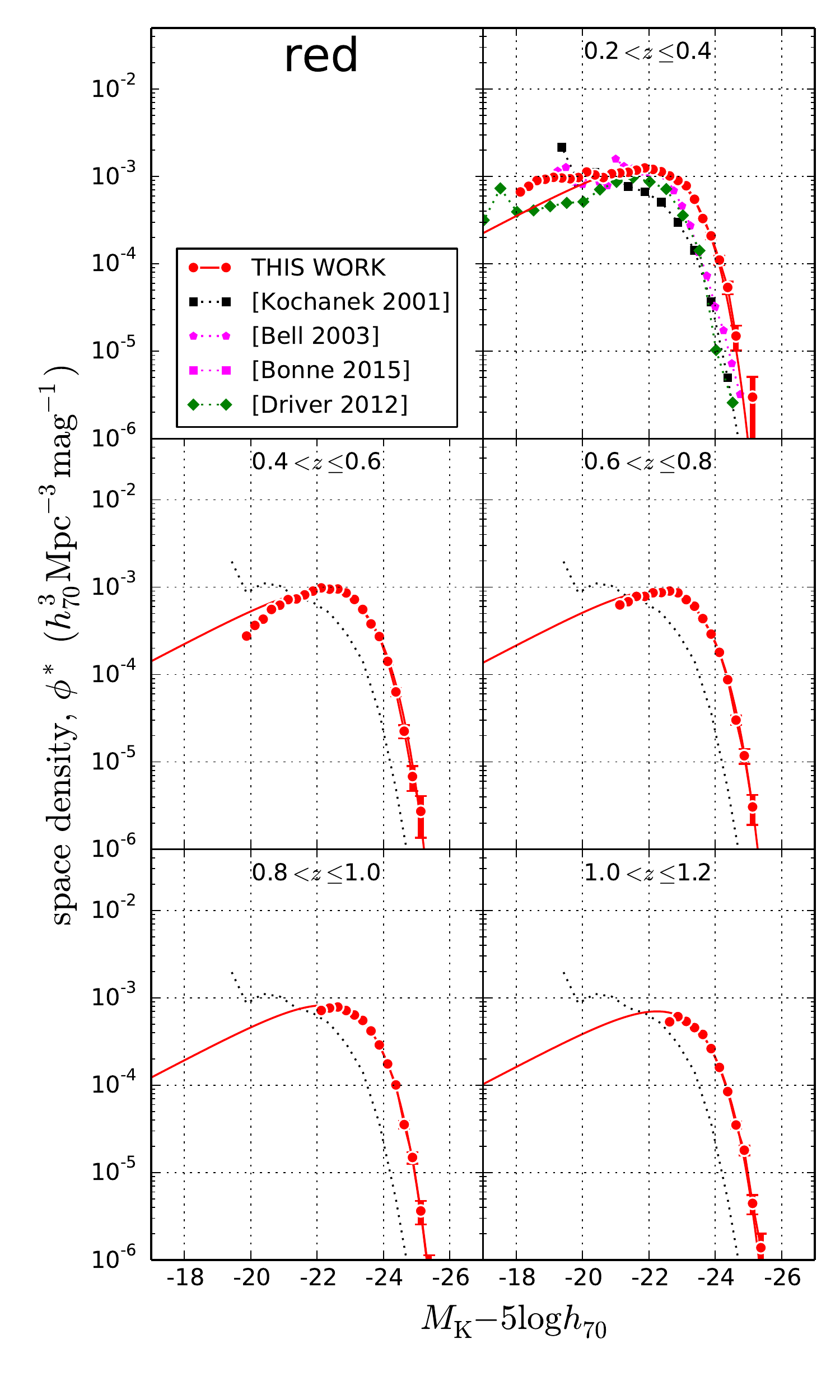}
	\caption{Binned $K$-band LFs for red galaxies with $1-\sigma$ Poisson uncertainties shown for the \bootess data. Overplotted in red are maximum likelihood fits to the (unbinned) data.  LFs for the low redshift Universe are labelled using square brackets. To provide a fixed reference, $z\sim0$ results from \citet{kocha01} are shown as black dotted lines in the lower four panels. The luminosity of the brightest red galaxies decreases with time, but not as fast as a passively evolving population, thus indicating a steady build up of mass through minor mergers.}
	\label{fig:K_LF_binned_red}
\end{figure}

\begin{figure}
 	\centering
	\includegraphics[width=0.49\textwidth]{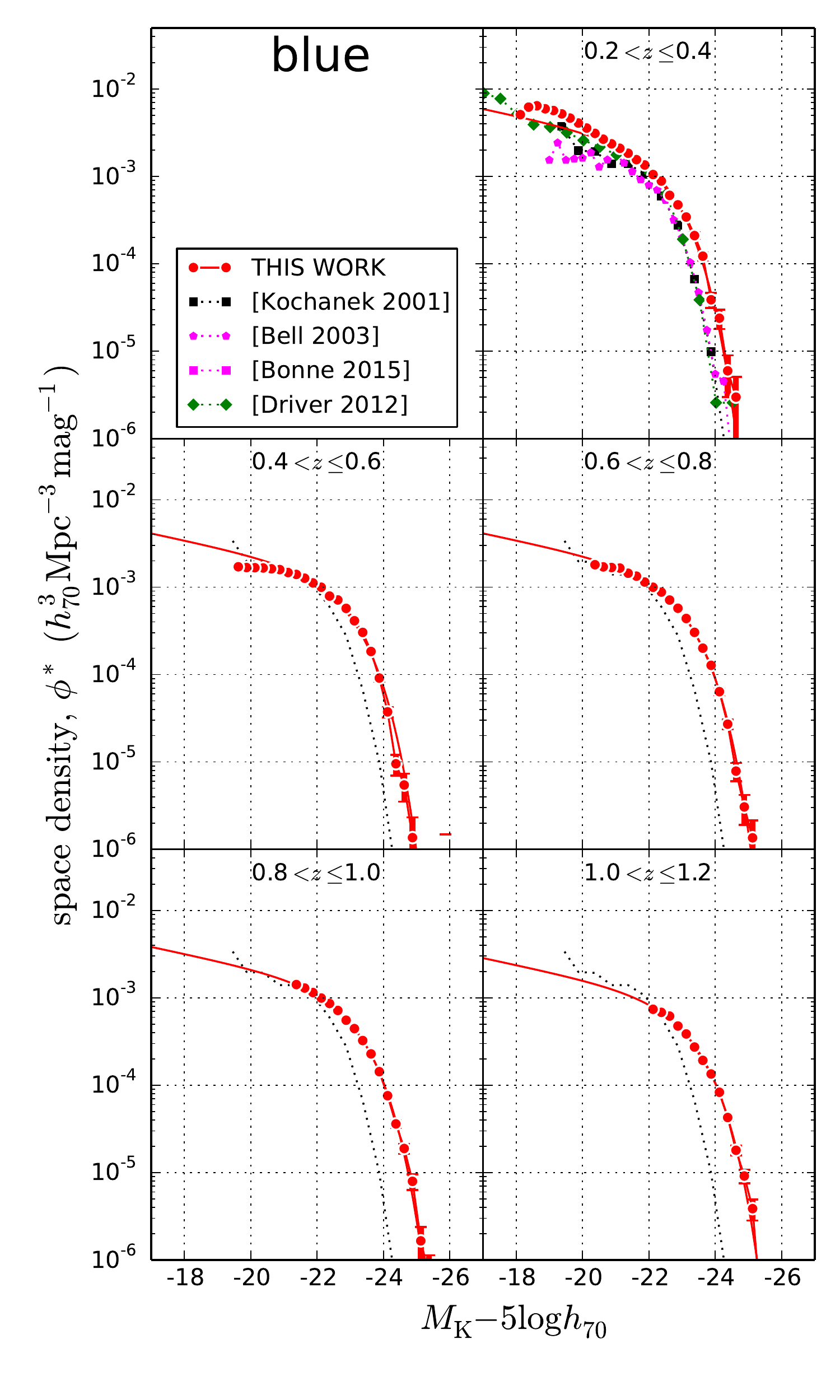}
	\caption{Binned $K$-band LFs for blue galaxies with $1-\sigma$ Poisson uncertainties shown for the \bootess data. Overplotted in red are maximum likelihood fits to the (unbinned) data. LFs for the low redshift Universe are labelled using square brackets.  To provide a fixed reference, $z\sim0$ results from \citet{kocha01} are shown as black dotted lines in the lower four panels.  The luminosity of the brightest blue galaxies decreases with time.}
	\label{fig:K_LF_binned_blue}
\end{figure}

\begin{figure}
 	\centering
	\includegraphics[width=0.49\textwidth]{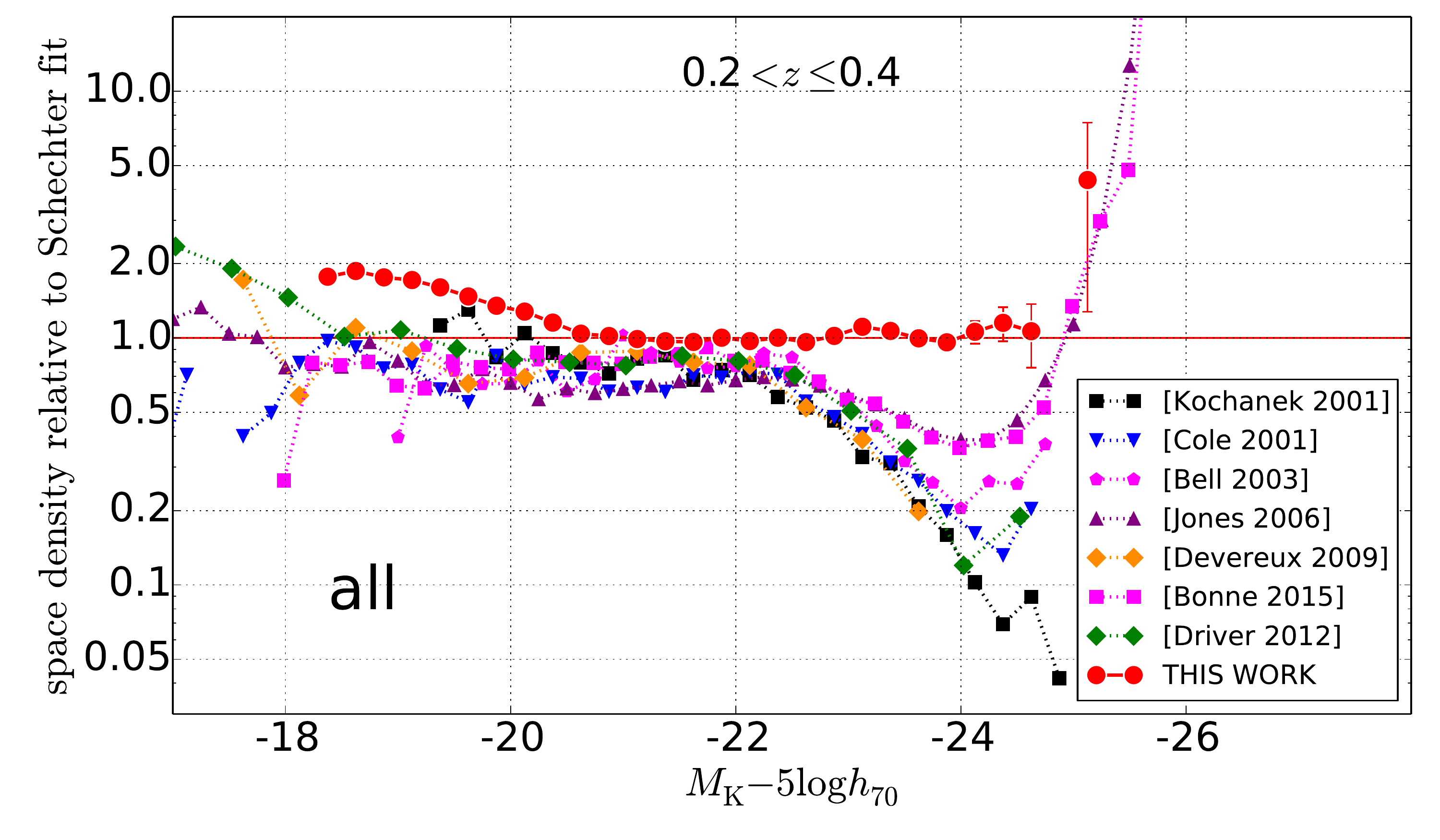}
	\caption{Detailed comparison of $K$-band space densities for all galaxies at $0.2 \leq z < 0.4$  with $1-\sigma$ Poisson uncertainties shown for the \bootess data.  This displays the same data as in the top right panel of Figure \ref{fig:K_LF_binned_redandblue} but plotting the ratio of binned space densities from the literature to the (unbinned) maximum likelihood Schechter function fit to our data.  LFs for the low redshift Universe are labelled using square brackets. We see a greater number of highly luminous galaxies than other studies. Our results show smoother variation due to our large sample size and area.}
	\label{fig:K_LF_binned_redandblue_closeup}
\end{figure}

\section{Determining the SMF}
\label{sec:M}

We measured binned SMFs of red, blue and all galaxies in the same manner as the binned LFs, including determining mass limits using the apparent magnitude limits and the mass-to-light ratios of passive galaxies. To parameterise our SMFs and measure their evolution we used the maximum likelihood method to fit Schechter functions to the (unbinned) galaxy masses over a mass range no wider than that over which the sample was complete to $I = 24.0$:
\begin{equation} \label{eq:schechter_mass}
\phi_M \left(M\right) dM = \left( \frac{\phi^*}{M^*}\right) \left(\frac{M}{M^*}\right)^\alpha  \exp\left( \frac{-M}{M^*} \right) dM.
\end{equation}
Here $\phi_M$ is the comoving space density per unit increment in stellar mass $M$, $\phi_M^*$ is a normalising factor, $M^*$ is a characteristic mass, and $\alpha$ determines the slope of the power law variation at the low mass end.  We used fixed $\alpha$ values of   of -0.5, -1.2 and -1.0 for red, blue and all galaxies respectively.

Given the large range of galaxy masses and to avoid confusion between mass and absolute magnitude, we rewrite Equation \eqref{eq:schechter_mass} in terms of logarithms of the mass $\mu = \log_{10} (M / M_{\sun})$, so the comoving density per unit $\mu$ is:
\begin{equation}  \label{eq:schechterM2}
	\phi_{\mu} (\mu) d\mu  = (\phi_{\mu}^* \ln{10}) \, 10^{(\alpha + 1) (\mu - \mu^*) } \exp ({ -10^{(\mu - \mu^*) }})d\mu.
\end{equation}
Here $\phi_{\mu}^*$ is the normalising factor and $\mu^* = \log_{10} M^*$. At the value  $\mu = \mu^*$, $\phi = \phi^* \ln{10} = 2.30 \phi^*$, so $\phi_{\mu}^*$ effectively measures the space density of galaxies per unit $\mu = \log_{10} M$ at the characteristic mass $M^*$.

Integrating Equation \ref{eq:schechter_mass} over all masses gives the total stellar mass density in galaxies:
\begin{equation} \label{eq:massdensity}
	\rho_M = \phi^* M^* \Gamma(\alpha + 2).
\end{equation}

To measure evolution of the mass of the most massive galaxies we determined how the value $\log_{10}\widetilde{M}$ corresponding to a fixed space density of $\widetilde{\phi} = 2.5 \times 10^{-4.0} {h_{70}}^3 {\rm{Mpc}}^{-3} \textrm{dex}^{-1}$ had evolved. In order to do this as precisely as possible, we fitted a Schechter function with variable $\alpha$ parameter just to the massive end of the SMF, i.e. using only data points for red (blue, all) galaxies for which $\log_{10} (M / M_{\sun}) > $ 11.0 (10.5, 10.5). (Because a luminosity ratio of 1.0 dex is equivalent to 2.5 mag, an LF space density of $10^{-4.0} {h_{70}}^3 {\rm{Mpc}}^{-3} {\rm{mag}}^{-1}$ is equivalent to an SMF space density of  $\widetilde{\phi} = 2.5 \times 10^{-4.0} {h_{70}}^3 {\rm{Mpc}}^{-3} \textrm{dex}^{-1}$, assuming a fixed stellar $M/L$ ratio. The fixed space densities used for the LF and SMF were therefore equivalent.)

\begin{deluxetable*}{cccccc}
\tablewidth{0pt}
\tablecolumns{6}
\tabletypesize{\scriptsize}
\tablecaption{$K$-band luminosity function Schechter parameters for fixed $\alpha$.\\}
\tablehead{
\colhead{z} & \colhead{$\alpha$} & \colhead{$\phi^*$} & \colhead{$M_K^* - 5\log h_{70}$} & \colhead{$M_K$ \rm{at fixed space density\tablenotemark{a}} (measures} & \colhead{$j_K$}
\\
\colhead{ } & \colhead{ } & \colhead{$(h_{70}^3 {\rm{Mpc}}^{-3} {\rm{mag}}^{-1})$} & \colhead{ } & \colhead{\rm{evolution of most luminous galaxies)}} & \colhead{$L_{\sun} \, {\rm{Mpc}}^{-3}$}
}
\startdata
\\
\multicolumn{6}{l}{Red galaxies - $\alpha = -0.5$}\\
0.3 & 	$-0.5$  & 	$3.14\pm 0.10 \times 10^{-3}$ & 	$-22.59\pm 0.04$ & 	$-24.13\pm 0.03$ & 	$3.50\pm 0.24 \times 10^{8}$ \\
0.5 & 	$-0.5$  & 	$2.29\pm 0.10 \times 10^{-3}$ & 	$-22.88\pm 0.08$ & 	$-24.24\pm 0.06$ & 	$3.33\pm 0.31 \times 10^{8}$ \\
0.7 & 	$-0.5$  & 	$2.25\pm 0.10 \times 10^{-3}$ & 	$-22.93\pm 0.05$ & 	$-24.31\pm 0.03$ & 	$3.43\pm 0.15 \times 10^{8}$ \\
0.9 & 	$-0.5$  & 	$2.01\pm 0.20 \times 10^{-3}$ & 	$-22.97\pm 0.05$ & 	$-24.35\pm 0.04$ & 	$3.19\pm 0.44 \times 10^{8}$ \\
1.1 & 	$-0.5$  & 	$1.75\pm 0.13 \times 10^{-3}$ & 	$-23.00\pm 0.05$ & 	$-24.32\pm 0.03$ & 	$2.85\pm 0.24 \times 10^{8}$ \\\\
\\
\multicolumn{6}{l}{Blue galaxies - $\alpha = -1.2$}\\
0.3 & 	$-1.2$  & 	$2.35\pm 0.06 \times 10^{-3}$ & 	$-22.51\pm 0.05$ & 	$-23.68\pm 0.05$ & 	$3.19\pm 0.46 \times 10^{8}$ \\
0.5 & 	$-1.2$  & 	$1.52\pm 0.16 \times 10^{-3}$ & 	$-22.88\pm 0.09$ & 	$-23.85\pm 0.08$ & 	$2.92\pm 1.42 \times 10^{8}$ \\
0.7 & 	$-1.2$  & 	$1.54\pm 0.02 \times 10^{-3}$ & 	$-22.99\pm 0.05$ & 	$-23.96\pm 0.06$ & 	$3.28\pm 0.77 \times 10^{8}$ \\
0.9 & 	$-1.2$  & 	$1.43\pm 0.11 \times 10^{-3}$ & 	$-23.11\pm 0.06$ & 	$-24.06\pm 0.04$ & 	$3.39\pm 0.46 \times 10^{8}$ \\
1.1 & 	$-1.2$  & 	$0.99\pm 0.14 \times 10^{-3}$ & 	$-23.24\pm 0.09$ & 	$-24.01\pm 0.07$ & 	$2.62\pm 0.55 \times 10^{8}$ \\\\
\\
\multicolumn{6}{l}{All galaxies - $\alpha = -1.0$}\\
0.3 & 	$-1.0$  & 	$4.34\pm 0.25 \times 10^{-3}$ & 	$-22.79\pm 0.06$ & 	$-24.19\pm 0.04$ & 	$6.58\pm 0.90 \times 10^{8}$ \\
0.5 & 	$-1.0$  & 	$3.15\pm 0.23 \times 10^{-3}$ & 	$-23.08\pm 0.08$ & 	$-24.31\pm 0.07$ & 	$6.25\pm 1.51 \times 10^{8}$ \\
0.7 & 	$-1.0$  & 	$3.14\pm 0.09 \times 10^{-3}$ & 	$-23.16\pm 0.04$ & 	$-24.40\pm 0.04$ & 	$6.64\pm 0.66 \times 10^{8}$ \\
1.1 & 	$-1.0$  & 	$3.11\pm 0.22 \times 10^{-3}$ & 	$-23.17\pm 0.05$ & 	$-24.47\pm 0.04$ & 	$6.67\pm 0.83 \times 10^{8}$ \\
1.1 & 	$-1.0$  & 	$2.67\pm 0.19 \times 10^{-3}$ & 	$-23.20\pm 0.06$ & 	$-24.46\pm 0.05$ & 	$5.89\pm 0.76 \times 10^{8}$ \\
\enddata
\label{tab:K_parameters}
\tablenotetext{a}{\footnotesize{$10^{-4.0} \, h_{70}^3 \, \rm{Mpc}^{-3} \, {\rm{mag}}^{-1}$}\, ($M_K$ \rm{at this fixed space density is measured by fitting a Schechter function with variable $\alpha$ to the bright end of the LF).}}
\end{deluxetable*}

\newpage

\section{$K$-band luminosity evolution - results and discussion}
\label{sec:K_results}

\subsection{$K$-band LF evolution - space density and characteristic magnitude} 
\label{sec:K_results_spacedens}

Figures \ref{fig:K_LF_binned_redandblue} to \ref{fig:K_LF_binned_blue} show our binned $K$-band LFs for all, red and blue galaxies, together with results from a variety of previous studies.  Maximum likelihood fits to our (unbinned) data are over-plotted as continuous red lines. To provide a fixed reference we show the local LFs of \citet{kocha01} for all, red and blue galaxies in each bin. We only plot bins for which 97.7\% (the $2-\sigma$ limit) of the measured absolute magnitudes have observed magnitudes brighter than our faint limit of $I = 24.0$. 

Our LFs are often brighter than the prior literature, with differences of up to \s0.5 mag evident for $z\sim0.3$, where bright galaxies are relatively well-resolved.  We attribute this to our careful accounting for light falling outside the photometric aperture by the use of magnitude dependent aperture sizes and corrections based on growth curves of measured magnitude with aperture diameter \citep{beare15}. Figure \ref{fig:K_LF_binned_redandblue_closeup} shows more clearly the difference between our LF and those of other authors for all galaxies at $0.2 < z \leq 0.4$. (Our binned data points lie above unity for the faintest galaxies, i.e. $M_K - 5\log h_{70} > -21$, because our maximum likelihood Schechter fit in Figure \ref{fig:K_LF_binned_redandblue} underestimates the space density at fainter magnitudes.) We measure significantly greater space densities for the brightest galaxies than other authors.

\begin{figure}
 	\centering
	\includegraphics[width=0.49\textwidth]{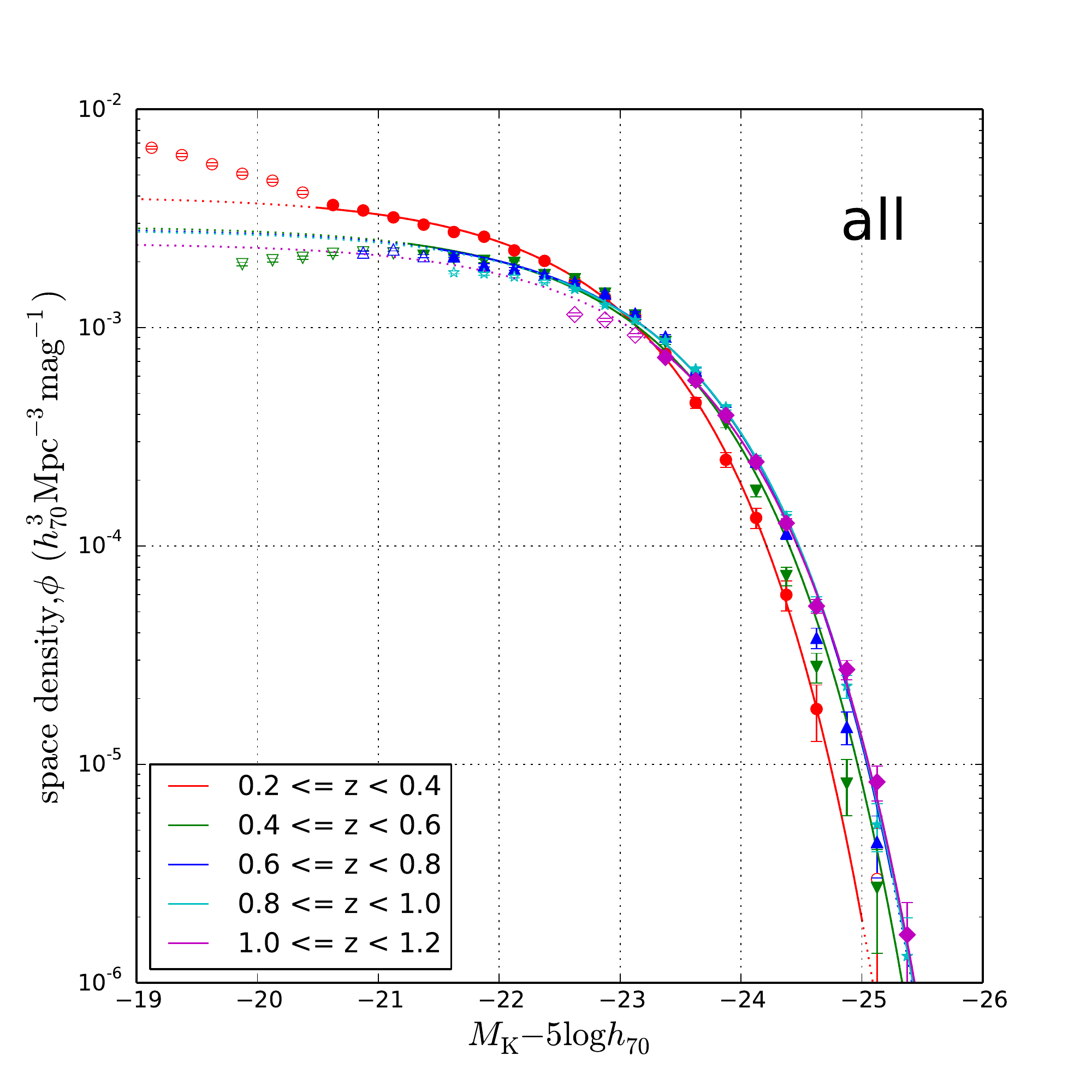}
	\caption{Evolution of the $K$-band Schechter function for all galaxies showing all five redshift ranges in one panel.   The symbols denote comoving space densities for the various absolute magnitude bins. Filled symbols denote the range of absolute magnitudes used to perform the maximum likelihood fit.  Continuous curves show maximum likelihood fits to the (unbinned) data. Open symbols denote data for very faint galaxies which were expected to be reliable on the basis of apparent $I$ and $[3.6 \mu \rm{m} \, ]$ magnitudes, but which were not represented adequately by a  Schechter function. The error bars show $1-\sigma$ Poisson uncertainties for the numbers in each bin.}
	\label{fig:K_evolution_redandblue}
\end{figure}

\begin{figure}
 	\centering
	\includegraphics[width=0.49\textwidth]{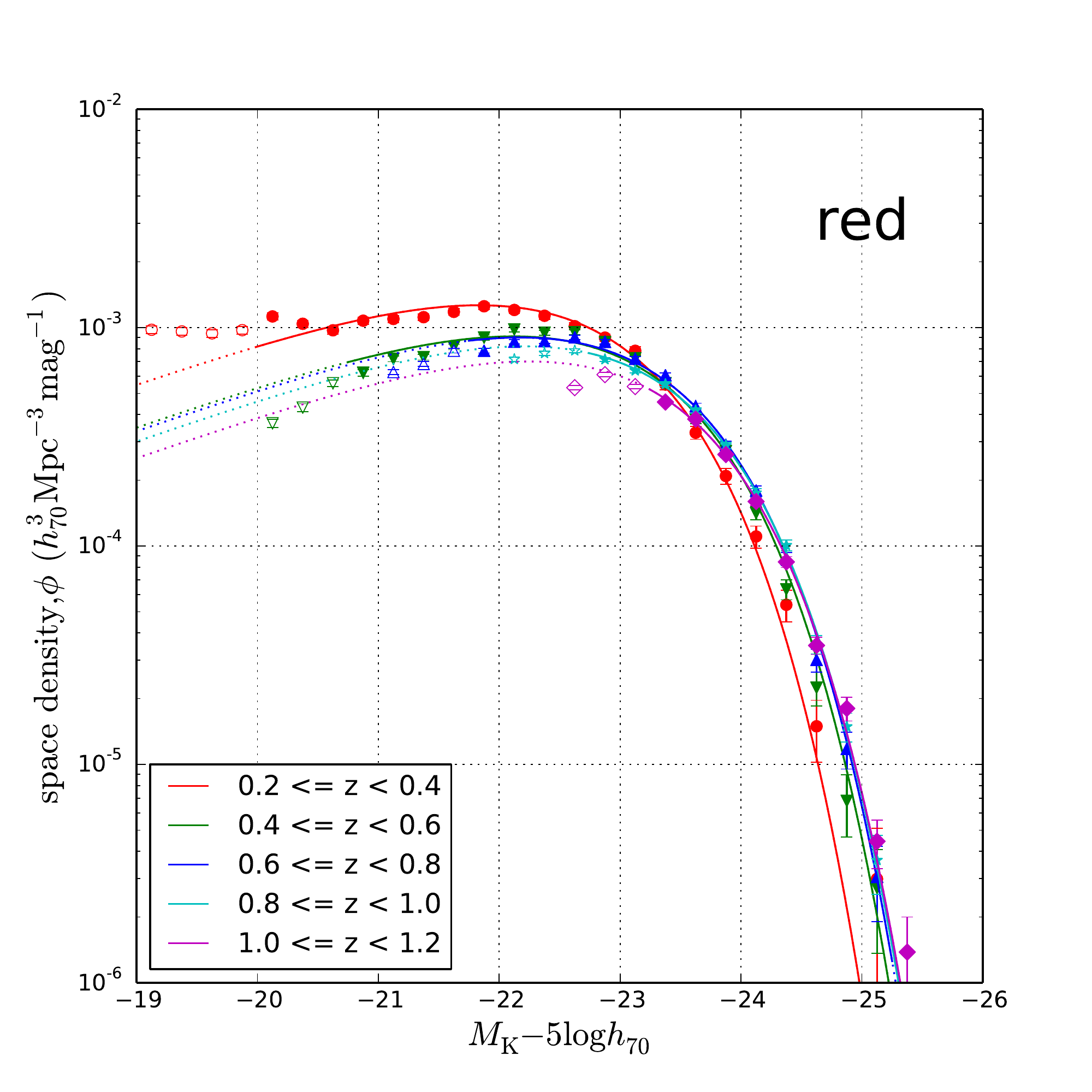}
	\caption{Evolution of the $K$-band Schechter function for red galaxies, showing all redshift ranges in one panel. Symbols are as in Figure \ref{fig:K_evolution_redandblue}.}
	\label{fig:K_evolution_red}
\end{figure}

\begin{figure}
 	\centering
	\includegraphics[width=0.49\textwidth]{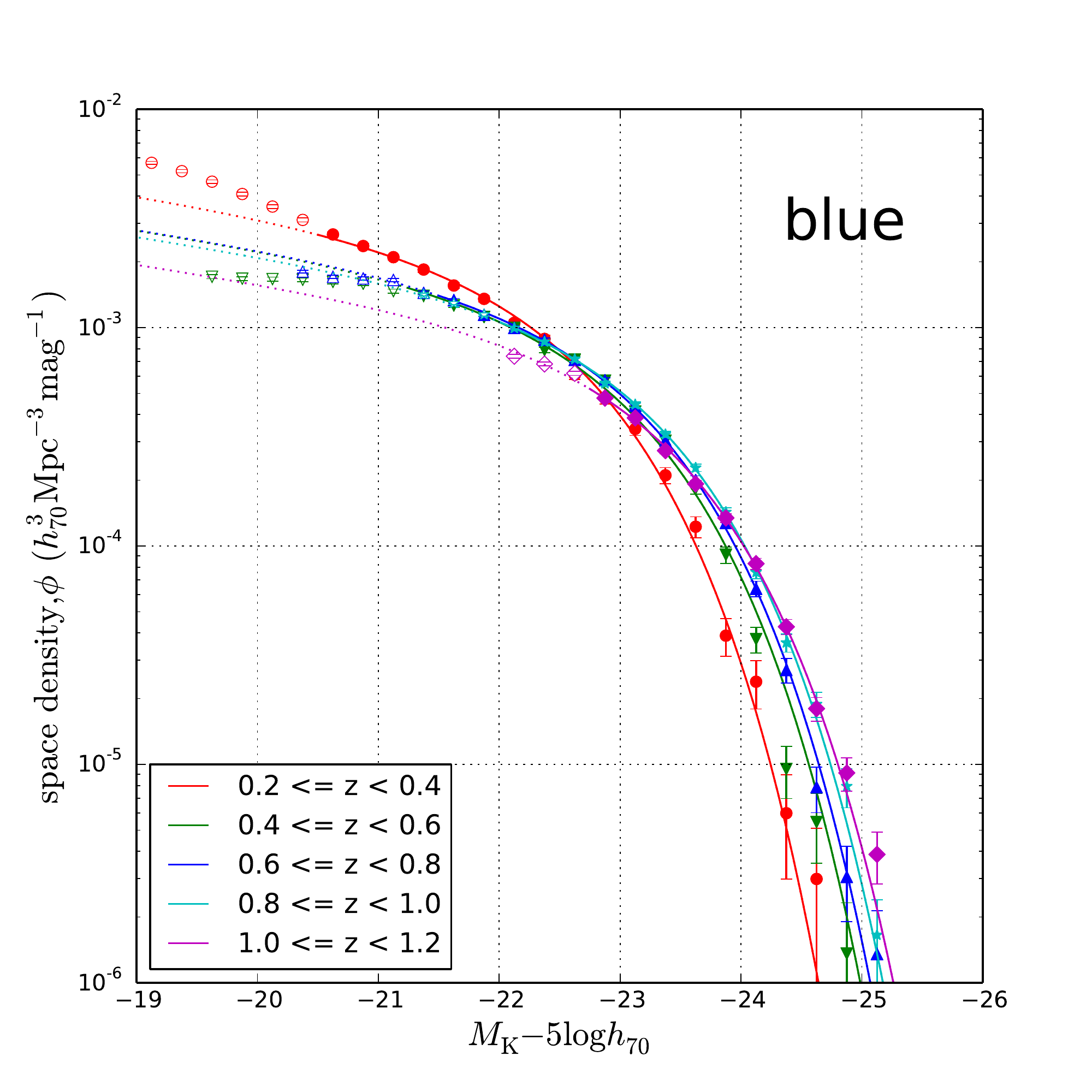}
	\caption{Evolution of the $K$-band Schechter function for blue galaxies, showing all redshift ranges in one panel.  Symbols are as in Figure \ref{fig:K_evolution_redandblue}.}
	\label{fig:K_evolution_blue}
\end{figure}

Figures \ref{fig:K_evolution_redandblue} to \ref{fig:K_evolution_blue} show our maximum likelihood Schechter fits (continuous lines) as well as our binned space densities (data points), with all redshift bins on a single plot to make the evolution of the LFs more apparent. 

We see from Figures \ref{fig:K_phi_star} and \ref{fig:K_M_star} and Table \ref{tab:K_parameters} that the characteristic space density $\phi_K^*$ approximately doubled for both red and blue galaxies from $z \sim 1.1$ to $z \sim 0.3$, while the characteristic magnitude $M_K^*$ of red galaxies faded by \s 0.4 mag and that of blue by \s 0.7 mag. The more rapid fading of $M_K^*$ for blue galaxies is consistent with downsizing \citep{cowie96}, i.e. the greater proportion of high mass galaxies seen in the star-forming population at $z\sim2$ than at $z\sim0$.

Luminosity density, as calculated by Equation \ref{eq:lumdens}, is a more physically meaningful quantity than the three individual Schechter parameters, because these relate to a specific functional representation of the LF rather than measurable physical quantities. Furthermore, luminosity density is relatively insensitive to degeneracy amongst the three Schechter parameters, because it is effectively the luminosity weighted area under the LF. This is illustrated by the fact that \citet[][]{jones06} obtained a value for $\phi_K^*$  for all galaxies which is half that of \citet[][]{bell03}, and an $\alpha$ value of -1.16 as compared with -0.77, but the two studies obtained luminosity densities within 4\% of  each other.

\subsection{$K$-band luminosity density evolution} 
\label{sec:K_results_lumdens}

\begin{figure}
 	\centering
		\includegraphics[width=0.49\textwidth]{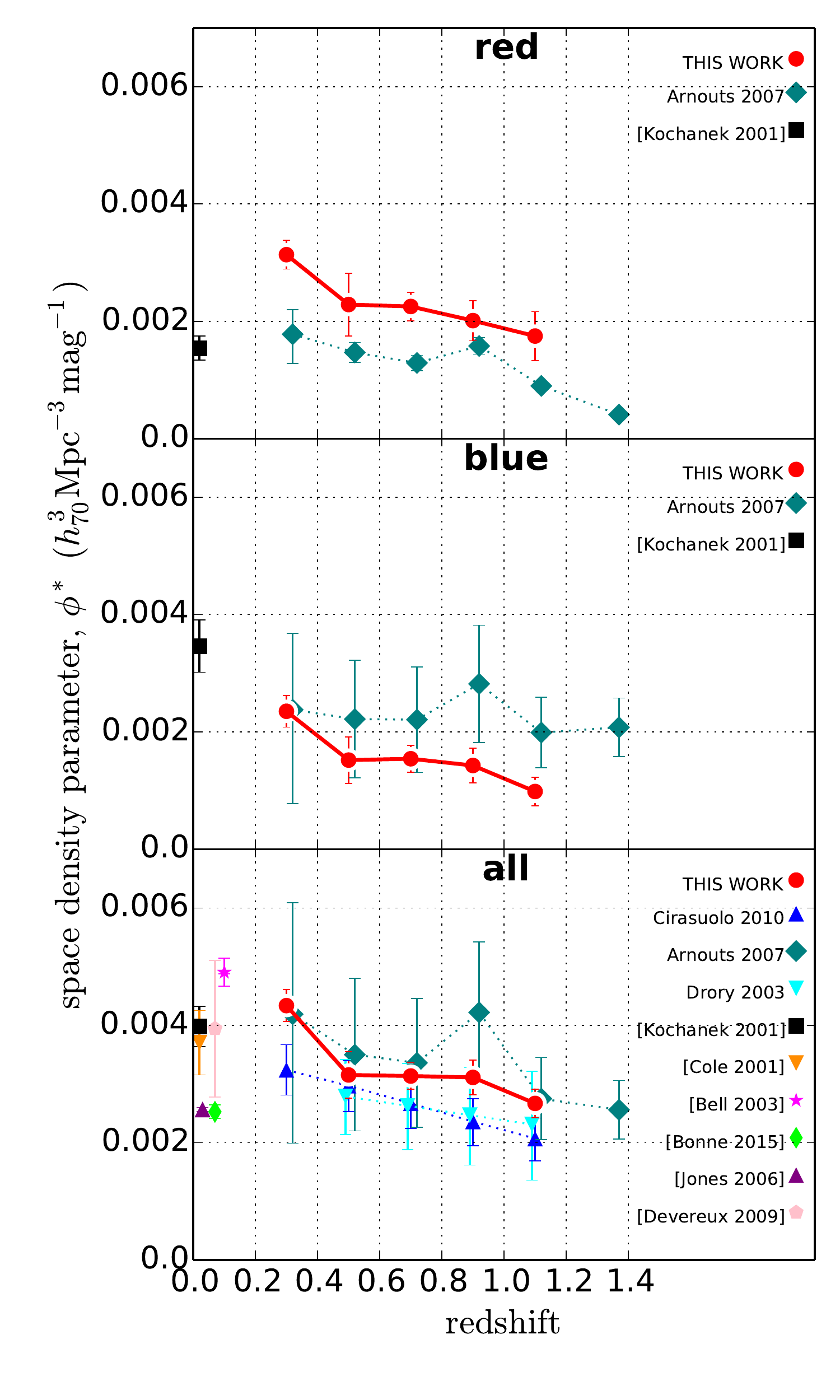}	
		\caption{Evolution from $z=1.1$ to $z=0.3$ of the $K$-band maximum likehihood Schechter parameter $\phi^*$ which normalises the space density. $\phi^*$ effectively measures the space density of $\sim L^*$ galaxies and for both red and blue $\sim L^*$ galaxies it approximately doubled from $z \sim 1.1$ to $z \sim 0.3$. Separate plots are shown for red, blue and all galaxies (red data points), assuming fixed alpha values of -0.5, -1.2 and -1.0 respectively.  Shown for comparison are the results from \citet[][]{arnou07}, \citet{drory03} and \citet[][]{ciras10}, (the latter two plots varying smoothly because they are best fits to an evolving functional form). LFs for the low redshift Universe are shown for  \citet[][]{cole01}, \citet[][]{kocha01},  \citet[][]{bell03},  \citet{jones06}, \citet[][]{dever09} and \citet[][] {bonne15} and labelled using square brackets. Error bars on our results show errors due to cosmic variance. Error bars on results from the literature are as published.}
		\label{fig:K_phi_star}
\end{figure}

\begin{figure}
 	\centering
		\includegraphics[width=0.49\textwidth]{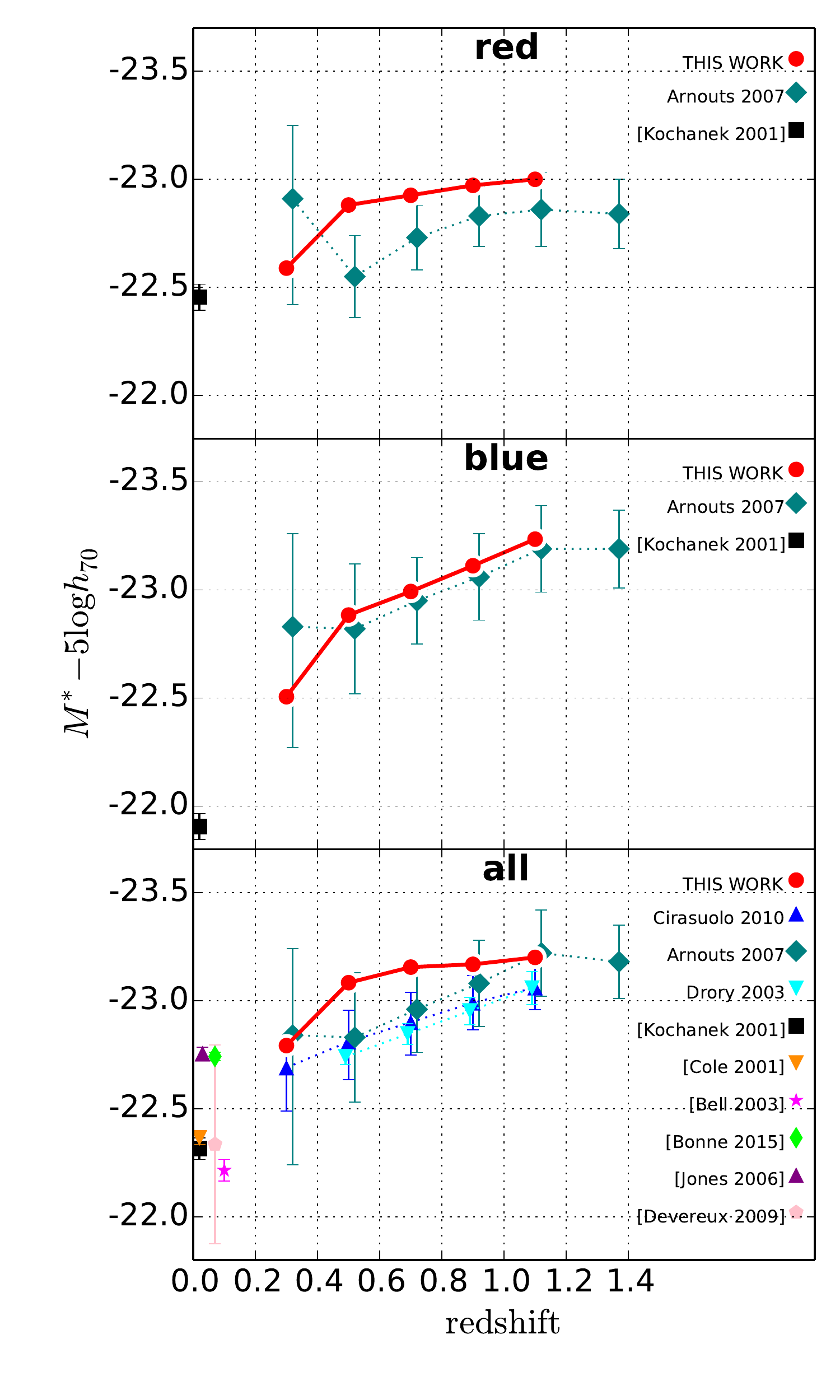}	
		\caption{Evolution from $z=1.1$ to $z=0.3$ of the $K$-band maximum likehihood Schechter characteristic magnitude parameter $M_K^* - 5 \log h_{70}$.   Separate plots are shown for red, blue and all galaxies (red data points), assuming fixed alpha values of -0.5, -1.2 and -1.0 respectively. $M_K^*$ faded faster from $z=1.1$ to $z=0.3$ for blue galaxies ($\Delta M_K^* \sim 0.7$ mag) than for red ($\Delta M_K^* \sim 0.4$ mag). Also shown are results from the literature as listed in each panel and referenced in the caption to Figure \ref{fig:K_phi_star}.}
		\label{fig:K_M_star}
\end{figure}

\begin{figure}
 	\centering
		\includegraphics[width=0.49\textwidth]{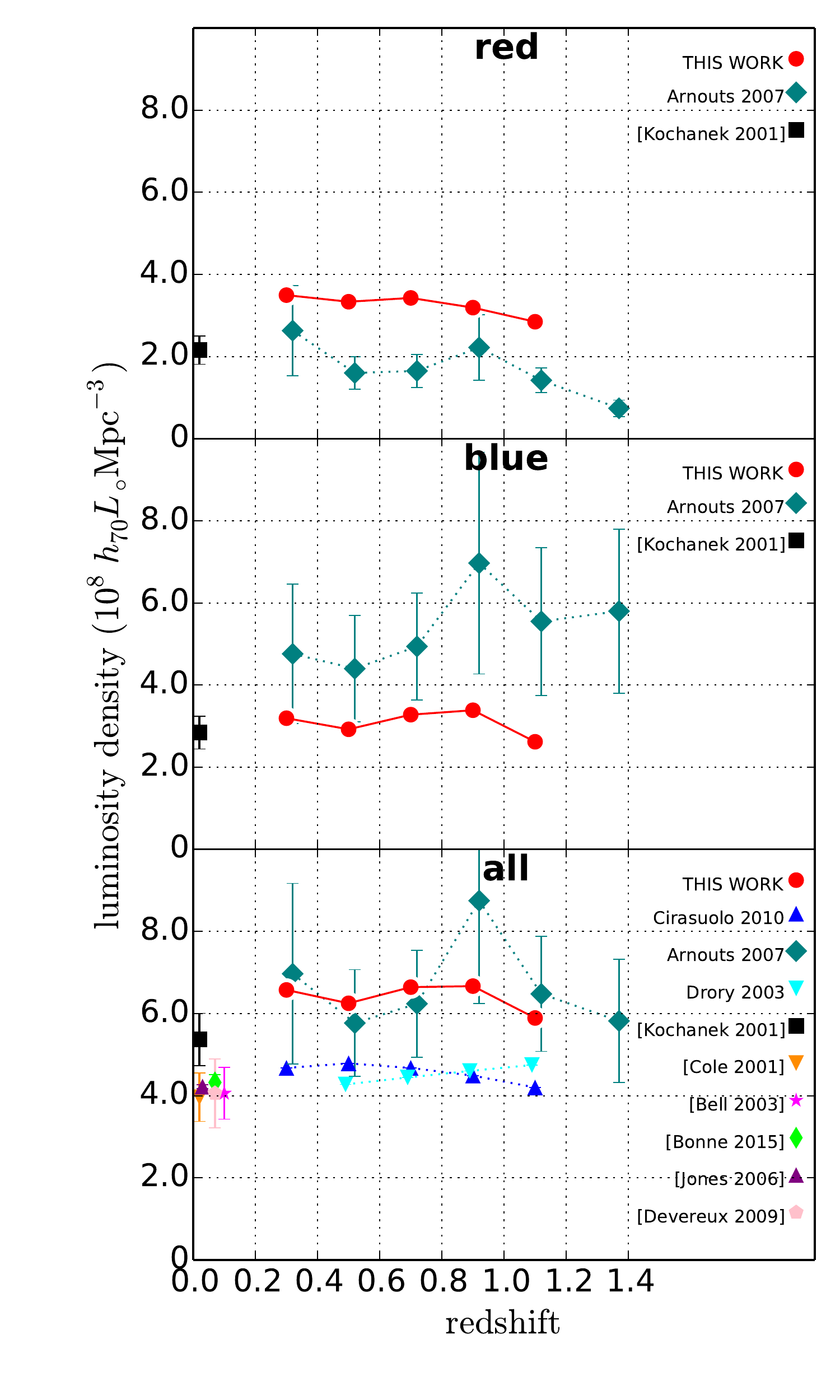}	
		\caption{The $K$-band luminosity density of both red and blue galaxies increased by a modest factor of \s1.2 from $z=1.1$ to $z=0.3$. Separate plots are shown for red, blue and all galaxies (red data points), assuming fixed alpha values of -0.5, -1.2 and -1.0 respectively. Also shown are results from the literature as listed in each panel and referenced in the caption to Figure \ref{fig:K_phi_star}. Red galaxy SMD grows at steady rate, and for red galaxies the increase in luminosity density with decreasing redshift implies a build up of stellar mass, because passively evolving stellar populations fade as they evolve.}
		\label{fig:K_lumdens}
\end{figure}

Figure \ref{fig:K_lumdens}  plots evolution of the total luminosity density $j_K$, showing that it increased by 20\% $\pm$10\% (0.08 dex) from $z \sim 1.1$ to $z \sim 0.3$ for all, red and blue galaxies. Also plotted are the results of \citet{ciras10}, \citet{arnou07} and \citet{drory03}, and luminosity densities for the same six low redshift studies as in Figures \ref{fig:K_phi_star} and \ref{fig:K_M_star}. 

For red galaxies, the measured luminosity density increase by a factor of $\sim1.2$ implies a build up of stellar mass, because a passively evolving stellar population \textit{fades} as it evolves. As in \citet{beare15}, we estimated the increase in SMD, $\rho$, by comparison with a passively fading quiescent population. Taking a single burst \citet[][]{bruzu03} SSP with formation redshift $z_f = 3.0$, Solar metallicity $Z = 0.02$ and Chabrier IMF to be representative of quiescent stellar populations, evolution of the stellar mass to light ratio $\Upsilon = \rho/j_K$ is given by $d\Upsilon/dz = -0.24$, resulting in passive fading of \s0.56 mag from $z = 1.1$ to $z = 0.3$. From this we deduce that red galaxy SMD increased by a factor of  \s2.1 from $z = 1.1$ to $z = 0.3$. Varying $z_f $ between 4.0 and 1.5 or adopting a very low metallicity of $Z = 0.0001$ alters the measured SMD increase by less than 20\%.

\subsection{$K$-band luminosity evolution of highly luminous galaxies} 
\label{sec:K_results_luminous}
	
\begin{figure}
 	\centering
	\includegraphics[width=0.49\textwidth]{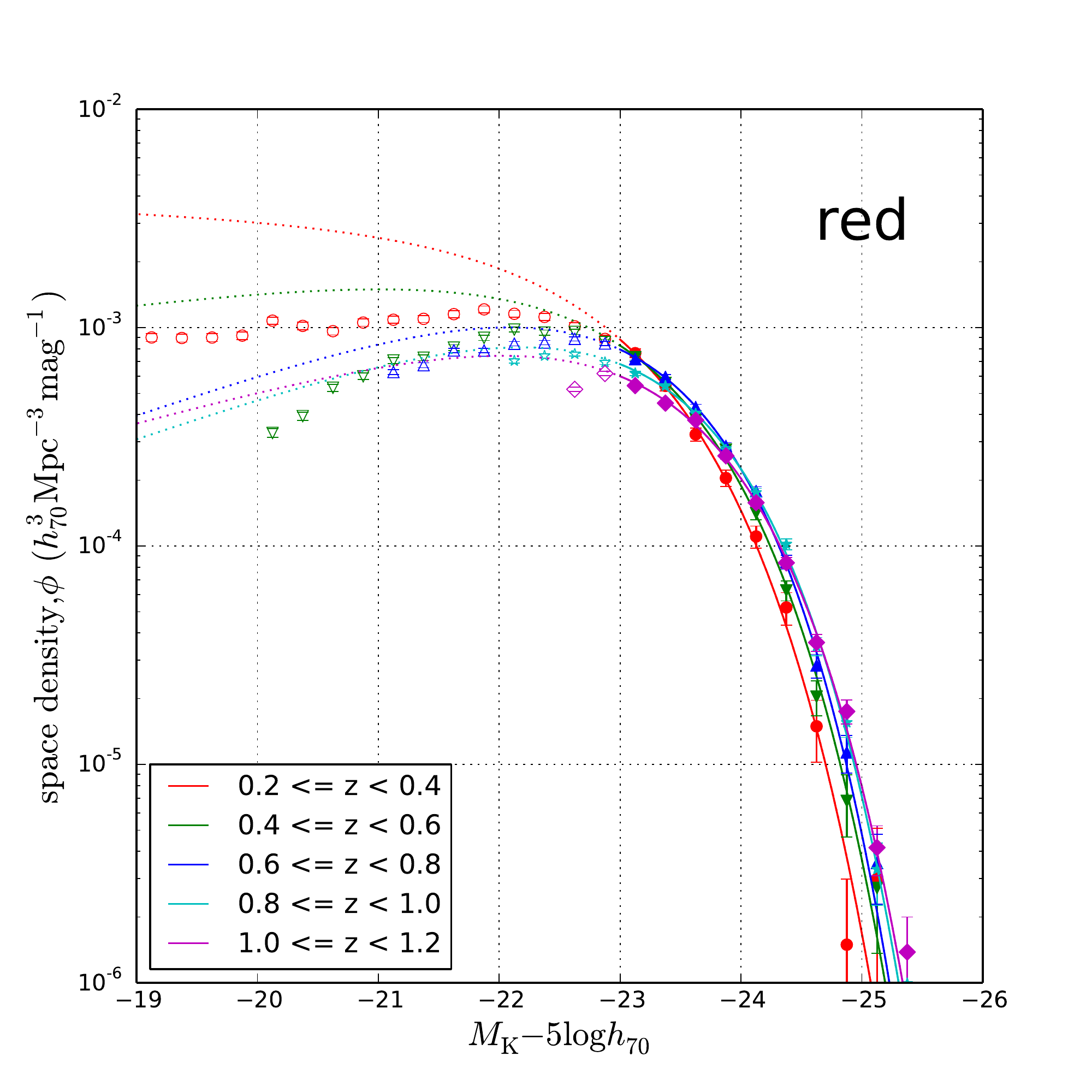}
	\caption{How evolution of the $K$-band highly luminous red galaxies was measured by fitting a Schechter function with variable $\alpha$ parameter to just the brightest part of the LF.   Symbols are as in Figure \ref{fig:K_evolution_redandblue}.}
	\label{fig:K_evolution_red_massive}
\end{figure}

As argued by \citet{bell04}, luminous red galaxies have evolved by a combination of passive stellar fading and the addition of stellar mass through mergers, because highly luminous blue galaxies are too rare to account for the growth in luminous red galaxy numbers via cessation of star formation. The evolution of the bright end of the red galaxy LF thus approximates the luminosity evolution of the most luminous red galaxies. Figure \ref{fig:K_evolution_red_massive}  shows how we used a Schechter function with variable $\alpha$ fitted to the bright end of the LF to measure evolution of the bright end of the red galaxy LF at a space density of $\widetilde{\phi}=10^{-4.0} \, h_{70}^3 \, \rm{Mpc}^{-3} \, {\rm{mag}}^{-1}$.

\begin{figure}
 	\centering
		\includegraphics[width=0.49\textwidth]{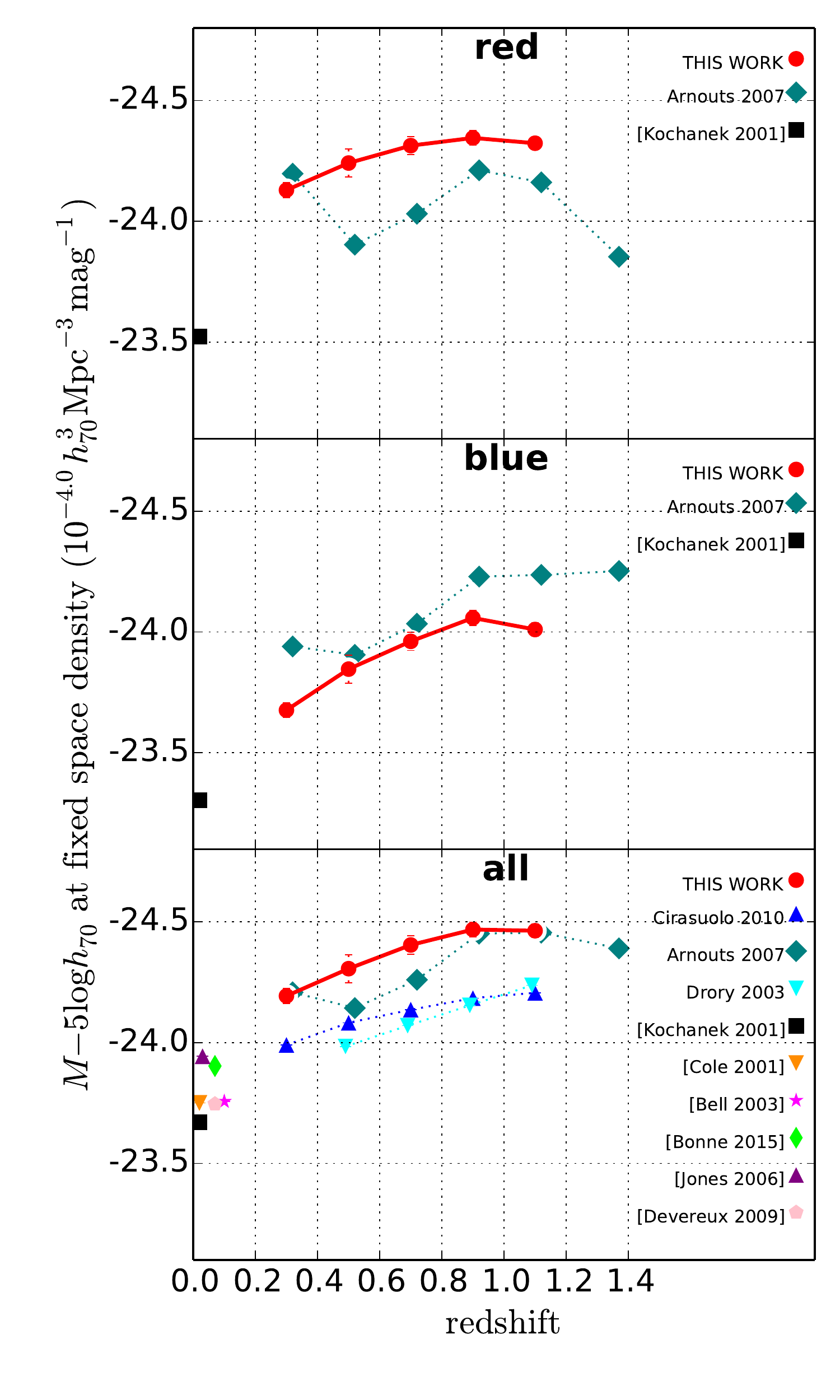}	
		\caption{Evolution of the bright end of the $K$-band LF.  The luminosity of the brightest blue galaxies decreases by \s0.33 mag from $z=1.1$ to $z=0.3$. For red galaxies the smaller decrease of \s0.19 mag represents the fading of individual highly luminous red galaxies offset by additional luminosity gained through minor mergers. The rate of fading increases with time. The value $\widetilde{M_{\rm{K}}} - 5 \log h_{70}$ corresponding to a space density of $10^{-4.0} h_{70}^3 \, \rm{ Mpc}^{-3} \, \rm{mag}^{-1}$ is used to measure evolution of the luminosity of the brightest galaxies from $z \sim 1.1$ to $z \sim 0.3$. The separate plots  shown for red, blue and all galaxies (red data points), assume fixed alpha values of -0.5, -1.2 and -1.0 respectively.  Also shown are results from the literature as listed in each panel and referenced in the caption to Figure \ref{fig:K_phi_star}.}
		\label{fig:K_Mfixed}
\end{figure}

Figure \ref{fig:K_Mfixed} plots the evolving absolute magnitude of galaxies having a space density of $\widetilde{\phi}=10^{-4.0} \, h_{70}^3 \, \rm{Mpc}^{-3} \, {\rm{mag}}^{-1}$, which measures the luminosity evolution of the most luminous galaxies. For red galaxies and blue galaxies there are decreases in luminosity of 0.19 mag and 0.33 mag (0.08 and 0.13 dex) respectively from $z \sim 1.1$ to $z \sim 0.3$.

Figure \ref{fig:K_Mfixed} indicates that highly luminous red galaxies fade at an ever-increasing rate from $z \sim 1.1$ to $z \sim 0.3$.  For a passively evolving model with a formation redshift of $z_f = 3.0$, we expect \s0.56 mag of fading in the $K$-band between $z = 1.1$ and $z = 0.3$. Given the \s0.1 mag fading we see in the absolute magnitudes of the brightest red galaxies, this implies that the most massive red galaxies  increased in mass by a factor of  \s1.4 between $z = 1.1$ and $z = 0.3$.

\begin{deluxetable*}{llll}							
\tablewidth{0pt}							
\tablecolumns{4}							
\tabletypesize{\scriptsize}							
\tablecaption{ORDER OF MAGNITUDE ESTIMATES OF RANDOM AND SYSTEMATIC ERRORS FOR RED GALAXIES - IN DEX}						
\startdata	
\hline\\																
Quantity	&	Details of error	&	Random	&	Systematic	\\						
\hline\\	
\multicolumn{4}{l}{OBSERVATIONAL ERRORS}\\										
\\							
Photometry errors	&	Aperture photometry errors ($0.02$ to $0.4$ mag)	&	$0.01$ to $0.16$\tablenotemark{a}	&	$0.01$	\\
	&	Total flux correction from growth curves ($0.1$ to $0.2$ mag)	&	$0.04$ to $0.1$	&	$0.01$	\\
	&	 $(1+z)^{-4}$ cosmological surface brightness dimming	&	$<0.01$	&	$<0.01$	\\
	&	TOTAL	&	$0.04$ to $0.19$		&	$0.02$	\\
Photometric redshift errors	&	$\lambda = (z_p-z_s)/(1+z_s) = $ 0.05 (random), 0.02 (systematic)	&	0.02	&	0.01	\\						
Space density uncertainties	&	Cosmic variance uncertainty - 8\% by subfields &	$0.03$		& \; -	 \\						
\hline\\	
\multicolumn{4}{l}{ERRORS IN DERIVED QUANTITIES FOR INDIVIDUAL GALAXIES}\\										
\\
Luminosity errors (U, B, V, K) (in dex, not mag)	&	Due to photometric redshift errors 	&	$0.04$	&	$0.02$\\			
	&	Due to photometry errors (as above)	&	$0.02-0.19$\tablenotemark{a}	&	$0.02$	\\
	&	Due to K-correction errors (Equation \ref{eq:absmag_polynomial})	&	$0.02$	&	$0.01$	\\
	&	TOTAL	&	$0.05$ to $0.20$\tablenotemark{a}	&	$0.05$	\\						
$M/L_K$ errors	&	Inaccuracy of $M/L_K$ formula (Equations \ref{eq:ML_G10} and \ref{eq:ML_G10_massive}) 	&	$0.10$	&	$0.03$	\\
	&	[Dependence  of $M/L_K$ formula on assumed 	\\
	&	\qquad SPS model, SFH, metallicity,  \& dust&	\: -	&	$0.30$		\\
	&	\qquad  (may not significantly affect relative mass growth)]	&	\: -	&	[$0.30$]	\\
	&	[Dependence of $M/L_K$ formula on assumed IMF	\\
	&	\qquad  (does not affect relative mass growth)]	&	\: -	&	[$0.30$]	\\				
	&	Impact of photometric redshift errors on evolving	\\				
	&	\qquad $M/L_K$ formula (Equations \ref{eq:ML_G10} and \ref{eq:ML_G10_massive})	&	$0.02$	&	$0.01$	\\
	&	TOTAL (ignoring SPS and IMF error) 	&	$0.10$	&	$0.04$	\\						
Stellar mass errors	&	Due to $M/L_K$ error (as above)	&	$0.10$	&	$0.04$	\\
	&	Due to $K$-band luminosity error (as above)	&	$0.05$ to $0.20$\tablenotemark{a}	&	$0.05$	\\
	&	TOTAL	&	$0.11$ to $0.22$\tablenotemark{a}	&	$0.09$	\\						
\hline\\	
\multicolumn{4}{l}{UNCERTAINTIES IN EVOLVING RED GALAXY LUMINOSITY DENSITIES AND STELLAR MASS DENSITIES,}\\						
\\	
Evolving $K$-band luminosity density errors for red galaxies	&	Due to $K$-band luminosity errors (as above)	& \; -	&	$0.05$	\\	
	&	Due to cosmic variance		&	$0.03$ & \; -	\\
	&	Due to inexactness of the Schechter parameterization	&	$0.03$	& \; -	\\
	&	Error due to $z_{\textrm{phot}}$ scattering across redshift bin boundaries\tablenotemark{b}	&	\; -	&	$0.01-0.07$\tablenotemark{c}	\\
	&	TOTAL	&  $0.04$ &	$0.06-0.12$	\\
Evolving SMD errors for red galaxies	&	Due to stellar mass errors (as above)	& \; -	&	$0.09$\tablenotemark{d}	\\	
	&	Due to cosmic variance	&	$0.03$ & \; -		\\
	&	Due to inexactness of the Schechter parameterization	&	$0.03$ & \; -		\\
	&	Error due to $z_{\textrm{phot}}$ scattering across redshift bin boundaries\tablenotemark{b}	&	\; -	&	$0.01-0.09$\tablenotemark{c}	\\	
	&	TOTAL	& $0.04$	&	$0.10-0.18$\tablenotemark{d}	\\
\hline\\	
\multicolumn{4}{l}{UNCERTAINTIES IN EVOLVING $K$-BAND LUMINOSITIES AND STELLAR MASSES OF LUMINOUS/MASSIVE RED GALAXIES}\\
\\	
$K$-band luminosity errors for highly luminous red galaxies	&	Due to $K$-band luminosity errors	&$<0.01$	&	$0.05$	\\	
	&	Eddington bias from $\sigma=0.03$ scatter in $K$-band luminosities	\\	
	&	\quad due to photometry and K-corrections (but not $z_{\textrm{phot}}$ errors)	&	\; -	&	$0.03$	\\	
	&	Eddington bias due to random $z_{\textrm{phot}}$ errors and bias due to\\		
	&	\quad	 $z_{\textrm{phot}}$ scattering across redshift bin boundaries\tablenotemark{b}	& $<0.01$ 	&	$0.001-0.04$\tablenotemark{c}	\\
	&	TOTAL	& $0.01$	&	$0.08-0.13$\tablenotemark{c}\\
	Stellar mass errors for massive red galaxies	&	Due to stellar mass errors	& $<0.01$ 	&	$0.09$\tablenotemark{d}	\\	
	&	Eddington bias from $\sigma=0.1$ scatter in  $\log_{10}(M/L_K)$ (Equation \ref{eq:ML_G10_massive})	&	\; -	&	$0.22$	\\			
	&	Eddington bias due to random $z_{\textrm{phot}}$ errors and bias due to\\		
	&	\quad	 $z_{\textrm{phot}}$ scattering across redshift bin boundaries\tablenotemark{b}	& $<0.01$ 	&	$0.001-0.04$\tablenotemark{c}	\\
	&	TOTAL	& $0.01$	&	$0.11-0.15$\tablenotemark{c} 	\\					
\hline\\													
\multicolumn{4}{l}{NET EFFECT ON RED GALAXY RESULTS OF RANDOM FRACTIONAL PHOTOMETRIC REDSHIFT ERROR $\sigma = 0.05$ (FROM SIMULATION)}	\\	Red galaxy $K$-band luminosity density			&	&  $\leq0.01$  &	$0.01-0.07$\tablenotemark{c}		\\	
$K$-band luminosity at fixed space density for luminous red galaxies			&	&  $\leq0.01$  &	$0.001-0.04$\tablenotemark{c}		\\	
Red galaxy stellar mass density			&	&  $\leq0.01$  &    $0.01-0.09$\tablenotemark{c}		\\	
Stellar mass $\widetilde{M}$  at fixed space density for massive red galaxies			&	&  $\leq0.01$  &	$0.001-0.04$\tablenotemark{c}			\\
\\
\multicolumn{4}{l}{NET EFFECT ON RED GALAXY RESULTS OF SYSTEMATIC FRACTIONAL PHOTOMETRIC REDSHIFT ERROR OF +0.01 (FROM SIMULATION)}	\\		Red galaxy $K$-band luminosity density			&	&	\; -	&	$-0.02$		\\	
$K$-band luminosity at fixed space density for luminous red galaxies			&	&	\; -	&	$-0.01$		\\	
Red galaxy stellar mass density			&	&	\; -	&    $-0.03$		\\	
Stellar mass $\widetilde{M}$  at fixed space density for massive red galaxies			&	&	\; -	&	$-0.01$	
\enddata			
\label{tab:errors}
\tablenotetext{a}{\footnotesize{Aperture photometry errors depend on the waveband and increase with apparent magnitude.}}		
\tablenotetext{b}{\footnotesize{From Monte Carlo simulations, as shown below'}}
\tablenotetext{c}{\footnotesize{Error increases with redshift, largely due to decreasing proportion of accurate spectroscopic redshifts.}}	
\tablenotetext{d}{\footnotesize{Assuming no uncertainty due to variation with redshift of SPS model predictions of stellar mass.}}		
\end{deluxetable*}																			
																		
\newpage											

\subsection{$K$-band luminosity evolution - errors}
\label{sec:luminosity_evolution_errors}

Random and/or systematic uncertainties arise from the aperture photometry, photometric redshifts, cosmic variance, absolute magnitude determinations (K-corrections), Schechter parameterization of the LF, choice of red/blue cut, and Eddington bias at the bright end. These uncertainties impact our measurements of evolution of the LF, the red galaxy luminosity density and the luminosity of highly luminous red galaxies.

As an important focus in this paper is to measure evolution of luminosity and stellar mass in red galaxies, we give order of magnitude estimates for uncertainties for red galaxies in Table \ref{tab:errors}. The table indicates how uncertainties propagate through our calculations of LF and red galaxy stellar mass evolution, and shows which uncertainties are of greatest significance. Note that all uncertainties in the table have been given in dex for consistency (e.g. magnitudes have been multiplied by 0.4).  Most of the uncertainties are discussed in detail in \citet{beare14} and \citet{beare15} but we now discuss each of them in the context of the present paper.

\textit{* Photometry}  We discuss photometry first as it forms the observational basis of all our work, enabling us to determine both galaxy luminosities and the photometric redshifts we used for the majority of galaxies. As described in \citet{beare15}, our apparent magnitudes were measured using variable size photometric apertures and corrections for total flux. These aperture sizes and corrections were based on analysis of growth curves of measured magnitude with aperture diameter for isolated galaxy images (i.e. those that did not overlap other objects). This method has the advantage that it is empirically based, and does not assume any specific light profile (e.g. Sersic, de Vaucoleurs, Petrosian, SDSS modelmag). Further, as described in \citet{brown07}, the use of SExtractor segmentation maps largely eliminates flux from neighbouring objects and corrects for the excluded flux.

We do not consider $(1+z)^{-4}$ cosmological surface brightness dimming (e.g. Calvi et al. 2014) to be a significant issue as the angular size at $z\sim 1$ of a large galaxy with half-light radius of 10 kpc is only 0.6 arcsec, and this is less than the FWHM of our 1.35 arcsec point spread function. Such a galaxy would only be partially resolved and be approximately a point source, which largely mitigates the effect of cosmological surface brightness dimming.

Random uncertainties in measured apparent magnitudes arise from the aperture photometry and from the total flux corrections we make. Aperture photometry uncertainties are greater in some wavebands than others. Overall, averaged over all the galaxies with different apparent magnitudes in our sample, 1-$\sigma$ uncertainties are \s0.1 mag for $B_W$, $R$, $I$ and $([3.6 \mu m]$, \s0.2 mag for $J$ and $[4.5 \mu m])$, \s0.4 mag for $K$. Uncertainties are significantly less for galaxies with brighter apparent magnitudes, as used in measuring the bright end of the $K$-band LF (and the massive end of the SMF). For example, 1-$\sigma$ $K$-band uncertainties for galaxies with reliable photometry in different redshift bins (the filled circles in Figures \ref{fig:K_LF_binned_redandblue} to \ref{fig:K_LF_binned_blue} are \s0.3 mag, but for galaxies with $M_K<-23$, as used in measuring evolution of the bright end of the LF, $K$-band uncertainties increase from 0.04 mag at $z\sim0.3$ to 0.15 mag at $z\sim1.1$. From our growth curves of magnitude with aperture diameter we estimate that random uncertainties in the total flux corrections range from \s0.04 mag for the brightest objects  to \s0.1 mag for the faintest. Adding these in quadrature to the uncertainties in aperture photometry, we obtain total random photometry errors ranging from \s0.1 mag to \s0.54 mag (\s0.04 dex to \s0.22 dex in Table \ref{tab:errors}).

Systematic uncertainties in aperture photometry will vary between different observing campaigns and we do not attempt to estimate them here. However, we note that their major effect will be to shift computed luminosities by a constant factor without affecting evolutionary rates of change. We take systematic magnitude uncertainty due to our total flux corrections to be zero as these corrections are as likely to be overestimates as underestimates.

\textit{* Poisson uncertainties and cosmic variance} Poisson uncertainties in both the binned and the maximum likelihood space densities are greater at the bright end of the LF where galaxy numbers are fewer, as can be seen in Figures \ref{fig:K_LF_binned_redandblue} to \ref{fig:K_evolution_blue}. Cosmic variance produces additional random uncertainty in space densities, which \citet{beare15} found using subfields to be  $\sim$3\% for all galaxies (0.01 dex), and somewhat more for red galaxies that are more strongly clustered, i.e. $\sim$8\% (0.03 dex). In combination, these two random uncertainties dominate in all redshift bins over systematic uncertainties arising from our very small systematic fractional photometric redshift errors which are 0.02 at most. This size of systematic error changes the number of galaxies in a redshift bin of width 0.2 by less than 0.01 dex, (e.g. at $z \sim 1$, a shift of 0.02 in the upper boundary of the bin, changes the number in the bin by a factor of $\sim0.202/0.200$ and this equates to 0.0043 dex).

\textit{* Luminosity errors} Errors in $U, B, V$ and $K$-band luminosity measurements arise from photometric redshift uncertainties, photometry uncertainties, and uncertainties in K-correction calculations (Equation \ref{eq:absmag_polynomial}). All these uncertainties are discussed in detail in \citet{beare14}. Table \ref{tab:errors} indicates how the three sources of uncertainty combine in the case of red galaxy luminosity calculations. K-correction uncertainties are due to the scatter of individual galaxies about the polynomials used to measure absolute magnitudes (Equation \ref{eq:absmag_polynomial}). In Figure \ref{fig:calibration_example} the 1-$\sigma$ K-correction uncertainty is given in the top left hand corner of the plot.

\textit{* Photometric redshifts} Photometric redshift uncertainties impact our calculations in several ways: calculation of galaxy luminosities (K-corrections), random and systematic uncertainties in binned and maximum likelihood  LF numbers, Eddington bias at the bright end of the LF and scattering of galaxies from one redshift bin to another. 

As Figure \ref{fig:photoz} shows, our 1-$\sigma$ random fractional photometric redshift errors ($\lambda = [z_{\textrm{phot}}-z_{\textrm{spec}}] \, / \, [1 + z_{\textrm{phot}}]$) are $\sim0.05$ (0.02 dex) or less. Our systematic fractional $z_{\textrm{phot}}$ errors are very small, being $< 0.01$ (0.004 dex) for $0.2 < z_{\rm{phot}} < 1.0$ and  $< 0.02$ (0.008 dex) for  $1.0 < z_{\rm{phot}} < 1.2$. 

The impact of photometric redshift errors is mitigated, especially at the bright end of the LF, by the use of spectroscopic redshifts which are available for \s75\% of red and blue $I < 19.5$ galaxies and \s45\% of  $I < 20.5$ galaxies.

Photometric redshift uncertainties perturb measured $K$-band restrame magnitudes by altering distance moduli $D_M$ in Equation \eqref{eq:absmag_polynomial}. Assuming an approximate inverse square law, the fractional uncertainties in measured luminosities will be twice the fractional uncertainties in  $z_{\textrm{phot}}$ values, i.e. random fractional uncertainties of \s0.02 dex, and systematic fractional uncertainties of 0.01 dex for $0.2 < z_{\rm{phot}} < 1.0$ and \s0.02 dex for  $1.0 < z_{\rm{phot}} < 1.2$.

It is important to realise that these estimated errors arising from  $z_{\textrm{phot}}$ errors are for individual galaxies. In the case of systematic redshift errors, the impact on  \textit{evolutionary} measurements depends on the distribution of $z_{\textrm{phot}}$ errors with redshift. Figure \ref{fig:photoz} shows that photometric redshifts are very slightly overestimated at $z \sim 0.25$ and $z \sim 0.7$, and underestimated at $z \sim 1.0$. We do not pursue this further here, beyond noting that the effect of $z_{\textrm{phot}}$ errors on evolutionary measurements will potentially be comparable with that on individual values.

To gauge the  impact of systematic $z_{\textrm{phot}}$ errors, we repeated all our calculations twice using $z_{\textrm{phot}}$ values increased and decreased by the fractional systematic error over most of the redshift range (i.e. $0.2 < z \leq1.0$)  as shown in Figure \ref{fig:photoz}. This was $\lambda = [z_{\textrm{phot}}-z_{\textrm{spec}}] \, / \, [1 + z_{\textrm{phot}}] = 0.01$. For the increased (decreased) $z_{\textrm{phot}}$ values, red galaxy luminosity density values decreased (increased) by up to  \s0.02 dex while massive red galaxies showed a luminosity increase (decrease) of up to \s0.01 dex.

To measure the effect of the random $z_{\textrm{phot}}$ uncertainties $\lambda = [z_{\textrm{phot}}-z_{\textrm{spec}}] \, / \, [1 + z_{\textrm{phot}}] = 0.05$ we repeated our calculations ten times, each time applying normally distributed random fractional  errors ($\sigma = 0.05$) to individual $z_{\textrm{phot}}$ values. Our results are shown in the penultimate section of Table \ref{tab:errors}. We found that individual measured values of both red galaxy luminosity density and highly luminous red galaxy luminosity differed between simulations by less than 0.01 dex, indicating that random photometric redshift errors did not produce significant scatter in these two measurements.

However, due to random $z_{\textrm{phot}}$ uncertainties, the measured luminosities of luminous red galaxies (at fixed space density) showed a small systematic shift which increased from \s0.01 dex for $0.2 < z \leq 0.4$ to \s0.04 for $1.0 < z \leq 1.2$. This shift is partly an Eddington shift \citep{eddin13} caused by greater numbers of galaxies being randomly scattered to higher luminosities than to lower luminosities, on account of the steeply declining exponential shape of the LF at high luminosity. The remainder of the observed systematic shift is due to galaxies being scattered between redshift bins: if the LF evolves with redshift this scattering across redshift bin boundaries will modify the measured LFs in individual redshift bins. As the fraction of red galaxies with spectroscopic redshifts decreases with increasing redshift, the impact of redshift errors on the  bright end of the luminosity function increases with redshift. All these different effects of random redshift errors are highly correlated and very difficult to analyse analytically. This is the reason why we used Monte Carlo simulations to measure the overall effect of random $z_{\textrm{phot}}$ uncertainties.

Monte Carlo simulations also show that random $z_{\textrm{phot}}$ errors  give rise to a systematic decrease in measured luminosity density. This systematic error ranges from \s0.01 dex at $0.2<z\leq0.4$, where a significant proportion of galaxies have accurate spectroscopic redshifts, to 0.07 dex at $1.0<z\leq1.2$, where few galaxies have spectroscopic redshifts. The error in measured luminosity densities can be attributed to galaxies being scattered across redshift bin boundaries with changed measured luminosities. For example, a galaxy with $\Delta z_{\textrm{phot}} = 0.05$ and $z = 0.98$  will be scattered upwards across the $z = 1$ boundary into the next redshift bin and its measured luminosity will be increased by \s10\%, assuming an approximate inverse square law. The space density at the faint end of the LF \textit{increases}  with increasing luminosity, rather than decreases, so at this end there is a net flow of galaxies to lower redshift bins and lower luminosities. The observed systematic change in luminosity density in each redshift bin is the net result of perturbed luminosity values and of galaxies being scattered in and out of the redshift bin at the upper and lower bin boundaries.

In addition to random photometric redshift errors, random photometry and K-correction errors also contribute to the random errors in our $K$-band luminosity measurements. This increases the Eddington shift in our measurements of the bright end of the LF for red galaxies.  We modelled the photometric and K-correction errors by convolving the Schechter function for red galaxies with a Gaussian with $\sigma = 0.03$ dex (Table \ref{tab:errors}) and found that the resultant shift in the luminosity corresponding to our fixed space density of $\widetilde{\phi} = 10^{-4.0} {h_{70}}^3 {\rm{Mpc}}^{-3} {\rm{mag}}^{-1}$ was increased by 0.03 dex.

\textit{* Schechter parameterization} In measuring evolution of the $K$-band LF and the $K$-band luminosity density, uncertainty arises from parameterizing the LF using a single Schechter function and using this parameterization to compute these quantities.  \citet{beare15}  found that the $B$-band luminosity density varied by less than 0.03 dex when $\alpha$ was varied by $\pm0.1$, pointing out that for red galaxies it is insensitive to the exact value of $\alpha$ chosen, even though $M^*$ and $\phi^*$ vary considerably. We assume that similar conclusions will apply in the case of the $K$-band LF and take 0.03 dex as the random uncertainty in $K$-band luminosity density due to the inexactness of the Schechter parameterization.

\textit{* Red-blue cut} Our evolving red-blue cut is intended to separate quiescent and star forming galaxies, but the correspondence is not exact due to factors such as residual star formation in some quiescent galaxies, and dust obscuration which causes some star forming galaxies to appear redder than they really are. The best position to adopt for the red-blue cut involves some uncertainty. We investigated this in \citet{beare15} and found that moving the position of the cut up or down by 0.05 mag made 0.05 dex difference to the measured optical luminosity density of red galaxies and 0.02 dex difference to the measured luminosity of luminous red galaxies. We assume here that similar differences will apply to the $K$-band LF and the SMF. 

\textit{* Conclusion} Table \ref{tab:errors} shows that systematic errors in red galaxy luminosity density range from \s0.06 dex at $z = 0.3$ to \s0.12 dex at $z = 1.1$ and are larger than the random uncertainties of \s0.04 dex. Potentially, correcting for the change in systematic error with redshift could decrease the measured increase of luminosity density from $z = 1.1$ to $z = 0.3$ by \s0.06 dex, altering the luminosity density growth from $0.08 \pm0.04$ dex ($\times 1.20 \pm0.11$) to $0.02 \pm0.04$ dex ($\times 1.05 \pm0.10$).

Systematic errors in the measured luminosity of highly luminous red galaxies range from \s0.08 dex at $z = 0.3$ to \s0.13 dex at $z = 1.1$ and dominate the random uncertainties which Monte Carlo simulations show to be less than 0.01 dex.  Potentially, correcting for the change in systematic error with redshift could decrease the measured luminosity fading of highly luminous red galaxies from $z = 1.1$ to $z = 0.3$ by \s0.05 dex, altering the measured luminosity decrease from $0.08\pm0.01$ dex ($\times 0.83 \pm0.03$) to $0.03\pm0.01$ dex ($\times 0.93 \pm0.02$).

To summarise, systematic errors could significantly reduce the 20\%  growth in red galaxy luminosity density to 5\% and significantly reduce the 17\%  fading of luminous red galaxies to 7\%.

\begin{figure}
 	\centering
	\includegraphics[width=0.49\textwidth]{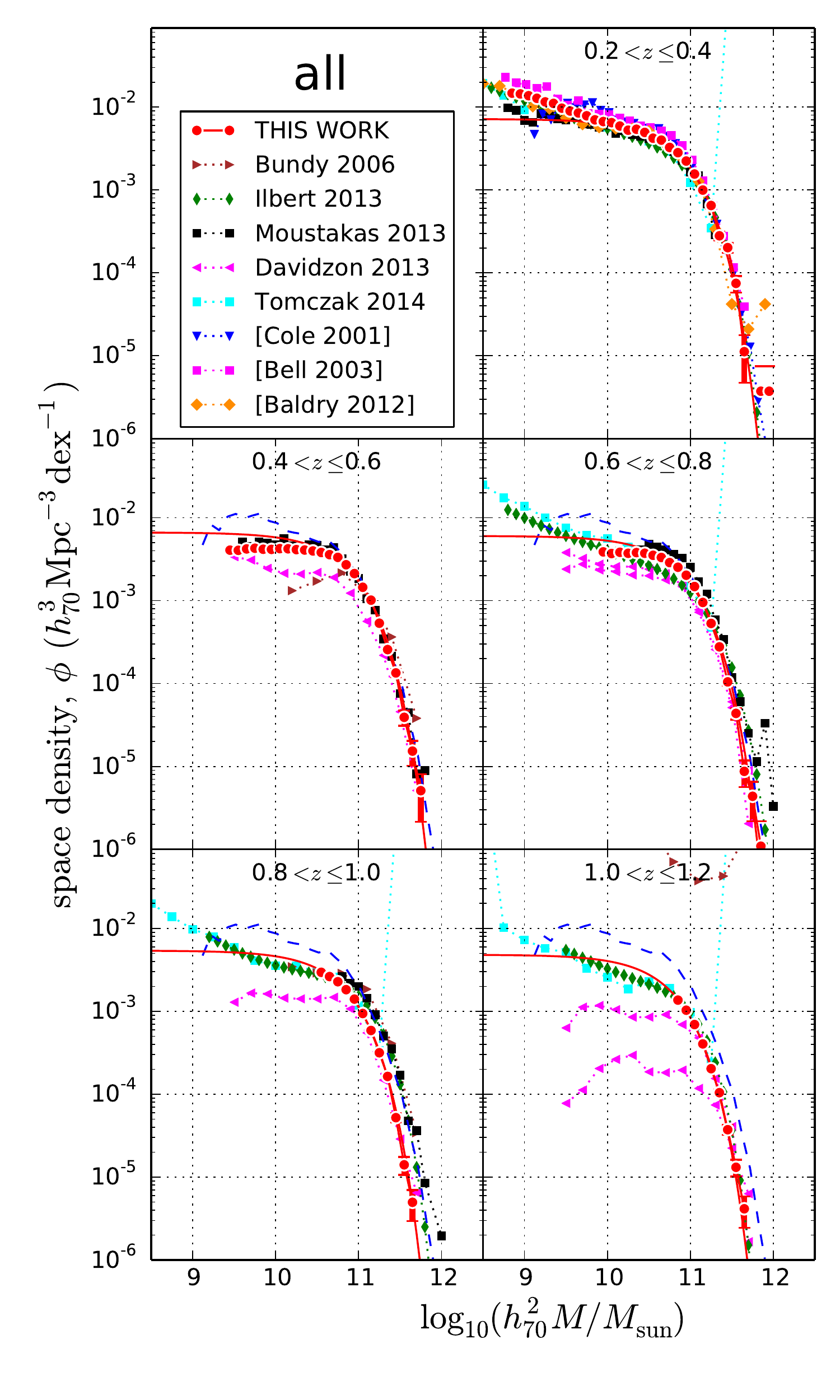}
	\caption{Binned SMFs for all galaxies based on $K$-band $M/L$ ratios  with $1-\sigma$ Poisson uncertainties shown for the \bootess data. Overplotted in red are maximum likelihood fits to the (unbinned) data. We show the evolving SMFs from \citet{bundy06, ilber13, moust13, david13} and \citet{tomcz14} for comparison, as well as SMFs for the low redshift Universe from \citet{cole01}, \citet{bell03} and \citet{baldr12}. The \citet{cole01} SMF for all galaxies in the low redshift Universe is shown in the lower four plots as a black dashed line in order to provide a fixed reference. The plots from \citet{david13} are for $z \sim 0.55$, 0.65, 0.75, 0.85, 1.0 and 1.2.}
	\label{fig:literature_mass_comparison_redandblue_log_K_G10}
\end{figure}

\begin{figure}
 	\centering
	\includegraphics[width=0.49\textwidth]{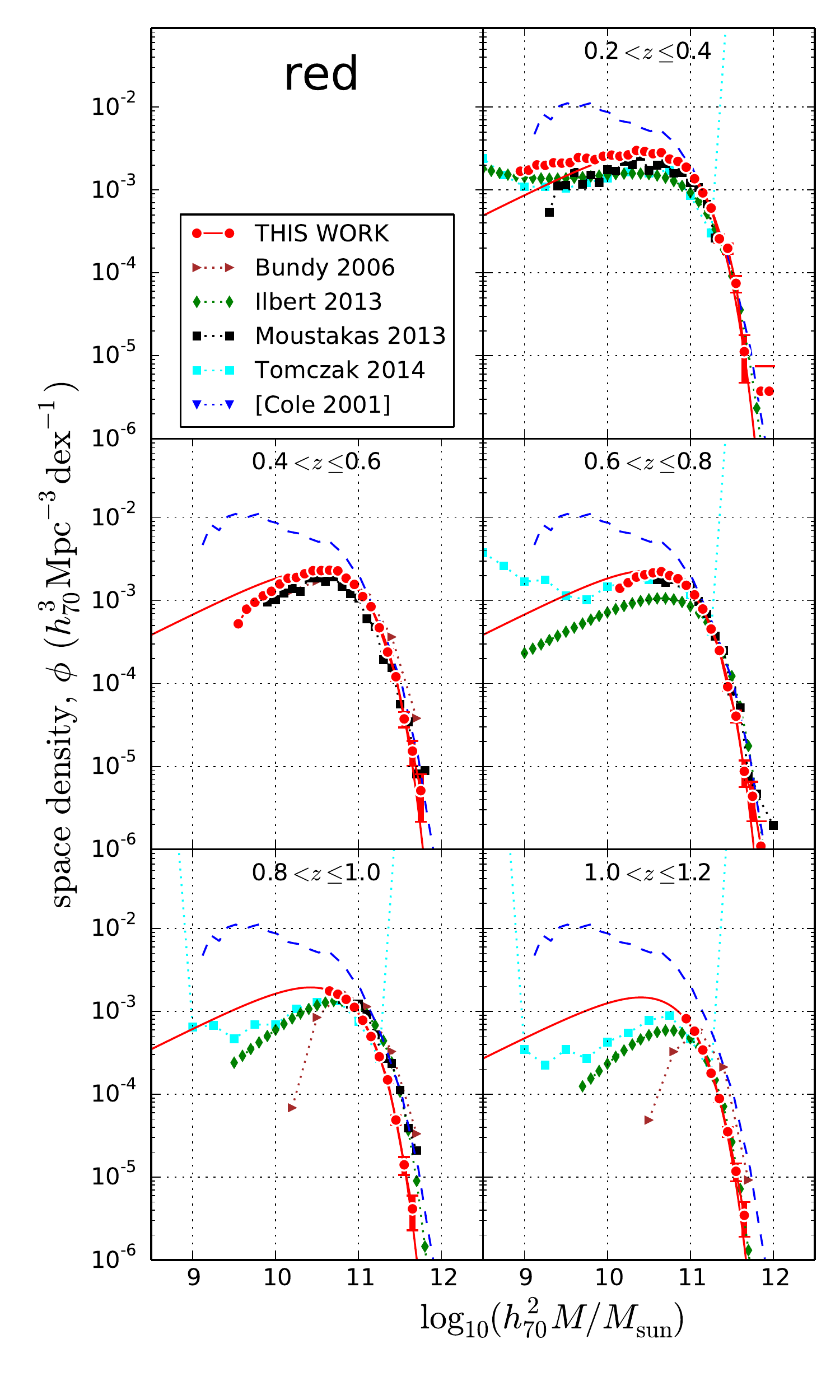}
	\caption{Binned SMFs for red galaxies based on $K$-band $M/L$ ratios  with $1-\sigma$ Poisson uncertainties shown for the \bootess data. Overplotted in red are maximum likelihood fits to the (unbinned) data. We show the evolving SMFs from \citet{bundy06, ilber13, moust13} and \citet{tomcz14} for comparison.  The \citet{cole01} SMF for all galaxies in the low redshift Universe is shown in the lower four plots as a black dashed line in order to provide a fixed reference.  The stellar masses of the most massive red galaxies increase with time due to minor mergers.}
	\label{fig:literature_mass_comparison_red_log_K_G10}
\end{figure}

\begin{figure}
 	\centering
	\includegraphics[width=0.49\textwidth]{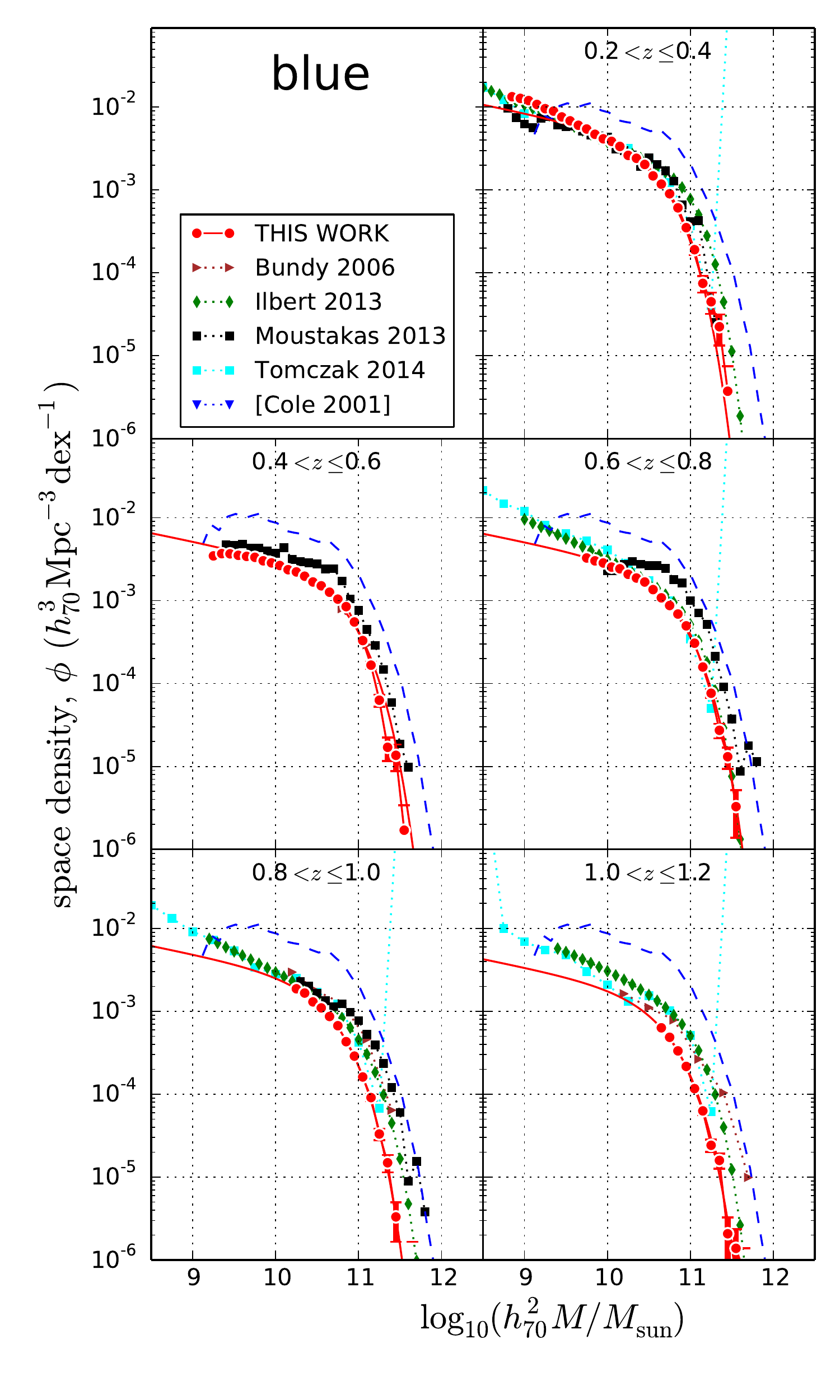}
	\caption{Binned SMFs for blue galaxies based on $K$-band $M/L$ ratios  with $1-\sigma$ Poisson uncertainties shown for the \bootess data.  Overplotted in red are maximum likelihood fits to the (unbinned) data.  We show the evolving SMFs from \citet{bundy06, ilber13, moust13} and \citet{tomcz14} for comparison. The \citet{cole01} SMF for all galaxies in the low redshift Universe is shown in the lower four plots as a black dashed line in order to provide a fixed reference.}
	\label{fig:literature_mass_comparison_blue_log_K_G10}
\end{figure}

\subsection{$K$-band luminosity evolution - comparison with the literature}
\label{sec:luminosity_evolution_comparison}

As Figures \ref{fig:K_LF_binned_redandblue} and \ref{fig:K_LF_binned_redandblue_closeup} show, the bright end of our LF for all galaxies at $0.2 \leq z < 0.4$ is \s 0.3 mag brighter than the LFs reported by several authors for the low redshift ($z < 0.2$) Universe.

A small part of this difference will be due to evolution of the LF between $z \sim 0.3$ and $z \sim 0.1$, but we  expect our measured magnitudes to be brighter than those from studies such as \citet{kocha01},  \citet{cole01},  \citet{bell03} and \citet {bonne15} which used the Two Micron All Sky Survey extended source catalog data \citep[2MASS, ][]{jarre00}. Because 2MASS is relatively shallow, the faint outer parts of extended sources become lost in the 2MASS sky background. \citet{andre02} demonstrated that the isophotal magnitude models used by authors such as \citet{kocha01} and \citet{cole01} underestimate the flux from extended sources by even more than the \s0.2 mag that these authors predicted, even in the case of bright galaxies. \citet{andre02} also showed that 2MASS failed to detect low surface brightness galaxies. 

By contrast, we accounted for light falling outside the photometric aperture by using growth curves of measured apparent magnitude with aperture size \citep{beare15}. E.g. for a galaxy with an apparent $I$-band magnitude of 20.5 mag we used an aperture of diameter 8 arcsec and made a correction for missing flux of -0.070 mag, while at magnitude -23.5 we used a 3 arcsec aperture and made a correction of -0.302.  In \citet{beare15} we compared our photometry with MAG\_AUTO and found it to be brighter by \s0.06 mag for $I < 20$, rising to \s0.13 at $I = 24$. These factors adequately account for the difference between our LFs and those based on 2MASS, as seen in Figures \ref{fig:K_LF_binned_redandblue} to \ref{fig:K_LF_binned_redandblue_closeup}. 

Local surveys do have a strong dependence on the type of photometry used, but an advantage of a deep survey like ours with partially resolved galaxies is a reduced dependence on the intrinsic light profiles of the galaxies. For example, when the galaxy is far smaller than the PSF, aperture photometry with PSF corrections is adequate.  A deep survey, with high signal-to-noise, can also afford to use oversized apertures for photometry of large galaxies whereas shallower surveys using comparable apertures could be swamped by noise.

Figure \ref{fig:K_LF_binned_redandblue} shows that at higher redshifts agreement with the evolving LFs for all galaxies of \citet{drory03} and \citet{ciras10} is somewhat uneven: it is closer at some redshifts than others.  Both these studies used photometric redshifts as we did, but their sample sizes were much smaller than ours (\s70 and \s10 times smaller respectively), so larger Poisson errors and cosmic variance can explain why there were greater differences from our work at some redshifts than others.

Figures \ref{fig:K_phi_star} and \ref{fig:K_M_star} compare our measurements of the evolution of  $\phi_K^*$ and $M_K^*$ with those of
 \citet{ciras10}, \citet{arnou07} and \citet{drory03} and  show values for the low redshift Universe from several authors.  \citet{ciras10} and \citet{arnou07} fitted an evolving functional form to their data so that their parameters were forced to vary smoothly with redshift.
 
It is difficult to compare $\phi_K^*$ and $M_K^*$ evolution measurements from different studies because the well-known degeneracy between the Schechter parameters means that different $\alpha$ value choices give rise to different measured values for $\phi^*$ and $M^*$, e.g. for the $B$-band LF of all galaxies at $0.2 \leq z < 0.4$, \citet{beare15} found that increasing $\alpha$ from -1.1 to -1.0 decreased $\phi_B^*$ by \s20\% while making $M_B^*$ \s0.1 mag brighter. The value $\alpha = -0.9$ adopted by \citet{ciras10} and \citet{drory03} for all galaxies is 0.1 greater than our value of -1.0 and this explains their fainter $M_K^*$  and smaller $\phi_K^*$ values. Similarly, at low redshift, discrepancies between different authors are accounted for by differences in the $\alpha$ values adopted, with \citet{jones06} and \citet{bonne15} using the largest (most negative) values \citep[see values in Table 6 of][]{bonne15}.

Taking into account the degeneracy between $\phi_K^*$, $M_K^*$ and $\alpha$, the lower faint end space densities and fainter magnitudes expected from studies based on 2MASS, and the different subsample criteria used by different authors, we conclude that our $\phi_K^*$ and $M_K^*$ measurements are not inconsistent with those from previous studies.
 
Figure \ref{fig:K_lumdens} shows that we measured a higher luminosity density at all redshifts for all galaxies than the studies of \citet{ciras10} and \citet{drory03}, while obtaining comparable values to \citet{arnou07}, but much smoother evolution. Our measured luminosity density at low redshift is higher than the literature, much of which utilises the relatively shallow 2MASS data.

As Figure \ref{fig:K_Mfixed} shows, bright end galaxies were 0.2 to 0.3 mag brighter than those in \citet{ciras10} and \citet{drory03} and varied more smoothly in luminosity than those of \citet{arnou07}. Recent studies \citep[e.g.][]{dsouz15,loved15, berna16} have shown that measurements of the bright end of the LF are highly sensitive to the photometric model used (e.g. Sérsic, Petrosian, SDSS cmodel) as this affects how much light from the faint outer edges of extended galaxies is measured. Our photometric corrections based on growth curves were model independent and should have provided good estimates of total galaxy light, even though we did not derive different corrections for quiescent and star-forming galaxies which generally exhibit de Vaucouleurs and Sérsic light profiles respectively. We therefore expect the bright end of our LFs to be brighter than much of the literature and Figures \ref{fig:K_LF_binned_redandblue} to \ref{fig:K_LF_binned_blue} show that this is the case by \s0.3 mag.

\begin{deluxetable*}{cccccc}
\tablewidth{0pt}
\tablecolumns{6}
\tabletypesize{\scriptsize}
\tablecaption{Stellar mass function Schechter parameters based on $K$, $V$ and $i$-band mass to light ratios.}
\tablehead{
\colhead{z} & \colhead{$\alpha$} & \colhead{$\phi^*$} & \colhead{$\log_{10} M^*/M_{\sun}$} & \colhead{$\log_{10}M/M_{\sun}$} & \colhead{$\log_{10}$stellar mass density}
\\
\colhead{ } & \colhead{ } & \colhead{$(h_{70}^3 {\rm{Mpc}}^{-3} {\log_{10}}M^{-1})$} & \colhead{ } & \colhead{at fixed space density\tablenotemark{a} (measures} & \colhead{($h_{70}^3 M_{\sun} \, {\rm{Mpc}}^{-3}$)}
\\
\colhead{ } & \colhead{ } & \colhead{} & \colhead{ } & \colhead{evolution of most massive galaxies)} & \colhead{}
}
\startdata
\\
\multicolumn{6}{l}{Red galaxies}\\
0.3 & 	$-0.5$  & 	$2.90\pm 0.11 \times 10^{-3}$ & 	$10.78\pm 0.07$ & 	$11.47\pm 0.07$ & 	$8.19\pm 0.02$ \tabularnewline
0.5 & 	$-0.5$  & 	$2.35\pm 0.16 \times 10^{-3}$ & 	$10.79\pm 0.04$ & 	$11.42\pm 0.08$ & 	$8.11\pm 0.03$ \tabularnewline
0.7 & 	$-0.5$  & 	$2.27\pm 0.11 \times 10^{-3}$ & 	$10.78\pm 0.04$ & 	$11.40\pm 0.06$ & 	$8.08\pm 0.02$ \tabularnewline
0.9 & 	$-0.5$  & 	$1.88\pm 0.04 \times 10^{-3}$ & 	$10.73\pm 0.06$ & 	$11.31\pm 0.05$ & 	$7.95\pm 0.01$ \tabularnewline
1.1 & 	$-0.5$  & 	$1.46\pm 0.13 \times 10^{-3}$ & 	$10.71\pm 0.05$ & 	$11.24\pm 0.10$ & 	$7.82\pm 0.04$ \tabularnewline
\\
\multicolumn{6}{l}{Blue galaxies}\\
0.3 & 	$-1.2$  & 	$1.78\pm 0.04 \times 10^{-3}$ & 	$10.59\pm 0.06$ & 	$11.10\pm 0.05$ & 	$7.90\pm 0.01$ \tabularnewline
0.5 & 	$-1.2$  & 	$0.99\pm 0.04 \times 10^{-3}$ & 	$10.79\pm 0.04$ & 	$11.17\pm 0.03$ & 	$7.85\pm 0.02$ \tabularnewline
0.7 & 	$-1.2$  & 	$0.97\pm 0.02 \times 10^{-3}$ & 	$10.79\pm 0.04$ & 	$11.16\pm 0.03$ & 	$7.85\pm 0.01$ \tabularnewline
0.9 & 	$-1.2$  & 	$1.04\pm 0.04 \times 10^{-3}$ & 	$10.67\pm 0.05$ & 	$11.04\pm 0.04$ & 	$7.75\pm 0.02$ \tabularnewline
1.1 & 	$-1.2$  & 	$0.74\pm 0.05 \times 10^{-3}$ & 	$10.66\pm 0.08$ & 	$10.95\pm 0.06$ & 	$7.59\pm 0.03$ \tabularnewline
\\
\multicolumn{6}{l}{All galaxies}\\
0.3 & 	$-1$  & 	$3.13\pm 0.06 \times 10^{-3}$ & 	$10.87\pm 0.08$ & 	$11.48\pm 0.07$ & 	$8.36\pm 0.01$ \tabularnewline
0.5 & 	$-1$  & 	$2.90\pm 0.13 \times 10^{-3}$ & 	$10.86\pm 0.03$ & 	$11.43\pm 0.05$ & 	$8.32\pm 0.02$ \tabularnewline
0.7 & 	$-1$  & 	$2.62\pm 0.06 \times 10^{-3}$ & 	$10.88\pm 0.03$ & 	$11.42\pm 0.04$ & 	$8.29\pm 0.01$ \tabularnewline
0.9 & 	$-1$  & 	$2.36\pm 0.08 \times 10^{-3}$ & 	$10.80\pm 0.05$ & 	$11.33\pm 0.04$ & 	$8.17\pm 0.01$ \tabularnewline
1.1 & 	$-1$  & 	$2.09\pm 0.16 \times 10^{-3}$ & 	$10.76\pm 0.05$ & 	$11.27\pm 0.05$ & 	$8.08\pm 0.03$ \tabularnewline
\\
\enddata
\label{tab:massfn_maxlhd}
\tablenotetext{a}{\footnotesize{$2.5 \times 10^{-4.0} \, h_{70}^3 \, \rm{Mpc}^{-3} \, \textrm{dex}^{-1}$}. \,\, ($\log_{10}M/M_{\sun}$ \rm{at this fixed space density is measured by fitting a Schechter function with variable $\alpha$ to the massive end of the SMF.)}}
\end{deluxetable*}

\begin{figure}
 	\centering
	\includegraphics[width=0.49\textwidth]{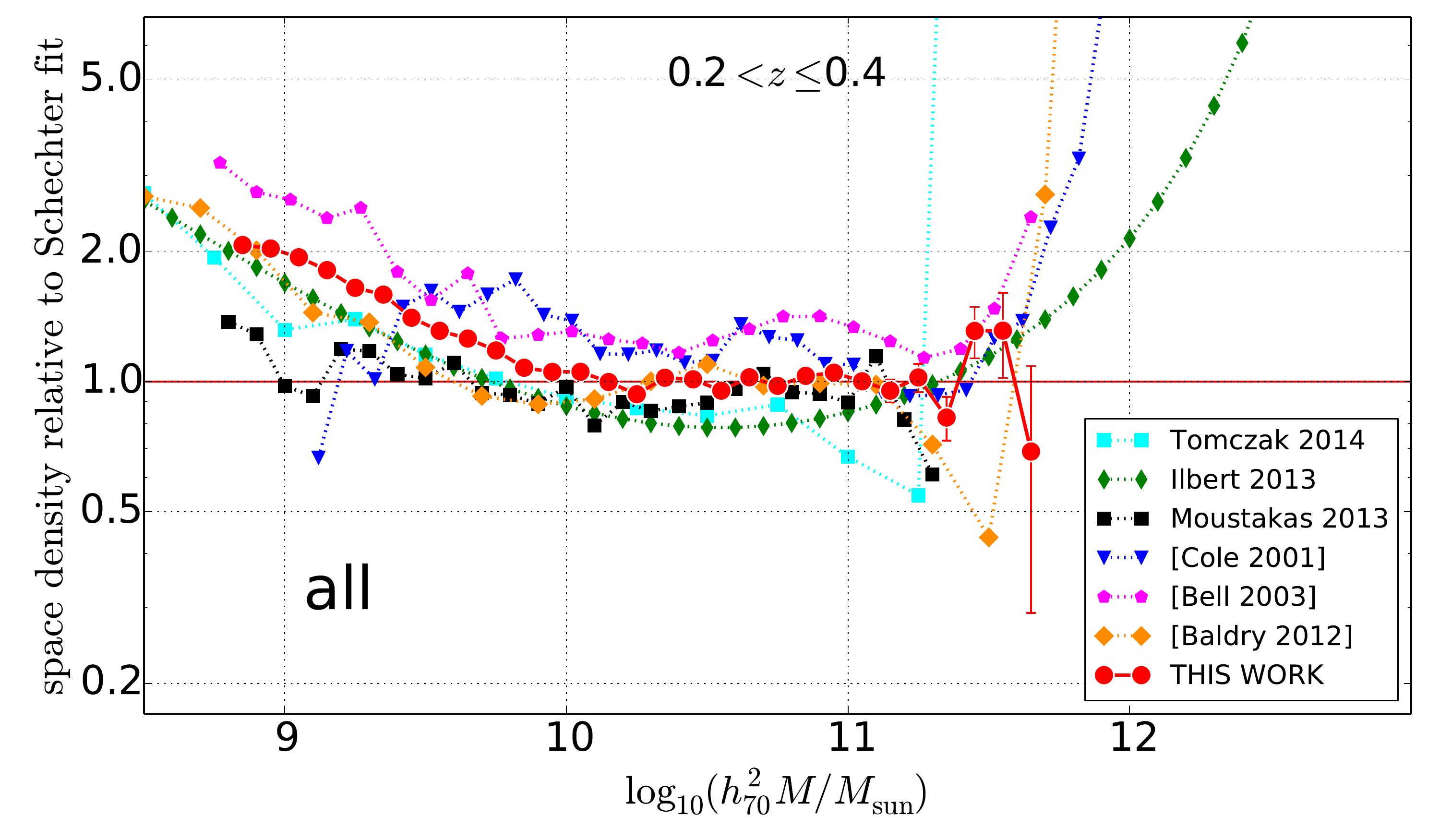}
	\caption{Detailed comparison of space densities for all galaxies at $0.2 \leq z < 0.4$  with $1-\sigma$ Poisson uncertainties shown for the \bootess data. This displays the same data as in the top right panel of Figure \ref{fig:literature_mass_comparison_redandblue_log_K_G10} but plotting the ratio of binned space densities from the literature to the (unbinned) maximum likelihood Schechter function fit to our (unbinned) space densities.  SMFs for the low redshift Universe are labelled using square brackets. There is good agreement over most of the mass range and we do not see a significantly greater density of very massive galaxies than other studies, even though we measure a greater density of highly luminous galaxies in the $K$-band (Figure \ref{fig:K_LF_binned_redandblue_closeup}).}
	\label{fig:MF_binned_redandblue_closeup}
\end{figure}

\section{Stellar mass evolution - results and discussion}
\label{sec:M_results}

\subsection{Evolution of the stellar mass function}

In order to compare our SMF results with the literature we plot our binned SMFs alongside those from a variety of previous studies in Figures \ref{fig:literature_mass_comparison_redandblue_log_K_G10} to  \ref{fig:MF_binned_redandblue_closeup}.  Maximum likelihood fits to our (unbinned) data are over-plotted as continuous red lines.  To provide a fixed reference in the plots we show the local SMF for all galaxies (i.e. red and blue combined) from \citet{cole01} in each bin. We only plot bins for which 97.7\% (the $2\sigma$ limit) or more of the measured masses correspond to magnitudes brighter than our faint limit of $I=24.0$.  

\begin{figure}
 	\centering
	\includegraphics[width=0.49\textwidth]{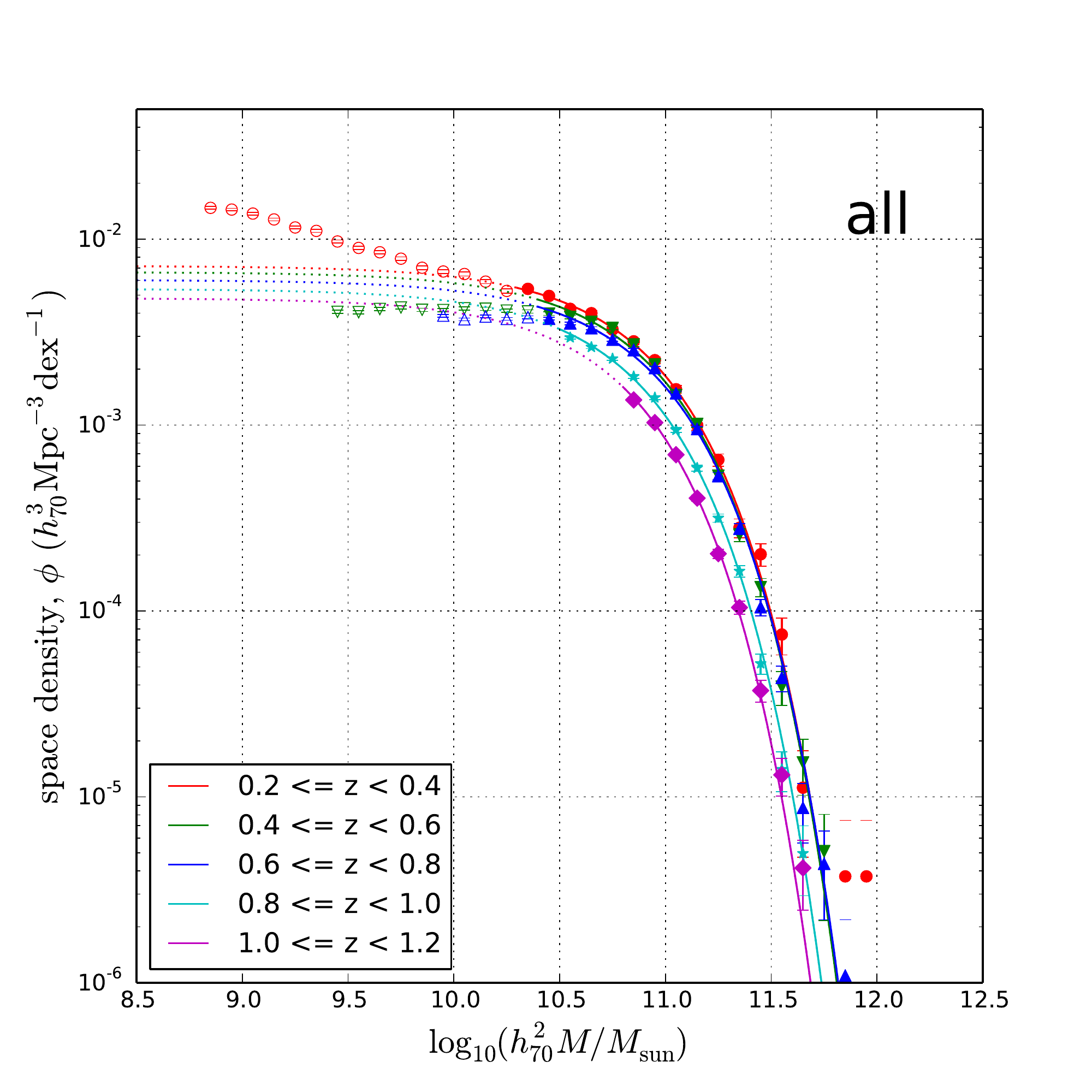}
	\caption{Evolution of the SMFs for all galaxies based on $K$-band $M/L$ ratios, showing all redshift bins in one panel.  Maximum likelihood fits to the (unbinned) data are shown by the continuous curves. The circles denote comoving space densities for the various mass bins. Filled circles denote the mass range used to perform the maximum likelihood fits.  Open circles denote data for very low mass galaxies which are expected to be reliable on the basis of apparent $I$ and $[3.6 \mu \rm{m} \, ]$ magnitudes, but which are not represented adequately by a  Schechter function. The error bars show $1-\sigma$ Poisson errors for the numbers in each bin.}
	\label{fig:MF_logmass_K_redandblue_specified_log_maxlhd_evolution}
\end{figure}

\begin{figure}
 	\centering
	\includegraphics[width=0.49\textwidth]{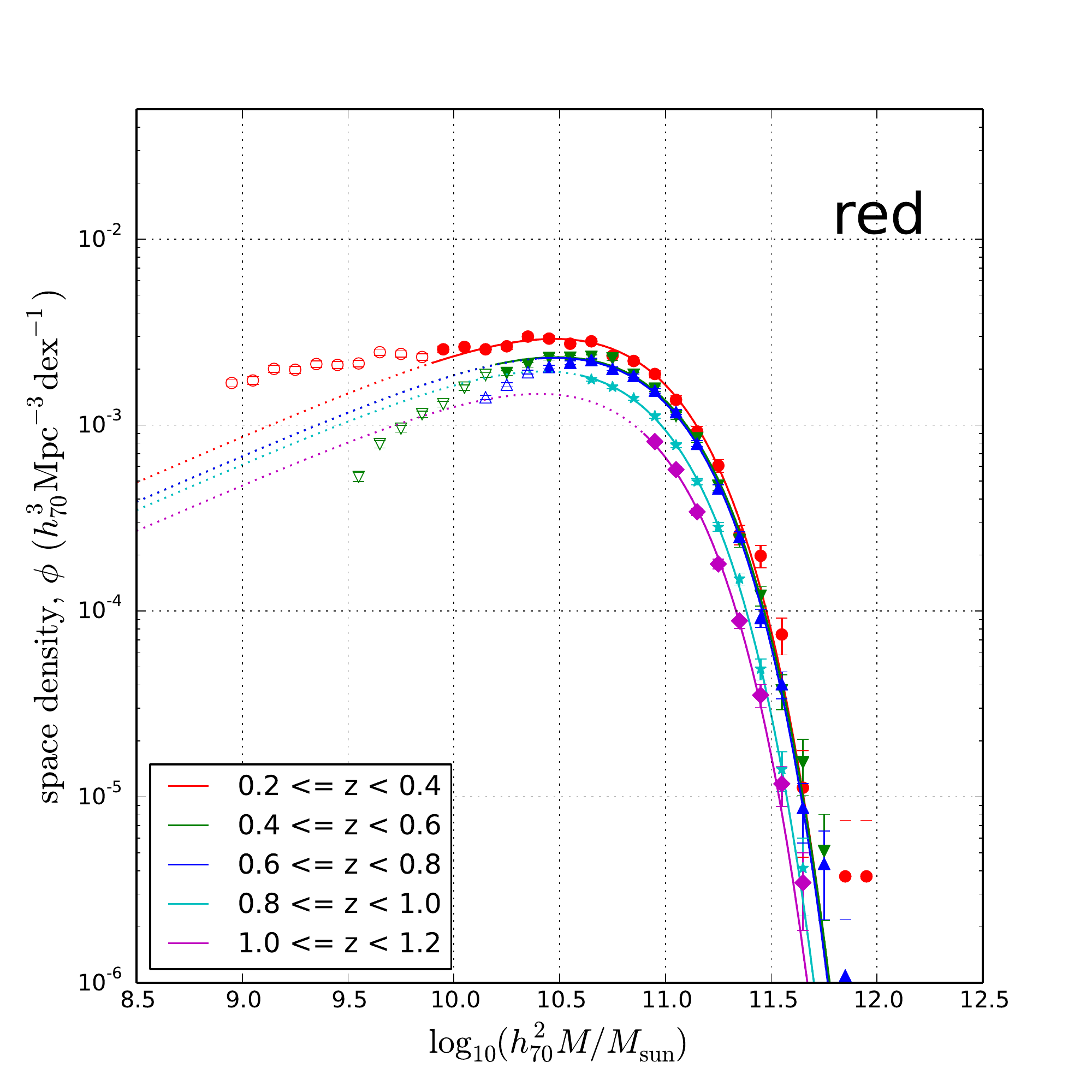}
	\caption{Evolution of the SMFs for red galaxies based on $K$-band $M/L$ ratios, showing all redshift bins in one panel. Symbols are as in Figure \ref{fig:MF_logmass_K_redandblue_specified_log_maxlhd_evolution}. Build up of stellar mass is evident within the red galaxy population as a whole and growth is visible in the stellar mass of the most massive red galaxies.}
	\label{fig:MF_logmass_K_red_specified_log_maxlhd_evolution}
\end{figure}

\begin{figure}
 	\centering
	\includegraphics[width=0.49\textwidth]{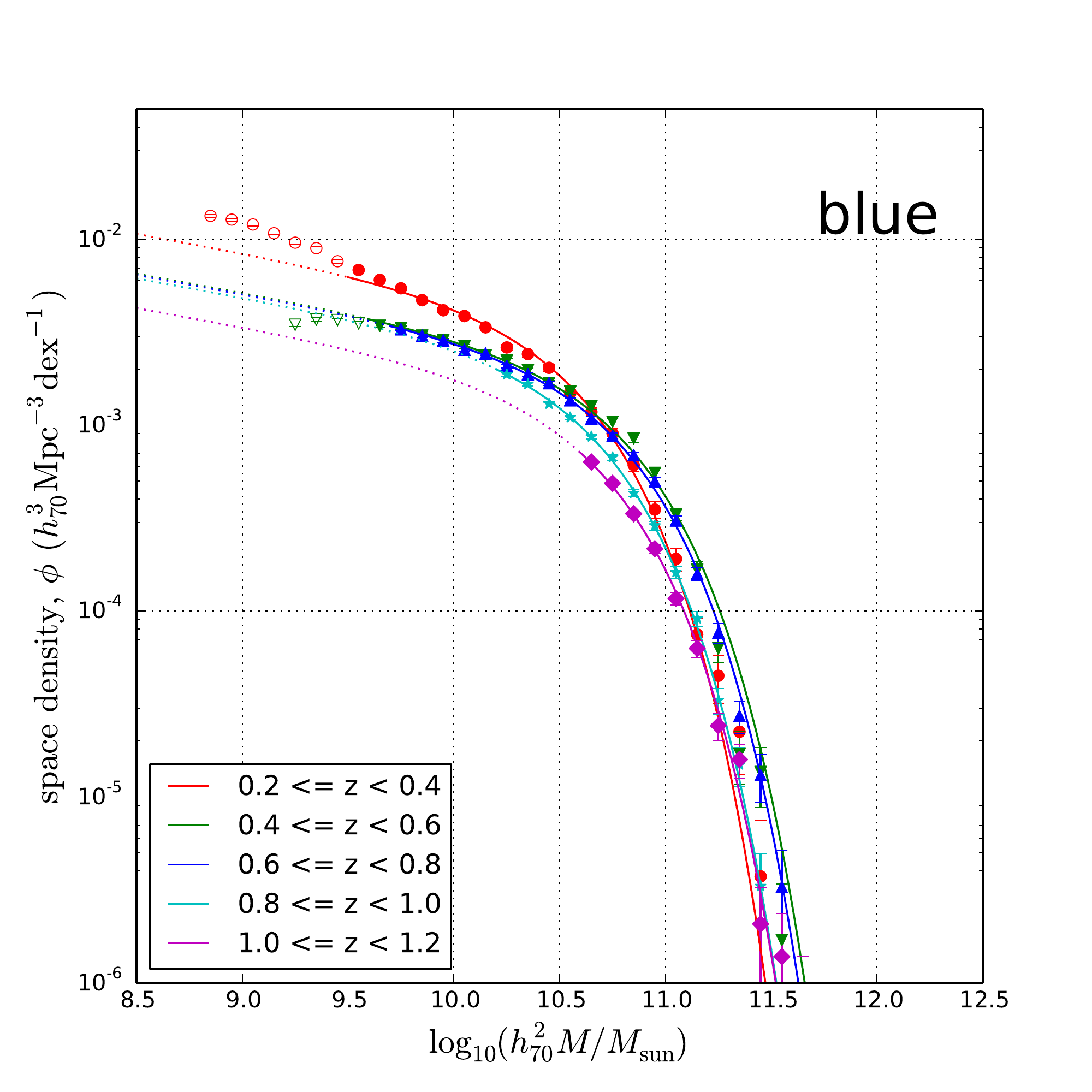}
	\caption{Evolution of the SMFs for blue galaxies based on $K$-band $M/L$ ratios, showing all redshift bins in one panel.  Symbols are as in Figure \ref{fig:MF_logmass_K_redandblue_specified_log_maxlhd_evolution}.}
	\label{fig:MF_logmass_K_blue_specified_log_maxlhd_evolution}
\end{figure}

In order to make evolution of our SMFs more apparent, we show our maximum likelihood Schechter fits (continuous lines) and  binned space densities (data points) in Figures \ref{fig:MF_logmass_K_redandblue_specified_log_maxlhd_evolution} to \ref{fig:MF_logmass_K_blue_specified_log_maxlhd_evolution} with all redshift bins on a single plot. 

Table \ref{tab:massfn_maxlhd} shows that from $z \sim 1.1$ to $z \sim 0.3$ the characteristic space density $\phi_M^*$ approximately doubled for red galaxies and increased by somewhat more ($\sim\times 2.4$) for blue galaxies. At the same time the characteristic mass $M^*$ of both red and blue galaxies changed by no more than 0.07 dex or $\times$ 1.17.

\subsection{Evolution of stellar mass density}
\label{sec:SMD evolution}

\begin{figure}
 	\centering
	\includegraphics[width=0.49\textwidth]{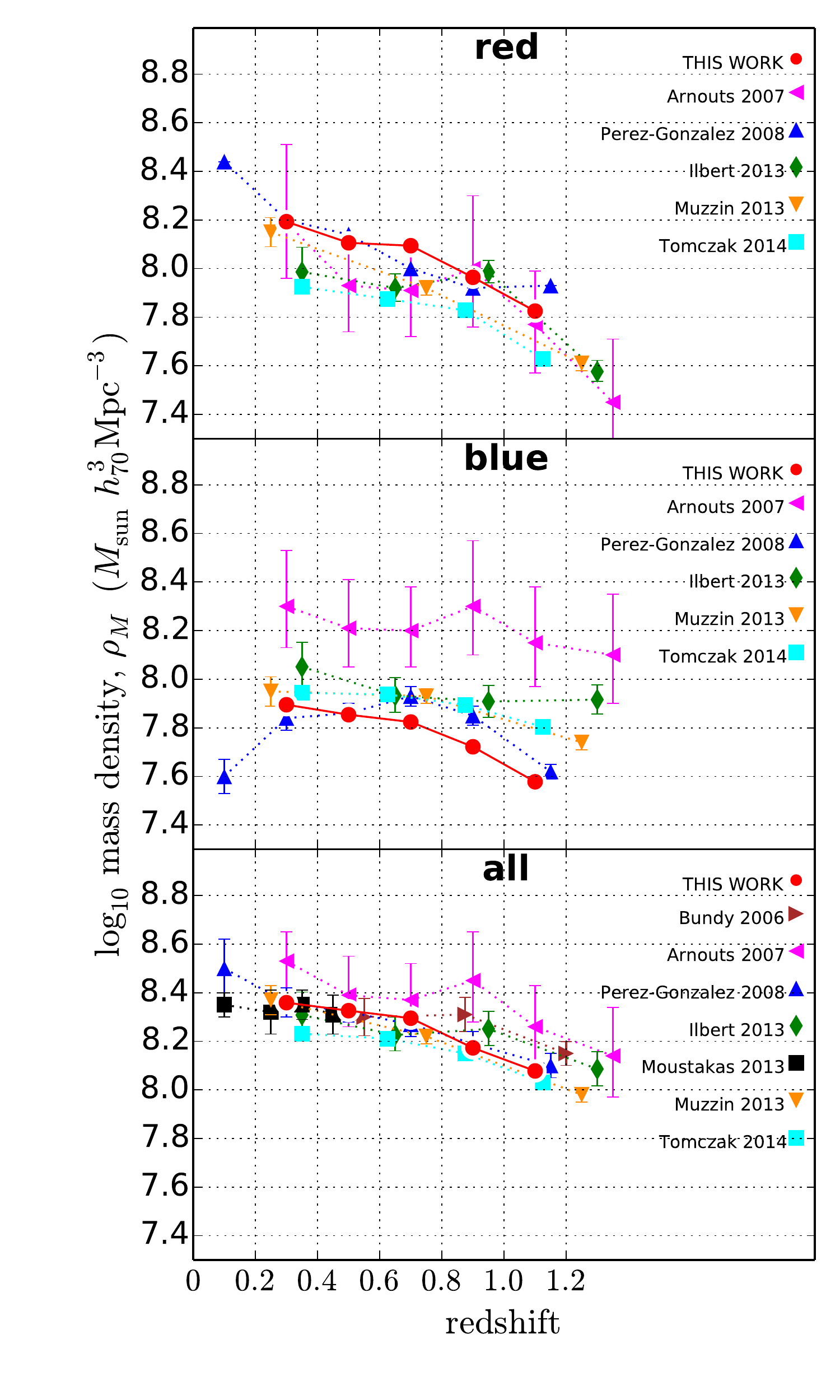}
	\caption{Evolution of the SMD based on $K$-band $M/L$ ratios. From $z \sim 1.1$ to $\sim 0.3$ the SMD for red (blue) galaxies increases by 0.37 (0.31) dex, i.e. a factor of 2.3 (2.1). Both red and blue galaxy SMD have grown at a steady rate. For red galaxies, this indicates that the rate at which blue galaxies move to the red sequence as they cease star formation varies little with time. For blue galaxies it indicates that new stellar mass from star formation is almost balanced by loss of stellar mass to the red sequence as star formation ceases. Error bars on our results show errors due to cosmic variance. Error bars on results from the literature are as published.}
	\label{fig:logmass_K_massdens_all_specified_maxlhd}
\end{figure}

We quantified the growth of stellar mass within the red and blue galaxy populations with the SMD, which has a clear physical meaning and is effectively the area under the SMF curve. Figure \ref{fig:logmass_K_massdens_all_specified_maxlhd} and Table \ref{tab:massfn_maxlhd} show our results based on $K$-band $M/L$ ratios. SMD is not as prone to degeneracies as the Schechter parameters are. For example, we showed in \citet{beare15} how varying the adopted fixed value for $\alpha$ significantly affected the measured values of $M^*$ and $\phi^*$ in the case of the $B$-band LF, but hardly  affected the measured luminosity density at all. Similar behaviour is to be expected in the equivalent case of SMFs and SMD. 

We found an increase of \s0.37 (\s0.31) dex in SMD for red (blue) galaxies from $z = 1.1$ to $z = 0.3$, i.e. a factor of \s2.3 (\s2.1). We note that the red galaxy SMD growth of  $\times2.1$ implied by comparison of $K$-band luminosity evolution with a passively evolving stellar population (Section \ref{sec:K_results_lumdens}) is very close to the $\times2.3$ growth deduced here from evolution of the SMF.  Figure \ref{fig:logmass_K_massdens_all_specified_maxlhd} shows that for red, blue and all galaxies the rate at which SMD is growing decreases slowly with time. For red galaxies, this indicates that the rate at which blue galaxies move to the red sequence as they cease star formation decreases slowly with time.

\subsection{Evolution of massive galaxies}
\label{sec:massive}

\begin{figure}
 	\centering
	\includegraphics[height=0.6\textheight]{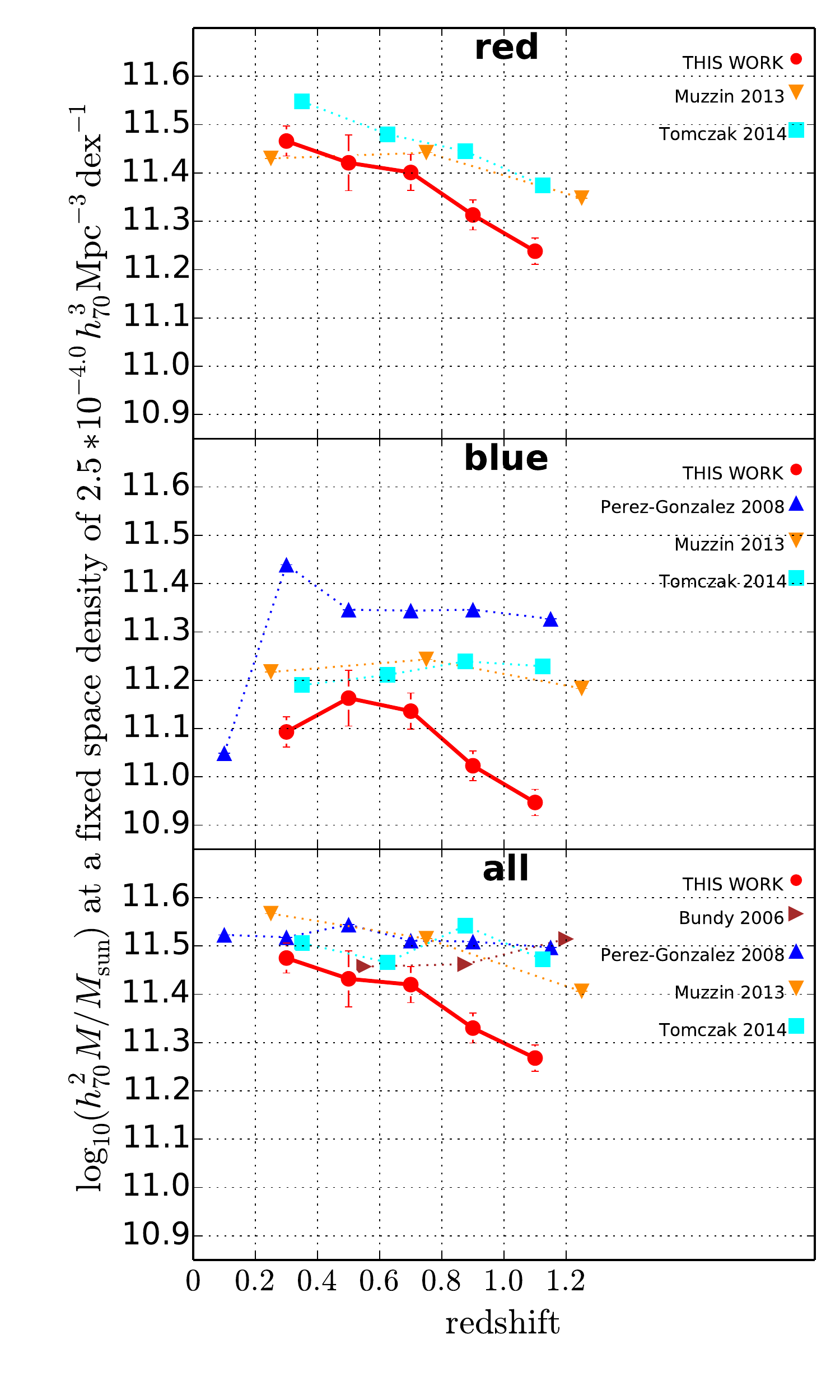}
	\caption{Evolution of very massive galaxies based on $K$-band $M/L$ ratios.  From $z \sim 1.1$ to $\sim 0.3$ the masses of the most massive red (blue) galaxies increased by 0.23 (0.16) dex, i.e. a factor of 1.7 (1.4).  The rate at which massive red galaxies increased in stellar mass through mergers with smaller galaxies appears to have slowed somewhat from \s0.4 dex per unit redshift for $z = 1.1$ to $z = 0.7$ to \s0.2 dex per unit redshift for $z = 0.7$ to $z = 0.3$. The value of $h_{70}^2 \, \log_{10} M/M_{\sun}$ corresponding to a fixed space density of $2.5\times10^{-4.0} h_{70}^3 {\rm{Mpc}}^{-3} {[\log_{10}}M]^{-1}$ effectively measures evolution of the most massive galaxies. Results from the literature have been computed using the published Schechter parameters when available.  Error bars on our results show errors due to cosmic variance. Error bars on results from the literature are as published.}
	\label{fig:logmass_K_Mfixed_all_specified_maxlhd}
\end{figure}

In order to look quantitatively at the mass growth of the most massive galaxies, in Figure \ref{fig:logmass_K_Mfixed_all_specified_maxlhd} we show the redshift evolution of stellar mass at a fixed comoving space density of $\widetilde{\phi} = 2.5 \times 10^{-4} {h_{70}}^3 {\rm{Mpc}}^{-3} \textrm{dex}^{-1}$.  This figure is directly comparable to the evolving luminosity at fixed density shown in Figure  \ref{fig:K_Mfixed}.  These results indicate that these most massive red galaxies grew in stellar mass from $z = 1.1$ to $z = 0.3$ by 0.23 dex, i.e. a factor of \s1.7. Our results also indicate stellar mass growth of 0.15 dex ($\times 1.4$) for massive blue galaxies and 0.21 dex ($\times 1.6$) for all massive galaxies, but it must be remembered that only for red galaxies is the measured stellar mass growth that for individual massive galaxies.

 At a fixed space density threshold of density of $\widetilde{\phi} = 2.5 \times 10^{-4} {h_{70}}^3 {\rm{Mpc}}^{-3} \textrm{dex}^{-1}$, the most massive blue galaxies are \s0.3 dex lower in stellar mass than the most massive red galaxies, i.e. half the mass. While we expect the most massive red galaxies to remain on the red sequence and increase in mass via mergers, it is likely that the most massive blue galaxies have star formation quenched and then move onto the red sequence.
 
The rate at which massive red galaxies increase in stellar mass through mergers with smaller galaxies appears to slow somewhat from \s0.4 dex per unit redshift for $z = 1.1$ to $z = 0.7$ to \s0.2 dex per unit redshift for $z = 0.7$ to $z = 0.3$. 

We know that massive early-type galaxies must have grown in stellar mass because observations clearly show them experiencing mergers sufficient in frequency and mass ratio to give rise to a significant increase in stellar mass. In fact merger studies have produced a  range of estimates of stellar mass growth. For example, \citet[][]{dokku05} found a stellar mass increase due to mergers in massive red galaxies of $\Delta M/M = 0.09 \pm 0.04$ per Gyr, which implies a stellar mass increase over the 4.7 Gyr from $z = 1.1$ to $z = 0.3$ of $\times \sim 1.4$. Similarly, \citet{lopez12} found stellar mass growth in massive early type galaxies due to mergers of $\times \sim 1.3$ from $z = 1$ to the present. On the other hand, \citet[][]{masje08} found that luminous red galaxies were growing due to merger activity at a much slower rate: at least $1.7 \pm0.1 h$ per cent per Gyr on average at redshift \s0.25, implying at least 8\% growth from  $z = 1.3$ to $z = 0.3$. 

The lack of evolution (in comoving coordinates) of the spatial correlation function of massive red galaxies also clearly indicates that the most massive red galaxies must be undergoing mergers.  \citet{white07} used observations of the clustering of luminous red galaxies to show that about one third of the most luminous satellite galaxies appear to have undergone merging or disruption with massive halos between $z = 0.9$ and $z = 0.5$, while \citet{brown08} found that massive red galaxies grew by $\times1.3$ from $z = 1.0$ to $z = 0$.

As with SMD evolution, our stellar mass growth measurements for massive red galaxies based on SMF evolution ($\times 1.7$) are comparable with the those implied by comparison of the $K$-band LF with passive evolution ($\times 1.4$).

\newpage

\subsection{Stellar mass evolution - errors}
\label{sec:SM_evolution_errors}

In addition to the sources of error inherent in measuring $K$-band luminosity evolution (Section \ref{sec:luminosity_evolution_errors}), we have one very significant additional source of error in measuring stellar mass evolution, namely uncertainty in stellar mass to light ratios $M/L_K$, and we include this in Table \ref{tab:errors} for red galaxies. $M/L_K$ uncertainties arise from the evolving  $M/L_K$ - restframe $(B-V)$ relationships in Equations \ref{eq:ML_G10} and \ref{eq:ML_G10_massive}.

We take the intrinsic random variation in $\log_{10}M_K/L$  between galaxies  to be 0.1 dex (Section \ref{sec:masses}). Systematic uncertainties arise from the fact that Equations \ref{eq:ML_G10} and \ref{eq:ML_G10_massive}  are derived from SPS models, and differences occur between different SPS models and the parameters used in them, notably SFH, metallicity and dust obscuration. A number of authors have investigated the relative impact of these different factors and arrived at various conclusions. \citet{conro09} found differences in stellar mass estimates of up to 0.3 dex between different models and different parameter inputs. In contrast, \citet{moust13} found that varying the SPS model, the SFH and the metallicity had little effect, except that the inclusion of bursts of star formation in the SFH did have a significant impact on the derived SMF.   \citet{muzzi13b}, surveying redshifts up to $z = 4$, found that the precise SPS model used was significant, with \citet{maras05} models producing stellar masses that are 0.2 dex lower ($\times 0.65$) than those of \citet{bruzu03} models. They also found that metallicity and delayed bursts of star formation in the SFH made little difference.

Differences in stellar mass estimates at the 0.3 dex level arising from the use of different SPS models and different SFH, metallicity and dust model inputs constitute a very significant source of uncertainty in our measurements of stellar mass for individual galaxies. However, unless the stellar mass differences between models vary with redshift, they will not impact measurements of stellar mass evolution. We do not attempt to estimate how stellar mass differences between models might vary with redshift (if indeed they do) and therefore do not include them in total error budget for red galaxy stellar mass evolution in Table \ref{tab:errors}. Also implicit in the use of SPS models is the adoption of a specific stellar IMF, \citep[e.g.][]{salpe55, kenni83, chabr03}. However, different choices of IMF effectively only produce offsets in calculated values of $\log_{10}M_K/L$ (i.e. stellar masses differ by constant multiplying factors).  The choice of IMF does not therefore impact conclusions regarding the percentage stellar mass growth in galaxies \citep[][]{bell01, bell03} and we do not take them into account in the total error budget for red galaxy stellar mass evolution in Table \ref{tab:errors}. 

We note that, although the relationship between $\log_{10}M_K/L$ and restframe $(B-V)$ color is redshift dependent (Equations \ref{eq:ML_G10} and \ref{eq:ML_G10_massive}), the effect of $z_{\textrm{phot}}$ errors on $M/L_K$ is small (random error of $<0.02$ dex and systematic error of  $< 0.01$ dex).

It is important to realise that the estimated errors arising from  $z_{\textrm{phot}}$ errors are for individual galaxies. In the case of systematic redshift errors, the impact on  \textit{evolutionary} measurements depends on the distribution of $z_{\textrm{phot}}$ errors with redshift. Figure \ref{fig:photoz} shows that photometric redshifts are very slightly overestimated at $z \sim 0.25$ and $z \sim 0.7$, and underestimated at $z \sim 1.0$. We do not pursue this further here, beyond noting that the effect of systematic $z_{\textrm{phot}}$ errors on evolutionary measurements will potentially be comparable with that on individual values.

As with our measurements of $K$-band luminosity evolution, in order to gauge the overall potential impact of systematic photometric redshift errors, we repeated all our calculations twice using $z_{\textrm{phot}}$ values increased and decreased by the fractional systematic error over most of the redshift range (i.e. $0.2 < z \leq1.0$)  as shown in Figure \ref{fig:photoz}. This was $\lambda = [z_{\textrm{phot}}-z_{\textrm{spec}}] \, / \, [1 + z_{\textrm{phot}}] = 0.01$. The last section of Table \ref{tab:errors} shows that for the increased (decreased) $z_{\textrm{phot}}$ values, red galaxy stellar mass density values decreased (increased) by up to  \s0.03 dex while massive red galaxies showed a stellar mass decrease (increase) of up to \s0.01 dex.

Again, as with $K$-band luminosities, to measure the effect on stellar masses of random $z_{\textrm{phot}}$ errors of $\lambda = [z_{\textrm{phot}}-z_{\textrm{spec}}] \, / \, [1 + z_{\textrm{phot}}] = 0.05$ we repeated our calculations ten times, each time applying normally distributed random fractional  errors ($\sigma = 0.05$) to individual $z_{\textrm{phot}}$ values. We found that individual measured values of both red galaxy SMD and the stellar mass of massive red galaxies differed between simulations by less than 0.01 dex, indicating that random photometric redshift errors did not produce significant scatter in these two measurements.

As with the LF, random photometric errors shift the massive end of the SMF due to Eddington bias and scattering of galaxies across redshift bin boundaries. As the penultimate section of Table \ref{tab:errors} indicates, this shift was found to increase from \s0.001 at $0.2<z\leq0.4$ to 0.04 dex  at $1.0<z\leq1.2$. The increase with redshift is due to the fact that the proportion of massive red galaxies with accurate spectroscopic redshifts decreases from $z \sim 0.2$ to $z \sim 1.2$.

We assume that the difference between stellar masses derived from SED fitting and those derived from the $M_K/L$-color relation (Equation \ref{eq:ML_G10_massive}) has scatter $\sigma = 0.1$ dex. We can measure the effect of this scatter on our results by convolving a $\sigma = 0.1$ dex Gaussian with our measured Schechter functions. We find an additional contribution of \s0.09 dex to the Eddington shift in the stellar mass of massive red galaxies in all redshift bins.

Monte Carlo simulations also show that random $z_{\textrm{phot}}$ errors  give rise to a systematic decrease in measured SMD. This systematic error ranges from \s0.01 dex at $0.2<z\leq0.4$, where a significant proportion of galaxies have accurate spectroscopic redshifts, to 0.09 dex at $1.0<z\leq1.2$, where few galaxies have spectroscopic redshifts. As with luminosity density (Section \ref{sec:luminosity_evolution_errors}), the observed systematic change in SMD in each redshift bin is the net result of galaxies being scattered in and out of the bin at the upper and lower bin boundaries with perturbed stellar mass values.

\textit{* Conclusion} Table \ref{tab:errors} shows systematic errors in red galaxy SMD range from \s0.10 dex at $z = 0.3$ to \s0.18 dex at $z = 1.1$ and dominate  random uncertainties of \s0.04 dex. Potentially, correcting for the change in systematic error with redshift could decrease the measured SMD growth from $z = 1.1$ to $z = 0.3$ by \s0.08 dex, altering the SMD growth from $0.37\pm0.04$ dex ($\times 2.34 \pm0.22$) to $0.29\pm0.04$ dex ($\times 1.95 \pm0.18$).

Systematic errors in the measured stellar mass of massive red galaxies range from \s0.11 dex at $z = 0.3$ to \s0.15 dex at $z = 1.1$ and dominate over the random errors   which Monte Carlo simulations show to be less than 0.01 dex.  Potentially, correcting for the change in systematic error with redshift could increase the measured stellar mass growth of massive red galaxies from $z = 1.1$ to $z = 0.3$ by \s0.04 dex, altering the measured growth from $0.23\pm0.01$ dex ($\times 1.70 \pm0.04$) to $0.27\pm0.01$ dex ($\times 1.86 \pm0.04$).

To summarise, systematic errors could have only a small effect on our conclusions for red galaxies, slightly reducing the $\times2.3$ growth in red galaxy SMD, and slightly increasing the 70\% mass growth in massive red galaxies.
 
\subsection{Stellar mass evolution - comparison with the literature}
\label{sec:mass_evolution_comparison}

Figures \ref{fig:literature_mass_comparison_redandblue_log_K_G10} and \ref{fig:MF_binned_redandblue_closeup} show that over the mass range $9 < \log M < 11$, our SMF for all galaxies at $0.2 \leq z < 0.4$ differs by less than \s0.2 dex ($\sim50\%$) in space density from other evolutionary SMF studies  and from the $z < 0.2$ Universe  studies of \citet{cole01},  \citet{bell03} and \citet{baldr12}. As can be seen from Figure  \ref{fig:literature_mass_comparison_redandblue_log_K_G10}, the agreement with other studies is also very good at $z > 0.4$, apart from the much lower space densities seen in the $0.3 \leq z < 1.5$ VIPERS study of \citet{david13}. It is noticeable that our SMFs vary more smoothly with stellar mass than most other studies due to our very large sample size and area.

Figures \ref{fig:literature_mass_comparison_red_log_K_G10} and \ref{fig:literature_mass_comparison_blue_log_K_G10} show offsets of up to 0.3 dex with the quiescent and star-forming subsamples in other studies, which may well be due to the different subsample criteria they use. We used a restframe color criterion, as did \citet{bell03}, \citet{bundy06},  \citet{david13} and \citet{tomcz14} and this caused our red subsample to include dust reddened star forming galaxies. Such galaxies were explicitly rejected by the multi-wavelength color criterion of \citet{ilber13} and the SED modelling approach of \citet{moust13}. 

We note that there is no large difference between our SMFs and the low redshift SMFs of \citet{cole01} and \citet{bell03} despite these authors' use of 2MASS $K$-band luminosity values which, even with isophotal magnitude model corrections (Section \ref{sec:luminosity_evolution_comparison}), miss much of the light from fainter outer regions of galaxies \citep{andre02}. We attribute this agreement in measured SMFs between their work and ours to the fact that their $M/L_K$ relationship was also calibrated using their 2MASS $K$-band luminosity values which are lower than those that we would measure.

In Figure \ref{fig:logmass_K_Mfixed_all_specified_maxlhd}  we compare our measurements for the mass evolution of very massive galaxies with measurements calculated using the Schechter parameters given by \citet{bundy06}, \citet{perez08}, \citet{muzzi13b} and \citet{tomcz14}. For massive red galaxies we see stellar mass growth of 0.23 dex ($\times1.7$), a little more than \citet{tomcz14} and considerably more than \citet{muzzi13b}.

For blue galaxies, and red and blue galaxies combined, we see similar stellar mass growth (0.22 and 0.21 dex respectively) whereas many other studies see little change, except for the growth of \s0.16 dex seen by \citet{muzzi13b} for red and blue galaxies combined.

Similarly, using the Stripe 82 Massive Galaxy Catalog of 41\,770 massive galaxies with SDSS and UKIDSS data, \citet{bundy17} recently concluded that the stellar mass in all ($\log M >11.3$) massive galaxies changed by less than 9\% from $z = 0.65$ to $z = 0.3$. They investigated in detail several potential sources of random and systematic error in the determination of stellar masses from photometry, including photometric redshift errors, differences due to the use of different SPS models, and differences due to differing assumed star formation histories in SPS models. They concluded that the latter source of error was the most significant. However they also speculated missing light in their photometry could strongly impact their conclusions.

It must be remembered that the results of other studies of massive galaxies are based on Schechter fits to the whole of the measured SMF, whereas we have based our mass growth measurements on variable $\alpha$ Schechter fits to just the massive end of the SMF. This will be considerably more accurate as we have only been using the Schechter parameterization as a tool to produce a very close fit in order to measure a small section of the SMF at the high mass end (Section \ref{sec:M}).

\newpage

\section{Summary}
\label{sec:summary}

We measured evolution of the $K$-band LF and the galaxy SMF from $z = 1.2$ to $z = 0.2$ using a very large sample of  353\,594  galaxies covering a large area  of 8.26 deg$^2$ in \bootes, surveyed to a depth of $I = 24$.  The imaging, catalogs and photometry were identical to those in \citet{beare15} and derived from various optical and infrared surveys.  Our very large sample size and area minimised both Poisson errors and the effects of cosmic variance (\s3\% for all galaxies and \s8\% for red galaxies, which are more strongly clustered). 

We used a magnitude dependent aperture diameter and curve of growth analysis to measure precise photometry in 13 optical and infrared bands. Using this photometry and the 129 SED templates from \citet{brown14} we were able to precisely determine photometric redshifts (and luminosity distances) using the Bayesian EAZY code. 

Galaxy luminosities were derived from apparent magnitudes and redshifts using the method of \citet{beare14} and the 129 SED templates of  \citet{brown14}. We used GAMA/G10 COSMOS data to derive the evolving dependence of $K$-band mass to light ratios on restframe $(B-V)$ color and used this to measure stellar masses.

Binned $K$-band LFs and SMFs were produced for five redshift bins between  $z = 0.2$ and $z = 1.2$ and found to be consistent with LFs and SMFs from the literature. LF and SMF evolution were measured using maximum likelihood Schechter function fits within each redshift bin. Red and blue galaxies were differentiated using an evolving restframe $(U-B)$ color-magnitude cut, as in \citet{beare15}. 

Luminosity densities and stellar mass densities were calculated from the Schechter parameters and their evolution measured. Evolution of the bright end of the LF and the massive end of the SMF were measured by finding the luminosity and stellar mass (respectively) corresponding to a fixed space density. This was done by fitting a maximum likelihood Schechter function with variable $\alpha$ parameter to the luminous and massive ends of the LF and SMF (respectively).

As a main focus of our work has been to measure evolution of the $K$-band luminosity and the stellar mass of red galaxies, we made detailed estimates of the various random and systematic errors for red galaxies, tabulating these in Table \ref{tab:errors}.

The total luminosity density of both red and blue galaxies increased by a modest 0.08 dex, i.e. a factor of 1.2,  from $z = 1.1$ to $z = 0.3$. Over the same redshift range the luminosity of highly luminous red (blue) galaxies decreased by 0.19 (0.33) mag, which equates to 0.08 (0.13) dex or a factor of $\times0.83$ ($\times0.74$). Highly luminous red galaxies fade at an ever-increasing rate from $z \sim 1.1$ to $z \sim 0.3$. Comparison with a passively evolving  population implied a factor of \s2.1 growth in red galaxy stellar mass density, and a factor of \s1.4 growth in the stellar mass of highly luminous massive red galaxies.

Using our evolving SMFs, we found an increase of 0.37 (0.31) dex in SMD for red (blue) galaxies, i.e. a factor 2.3 (2.1), from $z = 1.1$ to $z = 0.3$.  The rate at which SMD is growing decreases slowly with time. For red galaxies, this indicates that the rate at which blue galaxies move to the red sequence as they cease star formation decreases slowly with time.

The mass of the most massive red (blue) galaxies increased by 0.23 (0.15) dex, i.e. a factor of $\times1.7$ ($\times1.4$), from $z = 1.1$ to $z = 0.3$. The rate at which massive red galaxies increase in stellar mass through mergers with smaller galaxies slows from \s0.4 dex per unit redshift for $z = 1.1$ to $z = 0.7$ to \s0.2 dex per unit redshift for $z = 0.7$ to $z = 0.3$.

\section{ACKNOWLEDGEMENTS}
\label{sec:acknowledgements}

We thank the anonymous referee for suggestions which have greatly improved the quality of our results and enhanced the clarity of this paper. Richard Beare wishes to thank Monash University for financial support from MGS and MIPRS postgraduate research scholarships. Michael Brown acknowledges financial support from The Australian Research Council (FT100100280) and the Monash Research Accelerator Program (MRA).  Kevin Pimbblet acknowledges the support of STFC, through the University of Hull's Consolidated Grant ST/R000840/1. We thank colleagues on the NDWFS, SDWFS, NEWFIRM \bootes, and AGES teams, in particular M. L. N. Ashby, R. J. Cool, A. Dey, P. R. Eisenhardt, D. J. Eisenstein, A. H. Gonzalez, B. T. Jannuzi, C. S. Kochanek and D. Stern. This work is based in part on observations made with the Spitzer Space Telescope, which is operated by the Jet Propulsion Laboratory, California Institute of Technology under a contract with NASA. This research was supported by the National Optical Astronomy Observatory, which is operated by the Association of Universities for Research in Astronomy (AURA), Inc., under a cooperative agreement with the National Science Foundation.

\begin{deluxetable*}{ccccccc}[!ht]
\tablewidth{0pt}
\tablecolumns{7}
\tabletypesize{\scriptsize}
\tablecaption{Binned $K$-band luminosity function for all galaxies.}
\tablehead{
\multicolumn {2}{c}{$M_K - 5\log h_{70}$} & \multicolumn {5}{c}{Luminosity Function ($10^{-3} h_{70}^3 \, \rm{ Mpc}^{-3} \, \rm{mag}^{-1}$)}\\
\colhead{Min} & \colhead{Max} &\colhead{$0.2 \leq z < 0.4 $} & \colhead{$0.4 \leq z < 0.6 $} & \colhead{$0.6 \leq z < 0.8 $} & \colhead{$0.8 \leq z < 1.0 $} & \colhead{$1.0 \leq z < 1.2 $}}
\startdata
$   -26.00$ & $   -25.75$ &    -  & $   0.001\pm  0.001$ &    -  &    -  &    - \\
$   -25.75$ & $   -25.50$ &    -  &    -  &    -  &    -  & $   0.001\pm  0.001$\\
$   -25.50$ & $   -25.25$ &    -  &    -  &    -  & $   0.001\pm  0.001$ & $   0.001\pm  0.001$\\
$   -25.25$ & $   -25.00$ & $   0.003\pm  0.002$ & $   0.003\pm  0.001$ & $   0.004\pm  0.001$ & $   0.005\pm  0.001$ & $   0.008\pm  0.001$\\
$   -25.00$ & $   -24.75$ & $   0.001\pm  0.001$ & $   0.007\pm  0.002$ & $   0.015\pm  0.003$ & $   0.022\pm  0.003$ & $   0.026\pm  0.003$\\
$   -24.75$ & $   -24.50$ & $   0.018\pm  0.005$ & $   0.027\pm  0.004$ & $   0.037\pm  0.004$ & $   0.055\pm  0.004$ & $   0.052\pm  0.004$\\
$   -24.50$ & $   -24.25$ & $   0.060\pm  0.009$ & $   0.072\pm  0.007$ & $   0.113\pm  0.007$ & $   0.137\pm  0.007$ & $   0.125\pm  0.006$\\
$   -24.25$ & $   -24.00$ & $   0.134\pm  0.014$ & $   0.176\pm  0.011$ & $   0.241\pm  0.010$ & $   0.254\pm  0.009$ & $   0.241\pm  0.008$\\
$   -24.00$ & $   -23.75$ & $   0.248\pm  0.019$ & $   0.365\pm  0.016$ & $   0.420\pm  0.014$ & $   0.436\pm  0.012$ & $   0.396\pm  0.011$\\
$   -23.75$ & $   -23.50$ & $   0.453\pm  0.026$ & $   0.561\pm  0.020$ & $   0.646\pm  0.017$ & $   0.657\pm  0.015$ & $   0.573\pm  0.013$\\
$   -23.50$ & $   -23.25$ & $   0.759\pm  0.034$ & $   0.860\pm  0.024$ & $   0.919\pm  0.020$ & $   0.894\pm  0.017$ & $   0.729\pm  0.015$\\
$   -23.25$ & $   -23.00$ & $   1.122\pm  0.041$ & $   1.137\pm  0.028$ & $   1.183\pm  0.023$ & $   1.101\pm  0.019$ & $   0.924\pm  0.017$\\
$   -23.00$ & $   -22.75$ & $   1.369\pm  0.045$ & $   1.428\pm  0.031$ & $   1.458\pm  0.025$ & $   1.293\pm  0.021$ & $   1.085\pm  0.018$\\
$   -22.75$ & $   -22.50$ & $   1.617\pm  0.049$ & $   1.665\pm  0.034$ & $   1.645\pm  0.027$ & $   1.524\pm  0.023$ & $   1.149\pm  0.019$\\
$   -22.50$ & $   -22.25$ & $   2.017\pm  0.055$ & $   1.737\pm  0.034$ & $   1.775\pm  0.028$ & $   1.648\pm  0.024$ &    -    \\
$   -22.25$ & $   -22.00$ & $   2.255\pm  0.058$ & $   1.969\pm  0.037$ & $   1.892\pm  0.029$ & $   1.745\pm  0.025$ &    -    \\
$   -22.00$ & $   -21.75$ & $   2.602\pm  0.062$ & $   2.012\pm  0.037$ & $   1.960\pm  0.030$ & $   1.800\pm  0.025$ &    -    \\
$   -21.75$ & $   -21.50$ & $   2.738\pm  0.064$ & $   2.074\pm  0.038$ & $   2.152\pm  0.031$ & $   1.827\pm  0.025$ &    -    \\
$   -21.50$ & $   -21.25$ & $   2.957\pm  0.066$ & $   2.121\pm  0.038$ & $   2.150\pm  0.031$ &    -     &    -    \\
$   -21.25$ & $   -21.00$ & $   3.195\pm  0.069$ & $   2.176\pm  0.039$ & $   2.298\pm  0.033$ &    -     &    -    \\
$   -21.00$ & $   -20.75$ & $   3.437\pm  0.072$ & $   2.191\pm  0.039$ & $   2.226\pm  0.032$ &    -     &    -    \\
$   -20.75$ & $   -20.50$ & $   3.638\pm  0.074$ & $   2.153\pm  0.039$ &    -     &    -     &    -    \\
$   -20.50$ & $   -20.25$ & $   4.152\pm  0.079$ & $   2.053\pm  0.038$ &    -     &    -     &    -    \\
$   -20.25$ & $   -20.00$ & $   4.707\pm  0.084$ & $   2.001\pm  0.038$ &    -     &    -     &    -    \\
$   -20.00$ & $   -19.75$ & $   5.062\pm  0.087$ & $   1.917\pm  0.037$ &    -     &    -     &    -    \\
$   -19.75$ & $   -19.50$ & $   5.599\pm  0.092$ &    -     &    -     &    -     &    -    \\
$   -19.50$ & $   -19.25$ & $   6.164\pm  0.097$ &    -     &    -     &    -     &    -    \\
$   -19.25$ & $   -19.00$ & $   6.661\pm  0.101$ &    -     &    -     &    -     &    -    \\
$   -19.00$ & $   -18.75$ & $   6.876\pm  0.103$ &    -     &    -     &    -     &    -    \\
$   -18.75$ & $   -18.50$ & $   7.338\pm  0.107$ &    -     &    -     &    -     &    -    \\
$   -18.50$ & $   -18.25$ & $   6.987\pm  0.105$ &    -     &    -     &    -     &    -   
\enddata

\label{tab:K_bin_densities_redandblue}
\end{deluxetable*}

\begin{deluxetable*}{ccccccc}
\tablewidth{0pt}
\tablecolumns{7}
\tabletypesize{\scriptsize}
\tablecaption{Binned $K$-band luminosity function for red galaxies.}
\tablehead{
\multicolumn {2}{c}{$M_K - 5\log h_{70}$} & \multicolumn {5}{c}{Luminosity Function ($10^{-3} h_{70}^3 \, \rm{ Mpc}^{-3} \, \rm{mag}^{-1}$)}\\
\colhead{Min} & \colhead{Max} &\colhead{$0.2 \leq z < 0.4 $} & \colhead{$0.4 \leq z < 0.6 $} & \colhead{$0.6 \leq z < 0.8 $} & \colhead{$0.8 \leq z < 1.0 $} & \colhead{$1.0 \leq z < 1.2 $}}
\startdata
$   -26.00$ & $   -25.75$ &    -  & $   0.001\pm  0.001$ &    -  &    -  &    - \\
$   -25.75$ & $   -25.50$ &    -  &    -  &    -  &    -  &    - \\
$   -25.50$ & $   -25.25$ &    -  &    -  &    -  & $   0.001\pm  0.001$ & $   0.001\pm  0.001$\\
$   -25.25$ & $   -25.00$ & $   0.003\pm  0.002$ & $   0.003\pm  0.001$ & $   0.004\pm  0.001$ & $   0.003\pm  0.001$ & $   0.004\pm  0.001$\\
$   -25.00$ & $   -24.75$ & $   0.001\pm  0.001$ & $   0.007\pm  0.002$ & $   0.011\pm  0.002$ & $   0.016\pm  0.002$ & $   0.017\pm  0.002$\\
$   -24.75$ & $   -24.50$ & $   0.015\pm  0.005$ & $   0.020\pm  0.004$ & $   0.028\pm  0.004$ & $   0.035\pm  0.003$ & $   0.036\pm  0.003$\\
$   -24.50$ & $   -24.25$ & $   0.052\pm  0.009$ & $   0.063\pm  0.007$ & $   0.085\pm  0.006$ & $   0.102\pm  0.006$ & $   0.084\pm  0.005$\\
$   -24.25$ & $   -24.00$ & $   0.111\pm  0.013$ & $   0.142\pm  0.010$ & $   0.178\pm  0.009$ & $   0.176\pm  0.008$ & $   0.158\pm  0.007$\\
$   -24.00$ & $   -23.75$ & $   0.205\pm  0.017$ & $   0.276\pm  0.014$ & $   0.286\pm  0.011$ & $   0.276\pm  0.010$ & $   0.259\pm  0.009$\\
$   -23.75$ & $   -23.50$ & $   0.324\pm  0.022$ & $   0.382\pm  0.016$ & $   0.432\pm  0.014$ & $   0.403\pm  0.012$ & $   0.377\pm  0.010$\\
$   -23.50$ & $   -23.25$ & $   0.541\pm  0.028$ & $   0.558\pm  0.019$ & $   0.594\pm  0.016$ & $   0.540\pm  0.014$ & $   0.451\pm  0.012$\\
$   -23.25$ & $   -23.00$ & $   0.764\pm  0.034$ & $   0.732\pm  0.022$ & $   0.717\pm  0.018$ & $   0.617\pm  0.015$ & $   0.543\pm  0.013$\\
$   -23.00$ & $   -22.75$ & $   0.888\pm  0.036$ & $   0.859\pm  0.024$ & $   0.846\pm  0.019$ & $   0.694\pm  0.015$ & $   0.616\pm  0.014$\\
$   -22.75$ & $   -22.50$ & $   1.013\pm  0.039$ & $   0.955\pm  0.025$ & $   0.888\pm  0.020$ & $   0.757\pm  0.016$ & $   0.522\pm  0.013$\\
$   -22.50$ & $   -22.25$ & $   1.119\pm  0.041$ & $   0.948\pm  0.025$ & $   0.854\pm  0.019$ & $   0.742\pm  0.016$ &    -    \\
$   -22.25$ & $   -22.00$ & $   1.157\pm  0.042$ & $   0.979\pm  0.026$ & $   0.844\pm  0.019$ & $   0.704\pm  0.016$ &    -    \\
$   -22.00$ & $   -21.75$ & $   1.212\pm  0.043$ & $   0.898\pm  0.025$ & $   0.785\pm  0.019$ &    -     &    -    \\
$   -21.75$ & $   -21.50$ & $   1.152\pm  0.041$ & $   0.810\pm  0.024$ & $   0.786\pm  0.019$ &    -     &    -    \\
$   -21.50$ & $   -21.25$ & $   1.097\pm  0.040$ & $   0.728\pm  0.022$ & $   0.672\pm  0.018$ &    -     &    -    \\
$   -21.25$ & $   -21.00$ & $   1.087\pm  0.040$ & $   0.707\pm  0.022$ & $   0.626\pm  0.017$ &    -     &    -    \\
$   -21.00$ & $   -20.75$ & $   1.055\pm  0.040$ & $   0.598\pm  0.021$ &    -     &    -     &    -    \\
$   -20.75$ & $   -20.50$ & $   0.963\pm  0.038$ & $   0.530\pm  0.019$ &    -     &    -     &    -    \\
$   -20.50$ & $   -20.25$ & $   1.018\pm  0.039$ & $   0.393\pm  0.017$ &    -     &    -     &    -    \\
$   -20.25$ & $   -20.00$ & $   1.076\pm  0.040$ & $   0.329\pm  0.016$ &    -     &    -     &    -    \\
$   -20.00$ & $   -19.75$ & $   0.919\pm  0.037$ &    -     &    -     &    -     &    -    \\
$   -19.75$ & $   -19.50$ & $   0.901\pm  0.037$ &    -     &    -     &    -     &    -    \\
$   -19.50$ & $   -19.25$ & $   0.896\pm  0.037$ &    -     &    -     &    -     &    -    \\
$   -19.25$ & $   -19.00$ & $   0.901\pm  0.037$ &    -     &    -     &    -     &    -    \\
$   -19.00$ & $   -18.75$ & $   0.867\pm  0.037$ &    -     &    -     &    -     &    -    \\
$   -18.75$ & $   -18.50$ & $   0.808\pm  0.036$ &    -     &    -     &    -     &    -    \\
$   -18.50$ & $   -18.25$ & $   0.720\pm  0.034$ &    -     &    -     &    -     &    -    \\
$   -18.25$ & $   -18.00$ & $   0.630\pm  0.032$ &    -     &    -     &    -     &    -    
\enddata
\label{tab:K_bin_densities_red}
\end{deluxetable*}

\begin{deluxetable*}{ccccccc}
\tablewidth{0pt}
\tablecolumns{7}
\tabletypesize{\scriptsize}
\tablecaption{Binned $K$-band luminosity function for blue galaxies.}
\tablehead{
\multicolumn {2}{c}{$M_K - 5\log h_{70}$} & \multicolumn {5}{c}{Luminosity Function ($10^{-3} h_{70}^3 \, \rm{ Mpc}^{-3} \, \rm{mag}^{-1}$)}\\
\colhead{Min} & \colhead{Max} &\colhead{$0.2 \leq z < 0.4 $} & \colhead{$0.4 \leq z < 0.6 $} & \colhead{$0.6 \leq z < 0.8 $} & \colhead{$0.8 \leq z < 1.0 $} & \colhead{$1.0 \leq z < 1.2 $}}
\startdata
$   -25.75$ & $   -25.50$ &    -  &    -  &    -  &    -  & $   0.001\pm  0.001$\\
$   -25.50$ & $   -25.25$ &    -  &    -  &    -  & $   - $ &    - \\
$   -25.25$ & $   -25.00$ &    -  &    -  & $   0.001\pm  0.001$ & $   0.001\pm  0.001$ & $   0.004\pm  0.001$\\
$   -25.00$ & $   -24.75$ &    -  & $   0.001\pm  0.001$ & $   0.004\pm  0.001$ & $   0.007\pm  0.001$ & $   0.008\pm  0.001$\\
$   -24.75$ & $   -24.50$ & $   0.003\pm  0.002$ & $   0.007\pm  0.002$ & $   0.008\pm  0.002$ & $   0.020\pm  0.003$ & $   0.016\pm  0.002$\\
$   -24.50$ & $   -24.25$ & $   0.007\pm  0.003$ & $   0.010\pm  0.003$ & $   0.028\pm  0.003$ & $   0.035\pm  0.003$ & $   0.042\pm  0.003$\\
$   -24.25$ & $   -24.00$ & $   0.024\pm  0.006$ & $   0.035\pm  0.005$ & $   0.063\pm  0.005$ & $   0.079\pm  0.005$ & $   0.083\pm  0.005$\\
$   -24.00$ & $   -23.75$ & $   0.043\pm  0.008$ & $   0.088\pm  0.008$ & $   0.134\pm  0.008$ & $   0.161\pm  0.007$ & $   0.137\pm  0.006$\\
$   -23.75$ & $   -23.50$ & $   0.129\pm  0.014$ & $   0.180\pm  0.011$ & $   0.215\pm  0.010$ & $   0.254\pm  0.009$ & $   0.197\pm  0.007$\\
$   -23.50$ & $   -23.25$ & $   0.218\pm  0.018$ & $   0.302\pm  0.014$ & $   0.324\pm  0.012$ & $   0.354\pm  0.011$ & $   0.278\pm  0.009$\\
$   -23.25$ & $   -23.00$ & $   0.359\pm  0.023$ & $   0.405\pm  0.017$ & $   0.466\pm  0.014$ & $   0.484\pm  0.013$ & $   0.382\pm  0.010$\\
$   -23.00$ & $   -22.75$ & $   0.481\pm  0.027$ & $   0.570\pm  0.020$ & $   0.612\pm  0.016$ & $   0.599\pm  0.014$ & $   0.468\pm  0.012$\\
$   -22.75$ & $   -22.50$ & $   0.604\pm  0.030$ & $   0.710\pm  0.022$ & $   0.757\pm  0.018$ & $   0.767\pm  0.016$ & $   0.628\pm  0.014$\\
$   -22.50$ & $   -22.25$ & $   0.898\pm  0.037$ & $   0.789\pm  0.023$ & $   0.921\pm  0.020$ & $   0.906\pm  0.018$ & $   0.701\pm  0.014$\\
$   -22.25$ & $   -22.00$ & $   1.098\pm  0.040$ & $   0.990\pm  0.026$ & $   1.048\pm  0.022$ & $   1.040\pm  0.019$ & $   0.760\pm  0.015$\\
$   -22.00$ & $   -21.75$ & $   1.390\pm  0.046$ & $   1.114\pm  0.028$ & $   1.175\pm  0.023$ & $   1.185\pm  0.020$ &    -    \\
$   -21.75$ & $   -21.50$ & $   1.585\pm  0.049$ & $   1.264\pm  0.029$ & $   1.365\pm  0.025$ & $   1.343\pm  0.022$ &    -    \\
$   -21.50$ & $   -21.25$ & $   1.860\pm  0.053$ & $   1.393\pm  0.031$ & $   1.478\pm  0.026$ & $   1.447\pm  0.023$ &    -    \\
$   -21.25$ & $   -21.00$ & $   2.109\pm  0.056$ & $   1.469\pm  0.032$ & $   1.672\pm  0.028$ &    -     &    -    \\
$   -21.00$ & $   -20.75$ & $   2.383\pm  0.060$ & $   1.592\pm  0.033$ & $   1.695\pm  0.028$ &    -     &    -    \\
$   -20.75$ & $   -20.50$ & $   2.675\pm  0.063$ & $   1.623\pm  0.034$ & $   1.734\pm  0.028$ &    -     &    -    \\
$   -20.50$ & $   -20.25$ & $   3.134\pm  0.068$ & $   1.659\pm  0.034$ & $   1.825\pm  0.029$ &    -     &    -    \\
$   -20.25$ & $   -20.00$ & $   3.632\pm  0.074$ & $   1.672\pm  0.034$ &    -     &    -     &    -    \\
$   -20.00$ & $   -19.75$ & $   4.143\pm  0.079$ & $   1.677\pm  0.035$ &    -     &    -     &    -    \\
$   -19.75$ & $   -19.50$ & $   4.698\pm  0.084$ & $   1.717\pm  0.035$ &    -     &    -     &    -    \\
$   -19.50$ & $   -19.25$ & $   5.268\pm  0.090$ &    -     &    -     &    -     &    -    \\
$   -19.25$ & $   -19.00$ & $   5.759\pm  0.094$ &    -     &    -     &    -     &    -    \\
$   -19.00$ & $   -18.75$ & $   6.009\pm  0.097$ &    -     &    -     &    -     &    -    \\
$   -18.75$ & $   -18.50$ & $   6.531\pm  0.101$ &    -     &    -     &    -     &    -    \\
$   -18.50$ & $   -18.25$ & $   6.268\pm  0.100$ &    -     &    -     &    -     &    -    \\
$   -18.25$ & $   -18.00$ & $   5.133\pm  0.090$ &    -     &    -     &    -     &    -   
\enddata
\label{tab:K_bin_densities_blue}
\end{deluxetable*}

\begin{deluxetable*}{ccccccc}
\tablecolumns{7}
\tabletypesize{\scriptsize}
\tablecaption{Binned SMF for all galaxies.}
\tablehead{
\multicolumn {2}{c}{$\log M$} & \multicolumn {5}{c}{SMF ($h_{70}^3 {\rm{Mpc}}^{-3} {\log_{10}}M^{-1}$)}\\
\colhead{Min} & \colhead{Max} &\colhead{$0.2 \leq z < 0.4 $} & \colhead{$0.4 \leq z < 0.6 $} & \colhead{$0.6 \leq z < 0.8 $} & \colhead{$0.8 \leq z < 1.0 $} & \colhead{$1.0 \leq z < 1.2 $}}
\startdata
$     8.80$ & $     8.90$ & $  14.734\pm  0.235$ &    -     &    -     &    -     &    -    \\
$     8.90$ & $     9.00$ & $  14.424\pm  0.232$ &    -     &    -     &    -     &    -    \\
$     9.00$ & $     9.10$ & $  13.725\pm  0.226$ &    -     &    -     &    -     &    -    \\
$     9.10$ & $     9.20$ & $  12.765\pm  0.218$ &    -     &    -     &    -     &    -    \\
$     9.20$ & $     9.30$ & $  11.559\pm  0.208$ &    -     &    -     &    -     &    -    \\
$     9.30$ & $     9.40$ & $  11.077\pm  0.203$ &    -     &    -     &    -     &    -    \\
$     9.40$ & $     9.50$ & $   9.709\pm  0.190$ & $   4.011\pm  0.083$ &    -     &    -     &    -    \\
$     9.50$ & $     9.60$ & $   8.966\pm  0.183$ & $   3.981\pm  0.082$ &    -     &    -     &    -    \\
$     9.60$ & $     9.70$ & $   8.491\pm  0.178$ & $   4.134\pm  0.084$ &    -     &    -     &    -    \\
$     9.70$ & $     9.80$ & $   7.856\pm  0.171$ & $   4.212\pm  0.085$ &    -     &    -     &    -    \\
$     9.80$ & $     9.90$ & $   7.016\pm  0.162$ & $   4.102\pm  0.084$ &    -     &    -     &    -    \\
$     9.90$ & $    10.00$ & $   6.698\pm  0.158$ & $   4.083\pm  0.083$ & $   3.919\pm  0.065$ &    -     &    -    \\
$    10.00$ & $    10.10$ & $   6.489\pm  0.156$ & $   4.187\pm  0.084$ & $   3.725\pm  0.064$ &    -     &    -    \\
$    10.10$ & $    10.20$ & $   5.906\pm  0.149$ & $   4.187\pm  0.084$ & $   3.849\pm  0.065$ &    -     &    -    \\
$    10.20$ & $    10.30$ & $   5.264\pm  0.140$ & $   4.107\pm  0.084$ & $   3.761\pm  0.064$ &    -     &    -    \\
$    10.30$ & $    10.40$ & $   5.391\pm  0.142$ & $   4.057\pm  0.083$ & $   3.812\pm  0.064$ &    -     &    -    \\
$    10.40$ & $    10.50$ & $   4.946\pm  0.136$ & $   3.954\pm  0.082$ & $   3.738\pm  0.064$ &    -     &    -    \\
$    10.50$ & $    10.60$ & $   4.214\pm  0.125$ & $   3.802\pm  0.080$ & $   3.512\pm  0.062$ & $   2.968\pm  0.050$ &    -    \\
$    10.60$ & $    10.70$ & $   3.982\pm  0.122$ & $   3.583\pm  0.078$ & $   3.324\pm  0.060$ & $   2.633\pm  0.047$ &    -    \\
$    10.70$ & $    10.80$ & $   3.250\pm  0.110$ & $   3.321\pm  0.075$ & $   2.878\pm  0.056$ & $   2.276\pm  0.043$ &    -    \\
$    10.80$ & $    10.90$ & $   2.817\pm  0.103$ & $   2.708\pm  0.068$ & $   2.524\pm  0.052$ & $   1.827\pm  0.039$ & $   1.360\pm  0.031$\\
$    10.90$ & $    11.00$ & $   2.223\pm  0.091$ & $   2.120\pm  0.060$ & $   2.022\pm  0.047$ & $   1.404\pm  0.034$ & $   1.029\pm  0.027$\\
$    11.00$ & $    11.10$ & $   1.558\pm  0.076$ & $   1.443\pm  0.050$ & $   1.473\pm  0.040$ & $   0.939\pm  0.028$ & $   0.690\pm  0.022$\\
$    11.10$ & $    11.20$ & $   0.997\pm  0.061$ & $   1.007\pm  0.041$ & $   0.945\pm  0.032$ & $   0.584\pm  0.022$ & $   0.402\pm  0.017$\\
$    11.20$ & $    11.30$ & $   0.650\pm  0.049$ & $   0.531\pm  0.030$ & $   0.530\pm  0.024$ & $   0.315\pm  0.016$ & $   0.203\pm  0.012$\\
$    11.30$ & $    11.40$ & $   0.280\pm  0.032$ & $   0.255\pm  0.021$ & $   0.276\pm  0.017$ & $   0.164\pm  0.012$ & $   0.103\pm  0.008$\\
$    11.40$ & $    11.50$ & $   0.202\pm  0.027$ & $   0.133\pm  0.015$ & $   0.105\pm  0.011$ & $   0.052\pm  0.007$ & $   0.037\pm  0.005$\\
$    11.50$ & $    11.60$ & $   0.075\pm  0.017$ & $   0.037\pm  0.008$ & $   0.044\pm  0.007$ & $   0.014\pm  0.003$ & $   0.013\pm  0.003$\\
$    11.60$ & $    11.70$ & $   0.011\pm  0.006$ & $   0.015\pm  0.005$ & $   0.009\pm  0.003$ & $   0.006\pm  0.002$ & $   0.004\pm  0.002$\\
$    11.70$ & $    11.80$ &    -  & $   0.005\pm  0.003$ & $   0.004\pm  0.002$ &    -  &    - \\
$    11.80$ & $    11.90$ & $   0.004\pm  0.004$ &    -  & $   0.002\pm  0.002$ &    -  &    - \\
$    11.90$ & $    12.00$ & $   0.004\pm  0.004$ &    -  &    -  &    -  &    -
\label{tab:bin_massfn_redandblue}
\end{deluxetable*}

\begin{deluxetable*}{ccccccc}
\tablecolumns{7}
\tabletypesize{\scriptsize}
\tablecaption{Binned SMF for red galaxies.}
\tablehead{
\multicolumn {2}{c}{$\log M$} & \multicolumn {5}{c}{SMF ($h_{70}^3 {\rm{Mpc}}^{-3} {\log_{10}}M^{-1}$)}\\
\colhead{Min} & \colhead{Max} &\colhead{$0.2 \leq z < 0.4 $} & \colhead{$0.4 \leq z < 0.6 $} & \colhead{$0.6 \leq z < 0.8 $} & \colhead{$0.8 \leq z < 1.0 $} & \colhead{$1.0 \leq z < 1.2 $}}
\startdata
$     8.90$ & $     9.00$ & $   1.547\pm  0.076$ &    -     &    -     &    -     &    -    \\
$     9.00$ & $     9.10$ & $   1.621\pm  0.078$ &    -     &    -     &    -     &    -    \\
$     9.10$ & $     9.20$ & $   1.887\pm  0.084$ &    -     &    -     &    -     &    -    \\
$     9.20$ & $     9.30$ & $   1.887\pm  0.084$ &    -     &    -     &    -     &    -    \\
$     9.30$ & $     9.40$ & $   2.055\pm  0.088$ &    -     &    -     &    -     &    -    \\
$     9.40$ & $     9.50$ & $   2.047\pm  0.087$ &    -     &    -     &    -     &    -    \\
$     9.50$ & $     9.60$ & $   2.081\pm  0.088$ & $   0.463\pm  0.028$ &    -     &    -     &    -    \\
$     9.60$ & $     9.70$ & $   2.428\pm  0.095$ & $   0.718\pm  0.035$ &    -     &    -     &    -    \\
$     9.70$ & $     9.80$ & $   2.413\pm  0.095$ & $   0.883\pm  0.039$ &    -     &    -     &    -    \\
$     9.80$ & $     9.90$ & $   2.279\pm  0.092$ & $   1.087\pm  0.043$ &    -     &    -     &    -    \\
$     9.90$ & $    10.00$ & $   2.544\pm  0.097$ & $   1.225\pm  0.046$ &    -     &    -     &    -    \\
$    10.00$ & $    10.10$ & $   2.604\pm  0.099$ & $   1.553\pm  0.051$ &    -     &    -     &    -    \\
$    10.10$ & $    10.20$ & $   2.548\pm  0.098$ & $   1.832\pm  0.056$ & $   1.414\pm  0.039$ &    -     &    -    \\
$    10.20$ & $    10.30$ & $   2.581\pm  0.098$ & $   1.887\pm  0.057$ & $   1.629\pm  0.042$ &    -     &    -    \\
$    10.30$ & $    10.40$ & $   2.944\pm  0.105$ & $   2.099\pm  0.060$ & $   1.916\pm  0.046$ &    -     &    -    \\
$    10.40$ & $    10.50$ & $   2.828\pm  0.103$ & $   2.303\pm  0.063$ & $   2.015\pm  0.047$ &    -     &    -    \\
$    10.50$ & $    10.60$ & $   2.720\pm  0.101$ & $   2.297\pm  0.063$ & $   2.098\pm  0.048$ &    -     &    -    \\
$    10.60$ & $    10.70$ & $   2.794\pm  0.102$ & $   2.343\pm  0.063$ & $   2.169\pm  0.049$ & $   1.681\pm  0.037$ &    -    \\
$    10.70$ & $    10.80$ & $   2.339\pm  0.093$ & $   2.280\pm  0.062$ & $   1.953\pm  0.046$ & $   1.545\pm  0.036$ &    -    \\
$    10.80$ & $    10.90$ & $   2.185\pm  0.090$ & $   1.875\pm  0.056$ & $   1.801\pm  0.044$ & $   1.347\pm  0.033$ &    -    \\
$    10.90$ & $    11.00$ & $   1.834\pm  0.083$ & $   1.580\pm  0.052$ & $   1.496\pm  0.040$ & $   1.068\pm  0.030$ & $   0.814\pm  0.024$\\
$    11.00$ & $    11.10$ & $   1.367\pm  0.071$ & $   1.116\pm  0.044$ & $   1.126\pm  0.035$ & $   0.758\pm  0.025$ & $   0.570\pm  0.020$\\
$    11.10$ & $    11.20$ & $   0.893\pm  0.058$ & $   0.845\pm  0.038$ & $   0.761\pm  0.029$ & $   0.490\pm  0.020$ & $   0.344\pm  0.015$\\
$    11.20$ & $    11.30$ & $   0.598\pm  0.047$ & $   0.470\pm  0.028$ & $   0.433\pm  0.022$ & $   0.282\pm  0.015$ & $   0.184\pm  0.011$\\
$    11.30$ & $    11.40$ & $   0.262\pm  0.031$ & $   0.235\pm  0.020$ & $   0.248\pm  0.016$ & $   0.147\pm  0.011$ & $   0.088\pm  0.008$\\
$    11.40$ & $    11.50$ & $   0.194\pm  0.027$ & $   0.116\pm  0.014$ & $   0.086\pm  0.010$ & $   0.048\pm  0.006$ & $   0.035\pm  0.005$\\
$    11.50$ & $    11.60$ & $   0.075\pm  0.017$ & $   0.034\pm  0.008$ & $   0.038\pm  0.006$ & $   0.014\pm  0.003$ & $   0.012\pm  0.003$\\
$    11.60$ & $    11.70$ & $   0.011\pm  0.006$ & $   0.015\pm  0.005$ & $   0.008\pm  0.003$ & $   0.005\pm  0.002$ & $   0.003\pm  0.002$\\
$    11.70$ & $    11.80$ &    -  & $   0.003\pm  0.002$ & $   0.002\pm  0.002$ &    -  &    - \\
$    11.80$ & $    11.90$ & $   0.004\pm  0.004$ &    -  & $   0.001\pm  0.001$ &    -  &    - \\
$    11.90$ & $    12.00$ & $   0.004\pm  0.004$ &    -  &    -  &    -  &    - 
\enddata
\label{tab:bin_massfn_red}
\end{deluxetable*}

\begin{deluxetable*}{ccccccc}
\tablecolumns{7}
\tabletypesize{\scriptsize}
\tablecaption{Binned SMF for blue galaxies.}
\tablehead{
\multicolumn {2}{c}{$\log M$} & \multicolumn {5}{c}{SMF ($h_{70}^3 {\rm{Mpc}}^{-3} {\log_{10}}M^{-1}$)}\\
\colhead{Min} & \colhead{Max} &\colhead{$0.2 \leq z < 0.4 $} & \colhead{$0.4 \leq z < 0.6 $} & \colhead{$0.6 \leq z < 0.8 $} & \colhead{$0.8 \leq z < 1.0 $} & \colhead{$1.0 \leq z < 1.2 $}}
\startdata
$     8.80$ & $     8.90$ & $  13.438\pm  0.224$ &    -     &    -     &    -     &    -    \\
$     8.90$ & $     9.00$ & $  12.877\pm  0.219$ &    -     &    -     &    -     &    -    \\
$     9.00$ & $     9.10$ & $  12.104\pm  0.213$ &    -     &    -     &    -     &    -    \\
$     9.10$ & $     9.20$ & $  10.879\pm  0.202$ &    -     &    -     &    -     &    -    \\
$     9.20$ & $     9.30$ & $   9.672\pm  0.190$ & $   3.460\pm  0.077$ &    -     &    -     &    -    \\
$     9.30$ & $     9.40$ & $   9.022\pm  0.184$ & $   3.675\pm  0.079$ &    -     &    -     &    -    \\
$     9.40$ & $     9.50$ & $   7.662\pm  0.169$ & $   3.666\pm  0.079$ &    -     &    -     &    -    \\
$     9.50$ & $     9.60$ & $   6.885\pm  0.160$ & $   3.518\pm  0.077$ &    -     &    -     &    -    \\
$     9.60$ & $     9.70$ & $   6.063\pm  0.151$ & $   3.416\pm  0.076$ &    -     &    -     &    -    \\
$     9.70$ & $     9.80$ & $   5.443\pm  0.143$ & $   3.329\pm  0.075$ & $   3.299\pm  0.060$ &    -     &    -    \\
$     9.80$ & $     9.90$ & $   4.737\pm  0.133$ & $   3.015\pm  0.072$ & $   3.057\pm  0.058$ &    -     &    -    \\
$     9.90$ & $    10.00$ & $   4.154\pm  0.125$ & $   2.858\pm  0.070$ & $   2.860\pm  0.056$ &    -     &    -    \\
$    10.00$ & $    10.10$ & $   3.885\pm  0.120$ & $   2.633\pm  0.067$ & $   2.567\pm  0.053$ &    -     &    -    \\
$    10.10$ & $    10.20$ & $   3.358\pm  0.112$ & $   2.354\pm  0.063$ & $   2.434\pm  0.052$ &    -     &    -    \\
$    10.20$ & $    10.30$ & $   2.682\pm  0.100$ & $   2.220\pm  0.061$ & $   2.132\pm  0.048$ & $   1.956\pm  0.040$ &    -    \\
$    10.30$ & $    10.40$ & $   2.447\pm  0.096$ & $   1.958\pm  0.058$ & $   1.896\pm  0.045$ & $   1.726\pm  0.038$ &    -    \\
$    10.40$ & $    10.50$ & $   2.118\pm  0.089$ & $   1.650\pm  0.053$ & $   1.723\pm  0.043$ & $   1.394\pm  0.034$ &    -    \\
$    10.50$ & $    10.60$ & $   1.494\pm  0.075$ & $   1.506\pm  0.051$ & $   1.414\pm  0.039$ & $   1.206\pm  0.032$ &    -    \\
$    10.60$ & $    10.70$ & $   1.188\pm  0.067$ & $   1.240\pm  0.046$ & $   1.154\pm  0.035$ & $   0.952\pm  0.028$ & $   0.602\pm  0.020$\\
$    10.70$ & $    10.80$ & $   0.912\pm  0.058$ & $   1.041\pm  0.042$ & $   0.925\pm  0.032$ & $   0.732\pm  0.025$ & $   0.489\pm  0.018$\\
$    10.80$ & $    10.90$ & $   0.631\pm  0.049$ & $   0.834\pm  0.038$ & $   0.722\pm  0.028$ & $   0.480\pm  0.020$ & $   0.333\pm  0.015$\\
$    10.90$ & $    11.00$ & $   0.389\pm  0.038$ & $   0.539\pm  0.030$ & $   0.526\pm  0.024$ & $   0.336\pm  0.017$ & $   0.214\pm  0.012$\\
$    11.00$ & $    11.10$ & $   0.191\pm  0.027$ & $   0.327\pm  0.024$ & $   0.347\pm  0.019$ & $   0.181\pm  0.012$ & $   0.120\pm  0.009$\\
$    11.10$ & $    11.20$ & $   0.105\pm  0.020$ & $   0.162\pm  0.017$ & $   0.184\pm  0.014$ & $   0.094\pm  0.009$ & $   0.057\pm  0.006$\\
$    11.20$ & $    11.30$ & $   0.052\pm  0.014$ & $   0.061\pm  0.010$ & $   0.097\pm  0.010$ & $   0.033\pm  0.005$ & $   0.019\pm  0.004$\\
$    11.30$ & $    11.40$ & $   0.019\pm  0.008$ & $   0.020\pm  0.006$ & $   0.028\pm  0.006$ & $   0.017\pm  0.004$ & $   0.015\pm  0.003$\\
$    11.40$ & $    11.50$ & $   0.007\pm  0.005$ & $   0.017\pm  0.005$ & $   0.019\pm  0.004$ & $   0.004\pm  0.002$ & $   0.002\pm  0.001$\\
$    11.50$ & $    11.60$ &    -  & $   0.003\pm  0.002$ & $   0.005\pm  0.002$ &    -  & $   0.001\pm  0.001$\\
$    11.60$ & $    11.70$ &    -  &    -  & $   0.001\pm  0.001$ & $   0.001\pm  0.001$ & $   0.001\pm  0.001$\\
$    11.70$ & $    11.80$ &    -  & $   0.002\pm  0.002$ & $   0.002\pm  0.002$ &    -  &    - \\
$    11.80$ & $    11.90$ &    -  &    -  & $   0.001\pm  0.001$ &    -  &    - 
\enddata
\label{tab:bin_massfn_blue}
\end{deluxetable*}

\bibliographystyle{apj}

\end{document}